\documentclass[11pt,american]{article}
\usepackage{mathptmx}
\usepackage{helvet}

\usepackage[latin9]{inputenc}
\usepackage{color}
\usepackage{babel}
\usepackage{array}
\usepackage{cprotect}
\usepackage{booktabs}
\usepackage{url}
\usepackage{multirow}
\usepackage{amsmath}
\usepackage{amssymb}
\usepackage{stmaryrd}
\usepackage{graphicx}
\usepackage[letterpaper]{geometry}
\usepackage{setspace}
\usepackage[authoryear]{natbib}
\usepackage{microtype}
\usepackage[bookmarks=false,
 breaklinks=true,pdfborder={0 0 0},pdfborderstyle={},backref=false,colorlinks=false]
 {hyperref}

\makeatletter

\providecommand{\tabularnewline}{\\}

\usepackage{relsize}

\usepackage{natbib}
 \bibpunct[, ]{(}{)}{,}{a}{}{,}%

\usepackage{dcolumn}
\newcolumntype{d}[1]{D{.}{.}{#1}}

\usepackage[font=footnotesize,labelfont=bf]{caption}
\usepackage[noindentafter]{titlesec}

\titlespacing*{\section} {0pt}{2.5ex plus 1ex minus .2ex}{1.5ex plus .2ex}
\titlespacing*{\subsection} {0pt}{2.25ex plus 1ex minus .2ex}{1.5ex plus .2ex}
\titlespacing*{\subsubsection}{0pt}{2.25ex plus 1ex minus .2ex}{1.5ex plus .2ex}

\titlespacing*{\paragraph} {0pt}{1.5ex plus 1ex minus .2ex}{1em}

\makeatletter
\g@addto@macro\normalsize{%
  \setlength\abovedisplayskip{1ex plus .5ex minus .25ex}
  \setlength\belowdisplayskip{1ex plus .5ex minus .25ex}
  \setlength\abovedisplayshortskip{.75ex plus .25ex minus .125ex}
  \setlength\belowdisplayshortskip{.75ex plus .25ex minus .125ex}
}
\makeatother

\makeatother

\begin{document}
\title{Advertiser Learning in Direct Advertising Markets\thanks{The authors would like to thank seminar participants at the 9th Invitational
Choice Symposium (Noordwijk), the 2023 Conference on Artificial Intelligence,
Machine Learning, and Business Analytics, the 2024 Workshop on Platform
Analytics, Bocconi University, Cologne University, Duke University,
Erasmus University, Georgetown University, HEC, INSEAD, Rice University,
Santa Clara University, Stanford University, Tilburg University, the
University of British Columbia, the University of California Berkeley,
the University of Colorado, the University of Groningen, the University
of North Carolina, Charlotte, and Virginia Tech University as well
as Santiago Balseiro, Siddarth Prusty, Srinivas Tunuguntla, Nils Wernerfelt,
Boya Xu, and Levin Zhu for their comments and insights.}}
\author{Carl F. Mela\thanks{T. Austin Finch Foundation Professor of Business Administration at
the Fuqua School of Business, Duke University (email: mela@duke.edu).} \and Jason M.T. Roos\thanks{Associate Professor of Marketing at the Rotterdam School of Management,
Erasmus University (email: jroos@rsm.nl).}\and Tulio Sousa\thanks{Ph.D Candidate, Duke University Department of Economics (email: tuliosalvio.sousa@duke.edu).}}
\date{\today}

\maketitle
\thispagestyle{empty}
\begin{abstract}
Direct buy advertisers procure advertising inventory at fixed rates
from publishers and ad networks. Such advertisers face the complex
task of choosing ads amongst myriad new publisher sites. We offer
evidence that advertisers do not excel at making these choices. Instead,
they try many sites before settling on a favored set, consistent with
advertiser learning. We subsequently model advertiser demand for publisher
inventory wherein advertisers learn about advertising efficacy across
publishers' sites. Results suggest that advertisers spend considerable
resources advertising on sites they eventually abandon---in part
because their prior beliefs about advertising efficacy on those sites
are too optimistic. The median advertiser's expected CTR at a new
site is 0.177\%, four times higher than the true median CTR of 0.045\%.

We consider how an ad network's pooling of advertiser information
remediates this problem. As ads with similar visual elements garner
similar CTRs, the network's pooling of information enables advertisers
to better predict ad performance at new sites. Counterfactual analyses
indicate that gains from pooling advertiser information are substantial:
over six months, we estimate a median advertiser welfare gain of $\$3,621$
(an $18.3\%$ increase) and a median revenue gain of \$13,558 (a $77.7\%$
increase) among the 20 largest publishers.\textbf{\medskip{}
}

\end{abstract}
\thispagestyle{empty}

\newpage{}

\setcounter{page}{2}

\section{Introduction\label{sec:Introduction}}

Advertisers seek to place ads with publishers whose readers are potential
customers of the advertised goods. In the context of direct buy display
advertising, where advertisers buy ad placements directly from sites,
a myriad of sites from which to choose makes the publisher selection
problem daunting. Because sites differ in their readership and the
editorial context in which display ads are served, ad performance
can vary significantly, and advertisers lacking prior experience with
a given site are typically uncertain about the value of placing ads
on that site \citep{Perlich2012,tunuguntla_2022}. An uninformed advertiser
risks misallocating both its ad budget and its ad spend across sites.

In direct buy markets, advertisers pay a fixed price (as posted on
a rate card) to present an ad to all readers who visit the publisher's
site during a specified period lasting days or even weeks. \label{rev:frictions}Given
advertisers lack information about their advertising efficacy at specific
sites, and because this information is often expensive to attain,
advertisers are faced with costly learning. Advertisers in these markets
learn by doing \citep{Tadelis_et_al_2023}, and learning frictions
are induced by the cost of ad placements. In our data, for example,
a new advertiser who is ex-ante uncertain about the efficacy of an
ad buy at a site would typically have to spend \$800 to place the
ad on a publisher site, and only then would learn about how well its
ad performs on that site from the typical 821K impressions served.
Motivated by this problem, the key objective of this paper is to explore
the welfare implications of costly advertiser learning in direct buy
advertising markets. Our counterfactual analysis considers how the
ad network can mitigate these learning frictions by providing advertisers
forecasts of their ad performance at different sites.

Further, we measure the degree of advertiser learning and ascertain
the attendant welfare implications (for advertisers and publishers)
of potentially miscalibrated initial beliefs about the value or efficacy
of advertising across different sites. Although our emphasis is on
direct buy display markets, similar contexts include video, television,
retail media, print, radio, and other channels. In each of these settings,
advertisers often have little information about the efficacy of their
advertising in a particular channel prior to committing to a large
ad buy.

In this regard, this research makes several advances. First, much
of the canonical economic theory behind advertiser behavior assumes
that advertisers know a priori the value of ad impressions \citep{Varian2007,Athey2011,Choi_et_al_2020,Balseiro2015}.
In contrast, we do not presume that per-impression valuations are
known before buying ads on a site, but rather that they must be learned
based on ex post quality signals, such as the empirical click-through
rate (CTR). \label{rn:empirical-ctr-is-a-signal}If advertiser prior
beliefs are too optimistic (pessimistic), they will spend too much
(too little) on ads. With experience, however, their media buys should
become more efficient. A growing literature pertaining to automated
advertiser learning approaches has recently appeared \citep{Cai2017,Choi2019,Ren2018,Scott2010,Schwartz2017,Tunuguntlan_Hoban_2020,Waisman2019,Balseiro2019}.
Although these approaches imply advertisers should try to learn about
the value of ad impressions, these studies do not seek to i) demonstrate
whether advertisers do indeed learn, ii) measure advertisers' initial
beliefs about advertising efficacy, iii) assess the welfare implications
of those initial beliefs, or iv) explore the potential for an intermediary,
such as an ad network, to pool information across advertisers. Recently,
\citet{Tadelis_et_al_2023} address point i) cross-sectionally in
the context of an exchange network. They find that more experienced
advertisers enjoy greater lift from advertising, a result that is
consistent with advertiser learning. In contrast to their approach,
our research explicitly models the longitudinal learning process via
a structural learning model, explores the welfare implications of
learning, and suggests an approach to improve the efficiency of the
media buy. In general, there is a paucity of empirical evidence in
marketing and economics regarding advertisers learning by doing. \label{rn:learning-is-novel}As
noted by \citet{Tadelis_et_al_2023}, the learning by doing literature
has focused on the production of goods or services\emph{ }\citep[e.g.,][]{Benkard2000,Hendel2014,Levitt2013}.
Few, if any, papers explicitly measure the degree to which advertisers
learn and the attendant costs of learning.

\begin{figure}[t]
\begin{centering}
\includegraphics[width=0.6\columnwidth]{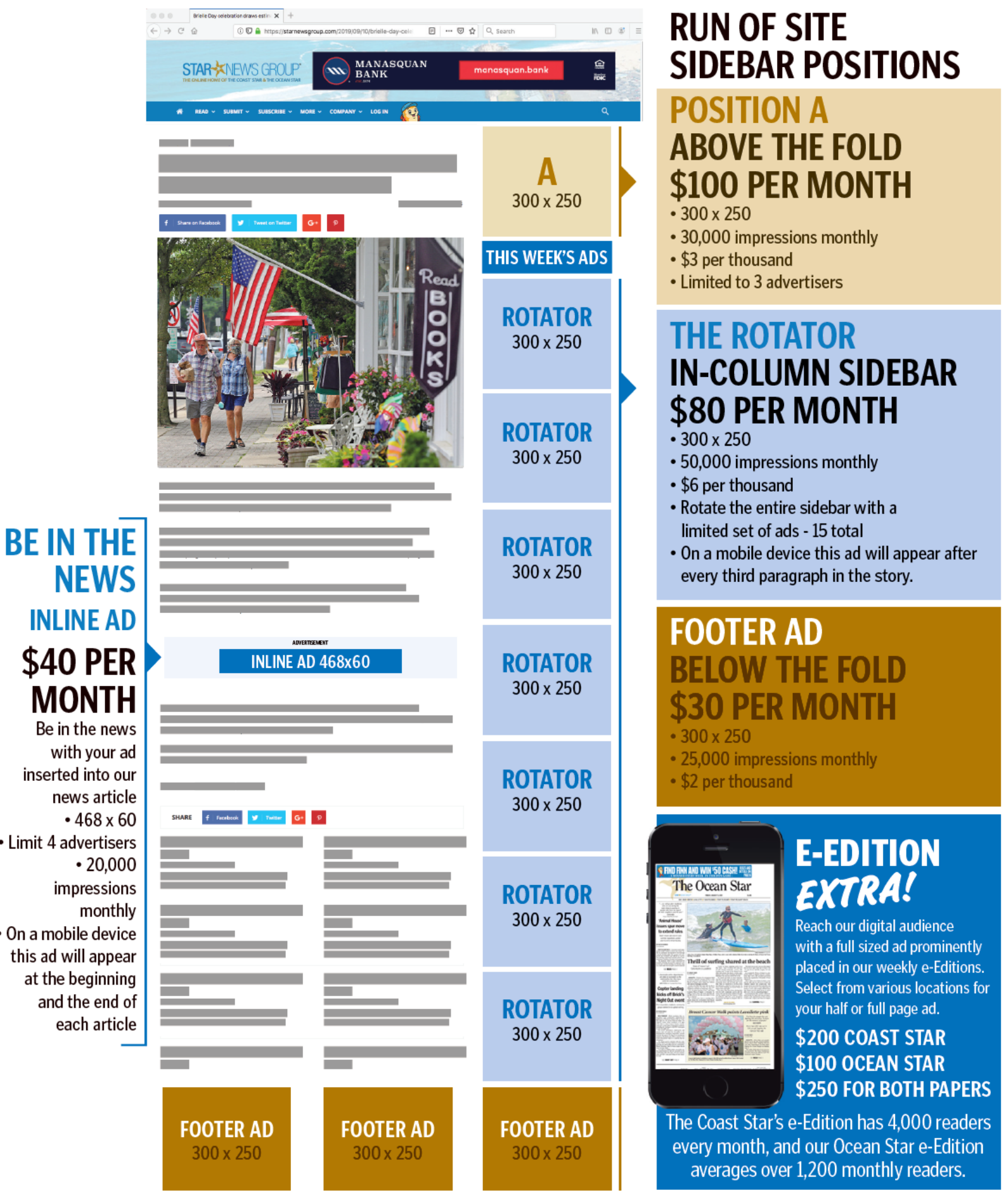}
\par\end{centering}
\cprotect\caption{2024 Star News Digital Advertising Rate Card. Source: Star News Group,
\protect\url{https://cdn.starnewsgroup.com/pdf/advertising/2024_SNG_Web_Advertising_Rates.pdf}
\protect\href{https://web.archive.org/web/20240318104528/https://cdn.starnewsgroup.com/pdf/advertising/2024_SNG_Web_Advertising_Rates.pdf}{(Archive.org)}.
\label{fig:2024-Star-News}}
\end{figure}

\label{rn:Our-second-advance}Our second advance is to consider the
setting of a direct buy ad network, as opposed to exchanges \citep{Allouah_Besbes_2017,Balseiro2017,Wu2015}.
According to eMarketer, direct buying represents 75\% of the annual
\$130B display ad market, with exchanges representing the balance.
In spite of this, most prior research in marketing has focused on
exchanges \citet{Choi_et_al_2020}.\footnote{\url{https://tinuiti.com/blog/performance-display/ana-programmatic-buying-guide-pov/}
\href{https://web.archive.org/web/20230929001048/https://tinuiti.com/blog/performance-display/ana-programmatic-buying-guide-pov/}{(Archive.org)},
\url{https://www.insiderintelligence.com/content/programmatic-ad-spending-forecast-h1-2024}
\href{https://web.archive.org/web/20240224101856/https://www.insiderintelligence.com/content/programmatic-ad-spending-forecast-h1-2024}{(Archive.org)}} In exchange markets, individual impressions are purchased in real
time and priced via an auction. In direct settings, impressions are
bundled and sold in bulk, often with a guarantee on the number of
days the ad will run, but not the precise number of impressions that
will eventually be served.\footnote{See \url{https://www.socialchimp.com/blog/direct-vs-programmatic-breakdown-media-buying/}\texttt{
}\href{https://web.archive.org/web/20230601095956/https://www.socialchimp.com/blog/direct-vs-programmatic-breakdown-media-buying/}{(Archive.org)}
and \url{https://newormedia.com/blog/ad-exchange-vs-ad-networks/}
\href{https://web.archive.org/web/20231202191412/https://newormedia.com/blog/ad-exchange-vs-ad-networks/}{(Archive.org)}.} Furthermore, direct inventory is often listed and sold at a fixed
price.\cprotect\footnote{See~\url{https://media-index.kochava.com/ad_partners?country=&channel\%5B\%5D=desktop-display&channel\%5B\%5D=native&category\%5B\%5D=ad-network&pricing_models\%5B\%5D=flat-rate}
\href{https://web.archive.org/web/20240318105311/https://media-index.kochava.com/ad_partners?country=&channel\%255B\%255D=desktop-display&channel\%255B\%255D=native&category\%255B\%255D=ad-network&pricing_models\%255B\%255D=flat-rate}{(Archive.org)}
for an example list of flat rate exchanges.} Figure \ref{fig:2024-Star-News} shows an example of the 2024 rate
card from the Star News Group, a small, local newspaper publisher
headquartered in Manasquan, New Jersey. An advertiser can purchase
a headline banner on the site for \$100 per month, and the Star News
Group estimates that this purchase would yield roughly 30,000 impressions
(larger papers, like The Guardian, charge upwards of \$100,000 for
a one-day take over of its homepage). A new advertiser at the Star
News Group must choose between not advertising with them, or spending
a relatively large sum of money (compared to a single impression sold
via exchange) in return for a relatively large number of impressions
of an uncertain value. In these environments, advertisers face an
overwhelming set of publishers of news, blogs, and other sites from
which to choose. After choosing to run an ad with a particular publisher,
advertisers effectively receive feedback on ad response in batches,
that is, only after large numbers of impressions are served. In our
data, the median ad buy is for 8 days, and the median number of impressions
from a single ad buy is about 820,000. The soonest that any information
about the value of ad impressions could impact subsequent choices
is after the initial ad buy.

The differences between fixed price contexts and ad exchanges are
consequential for how advertiser learning is modeled. In exchange
markets, feedback (in the form of KPIs such as clicks) arrives incrementally
by impression. In direct markets, a considerable amount of uncertainty
can be resolved between subsequent purchases, because each purchase
leads to many impressions, and the information about ad efficacy from
so many impressions overwhelms prior beliefs. \label{rn:also-friction}At
the same time, these large bundles of impressions make individual
transactions more costly, increasing frictions around learning. Under
these circumstances, obtaining better priors is more pressing than
test and learn, and there is little opportunity (if any) to explore
and exploit using exchange algorithms, such as those discussed in
\citet{Tunuguntlan_Hoban_2020}.\footnote{\label{fn:Because-direct-channels}Because direct channels, in contrast
to exchange markets, do not involve individual targeting of ads, there
are no ``test and learn'' algorithms used for targeting impressions
in direct buy markets. Therefore, advertiser learning behavior is
not algorithmically driven by direct exchanges.}

\label{rn:Hence,-our-third}Hence, our third contribution is to propose
a scalable approach to enhance advertiser priors about the value of
their impressions in order to improve their outcomes in the context
of direct ad networks. We propose a procedure whereby the ad network
pools its advertisers' information to improve advertisers' prior beliefs
about ad efficacy at new sites.\footnote{\label{fn:An-interesting-example}An interesting example of information
pooling in practice is Meta's training of models to forecast advertiser's
campaign outcomes such as reach and clicks. These tools pool information
across advertisers to generate forecasts to better inform advertiser
purchases. See \url{https://ai.meta.com/blog/ai-ads-performance-efficiency-meta-lattice/}
\href{https://web.archive.org/web/20240315073458/https://ai.meta.com/blog/ai-ads-performance-efficiency-meta-lattice/}{(Archive.org)}.} This raises the question of what information to use, as well as how
it should be shared with advertisers. Conceptually, similar advertisers
should evidence similar ad performance in the form of CTRs at the
same sites. However, a further consideration emerges about how to
measure similarity between advertisers (especially those who are new
to the ad network). \label{rn:bring-material-forward}One approach
to measuring similarity between advertisers is to consider advertising
copy or design, with the idea that any two advertisers producing similar
ad content should also generate similar ad responses \citep{Yao2023}.
Because our advertisements are images, it is possible to measure the
similarity of advertisers based on the image similarity of their display
ads. To accomplish this, we create a set of concept tags for each
ad in our data, and then compute a similarity score for each pair
of advertisers based on the number of tags they share in common. Then,
to improve advertisers' prior beliefs, we impute advertiser CTR at
each new site using a weighted average of \emph{other} advertisers'
CTRs at that site (with the weights determined by the advertiser similarity
scores). The approach thus builds on the cold start literature by
using machine learning image recognition techniques in the context
of an ad exchange \citep{Gope2017,Lika2014,Lam2008,Schein2002,Xu2022}.

Owing to our approach to sharing information, a fourth contribution
pertains to the literature on the role of network intermediaries in
advertising \citep{Choi_et_al_2020}. Typically, ad networks are construed
to reduce search frictions for advertisers, making it easier for them
to find publisher inventory. Yet ad networks can also improve match
by helping advertisers learn which sites are more effective. An open
question is whether publishers and the ad network have an incentive
to inform advertisers of prior match. On the one hand, if advertisers
are too optimistic in their initial beliefs about the efficacy of
sites they selected for advertising, better information would attenuate
advertiser spending and the network would have little incentive to
inform advertisers that they were overly optimistic. On the other
hand, to the extent substitute sites can be found with better match,
advertisers might increase spend. \label{rn:which-effect-dominates}Little
evidence exists to suggest which of the two effects dominates, and
whether networks have an incentive to share information on match values
with advertisers. Accordingly, we consider the role of information
sharing on advertiser, publisher, and ad network welfare.

We collect data from a direct sales ad network that consolidates direct
sales display ad inventory across multiple publishers (in this case,
the sites are blogs). As is common in the direct sales display advertising
channel, advertisers procure the publishers' ad inventory in advance
at a fixed rate. The data from this ad network are ideal for our empirical
context, because ad sales and ad prices are observed from the network's
inception and over a long duration (3 years). As a result, the behavior
of all advertisers is observed from the platform's infancy, providing
an ideal context in which to observe advertisers learning. Our empirical
strategy relies upon longitudinal changes in the advertisers' propensity
to choose particular sites, and this type of variation is common in
the data. The structural learning model we develop presumes that advertisers
choose the expected number of ad impressions across sites to maximize
their profits, conditioned on their (possibly incorrect) beliefs about
the efficacy of advertising on those sites. After an ad has run at
a site, the advertiser observes its CTR for that site. These CTRs
provide a noisy signal about the site's match with the advertiser's
ads. Sites with higher (lower) CTR's are more (less) likely to be
indicative of a good match, and thus sites with high CTRs are more
likely to be used again by the advertiser. When advertisers select
a site upon which to advertise, fail to obtain clicks, and then cease
to advertise, we reason they were too optimistic in their initial
beliefs about the efficacy of advertising on that site. We then use
the demand-side model estimates to gain insights about advertiser
conduct, and to simulate demand under counterfactual scenarios that
manipulate what advertisers know.

The data evidence patterns that are broadly consistent with learning,
such as advertisers initially trying many sites before settling on
a smaller number, and preferring to place additional ads at sites
that previously generated relatively higher CTRs. Results from a structural
model of advertiser demand suggest that advertisers are overly optimistic
in their initial beliefs about the efficacy of advertising. The median
advertiser's choices are consistent with an expected click through
rate of 0.177\% when, in practice, the median CTR is 0.045\%. The
finding suggests that advertisers often choose the wrong publisher
sites initially and overspend on them, while also potentially overlooking
sites that would have been a better match. With pooled information
provided by the platform, the median advertiser in the estimation
sample increases expected total spend by about $\$946$ (a $98.3\%$
increase over baseline spend) over six months, and generates an expected
$\$3,621$ ($18.3\%$ over baseline) in incremental value; 95\% of
advertisers see welfare gains. These welfare gains largely arise because
pooling allows advertisers to sort themselves into better matching
publisher sites. Owing to this better match, the median publisher
(among the top 20 by revenue) obtains an expected \$13,558 ($77.7\%$)
increase in revenue over six months. \label{rn:intro-platform-participation-incentive}Hence,
publishers and the ad network have an incentive to better inform advertisers
about match. Projecting these median effects across all advertisers
suggests an overall six-month welfare gain of approximately $\$7.5\mathrm{M}$.
Presumably, qualitatively similar gains could accrue to other direct
ad networks, as well as in related contexts such as retail media and
online video.

In what follows, we first overview the data and provide descriptive
evidence of advertiser learning. We then outline an advertiser learning
model, report the estimation results, detail the counterfactual analyses
related to information sharing, and conclude with a summary of our
findings and ideas for further research.

\section{Data}

\subsection{Data Overview}

To assess the effect of advertiser learning, it is necessary to observe
the advertising decisions of firms over time. To this end, we have
collected data provided by a fixed-price, direct-sales Internet ad
aggregator. The data span 3 years, starting with the aggregator's
inception in late July 2006 and ending in early December 2009 and
covers 8,000 advertisers and 3,200 publishers. These data record the
transactions between advertisers and publishers. Each transaction
corresponds with an individual advertiser's ad buy (commonly called
a \emph{subscription}), and specifies the number of consecutive days
(typically a week) the ad is to be shown to all visitors to the publisher's
website, and the price paid by the advertiser for that purchase. \label{rn:fixed-price}Notably,
the price paid depends on the length of the subscription, and not
on the eventual number of impressions served. Further, for each day
the ad is active, the data record the total number of \emph{impressions}---the
number of times the ad was served to a site visitor---and the number
of \emph{clicks}---the number of times a site visitor clicked on
the ad. Over the duration of the data, we are aware of no algorithmic
changes that might influence the nature of advertiser learning or
the allocation of advertisers to sites.

The estimation sample focuses on a subset of these data. \label{rev:closed-market}Specifically,
we consider a closed market, defined as a unique set of sites with
a unique set of advertisers. Using $K$-means clustering with a Jaccard
similarity metric computed from the overlap of firms advertising on
sites, we identify a subset of 165 politically liberal blogs and news
sites serving a similar set of advertisers that do not generally advertise
on other sites.\footnote{Similarity is higher between two sites if the number of advertisers
placing ads at both sites is higher. If no two advertisers have ever
placed ads at both sites, then their similarity is zero. The similarity
matrix is computed using i) sites with at least 9 advertisements (5th
percentile) and ii) advertisers with at least 20 advertisements placed
at 10 or more sites. Smaller advertisers and publishers are largely
inconsequential economically, and make the task of finding a unique
sets of adjacent sites impracticable.\label{fn:k-means-similarity}} This focal cluster of 165 publishers comprises 15.6\% of subscriptions
(ad buys) in the data, with 1801 (22.7\%) of the advertisers in the
data placing an ad at one or more of these sites. From this coherent
grouping of publishers, the top 20 are selected in terms of total
ad revenue using transactions conducted in the first half of 2007
(January 1--June 30). We then choose a random subset of 100 advertisers
who placed an ad at one or more of the top 20 sites, and aggregate
choices to the weekly level to comport with advertisers' typical purchase
frequency (rarely does interpurchase time fall short of a week). The
unit of analysis for our empirical model is thus at the advertiser-site-week
level. Over the 27 weeks in the first half of 2007, the estimation
sample comprises 547 ad subscriptions. We assume that advertisers
include the top 20 sites in their choice sets, unless a previously
purchased subscription is already running in a given week. Some advertisers
joined the ad network during the estimation window; for these we assume
choices begin with the week they first appear in the data. The estimation
sample thus comprises 36,911 choices leading to 547 purchases.

The daily number of impressions for each ad (in the period prior to
July 2007, including the second half of 2006) are used to impute the
daily average number of ad impressions at each site. This serves as
an approximation to sites' expected daily traffic. The site in the
estimation sample with the greatest number of daily visitors had a
peak audience of 910K daily visits in the first half of 2007, and
average daily traffic of 420K; the site with the least traffic peaked
at 43K, with average daily traffic of 20K. Prices vary accordingly,
with the most expensive placement garnering over \$755 per day (\$2265
for 3 days) and the least just \$5.50 per day (\$500 for 3 months).
A typical subscription in the estimation sample lasts one week, costs
\$800, and yields 821K impressions. We compute CTRs from the daily
number of clicks for each ad; the median click-through rate among
ads in the estimation sample is 0.045\%. As mentioned above, these
CTRs are presumed to provide noisy signals about the quality of ad
impressions.

\subsection{Descriptive Analyses\label{subsec:Descriptive-Analyses}}

\label{rev:descriptive-intro}To better frame our model development,
this subsection presents preliminary data analyses revealing patterns
in the data that are consistent with advertiser learning. First, we
document potential advertiser uncertainty about ad performance across
different sites. Uncertainty in ad performance implies that advertisers
have the opportunity to learn, which would not be the case if ad performance
were similar across sites. Second, we consider whether advertisers'
ex-ante ad placement decisions are ex-post optimal. To the extent
they are not, one would observe systematic differences in ad placement
decisions over time. Third, we consider whether advertisers resolve
uncertainty about ad performance from data generated by their own
campaigns. If they do, one would expect advertisers' decisions to
change in response to the performance of their most recent ad placements.

\subsubsection{Advertiser Uncertainty in Ad Placement Outcomes}

\begin{table}[!t]
\begin{centering}
\smaller%
\begin{tabular}{ld{4.0}d{0.4}d{2.1}}
\toprule 
Term & \multicolumn{1}{c}{DF} & \multicolumn{1}{c}{Sum of Squares} & \multicolumn{1}{c}{Variation Explained (\%)}\tabularnewline
\midrule 
Advertiser & 1783 & 0.0248 & 50.8\tabularnewline
Publisher & 164 & 0.0044 & 9.1\tabularnewline
Residuals & 8205 & 0.0196 & 40.1\tabularnewline
\bottomrule
\end{tabular}
\par\end{centering}
\caption{Variance Decomposition of Advertiser-Site Long-Run Click Through Rates.
Based on data from 1784 advertisers who placed one or more ads at
the focal set of 165 sites. The 40.1\% residual variance reflects
differences in advertisers' CTRs when placing ads at different sites.}
\label{tab:ctr_anova}
\end{table}

Our first conjecture is that there exists something for the advertiser
to learn; that is, they are uncertain about the true value of ad impressions
at a given site. If CTRs are homogeneous across sites, then ad effectiveness
is unlikely to vary across publisher sites, and there should be little
uncertainty about the value of ad impressions and therefore little
value to learning. Toward this end, Table \ref{tab:ctr_anova} reports
a variance decomposition of CTRs across advertisers and sites. This
analysis suggests that 9.1\% of the variation in CTRs can be apportioned
to publishers, 50.8\% to advertisers, with the remainder idiosyncratic
to the advertiser-site pair. The 40.1\% of variation in CTRs apportioned
to the advertiser-publisher residual suggests there exists ex-ante
uncertainty in match outcomes, and thus a potential benefit to advertisers
finding sites that yield better outcomes.

\subsubsection{Ex-ante Advertisers Decisions Evolve}

To the extent that advertisers' ex-ante decisions about ad placement
are not ex-post optimal, our second conjecture suggests that advertiser
placements should change over time. In the context of a learning model,
this would imply that the set of sites used by advertisers changes
with time. Two ways to measure this change include i) the number of
sites used by advertisers, and ii) the overlap in the set of sites
used by advertisers over time. A signal of advertiser learning would
be convergence to a narrower, stabler set of sites used for advertising.

\begin{figure}[!t]
\includegraphics[width=1\textwidth]{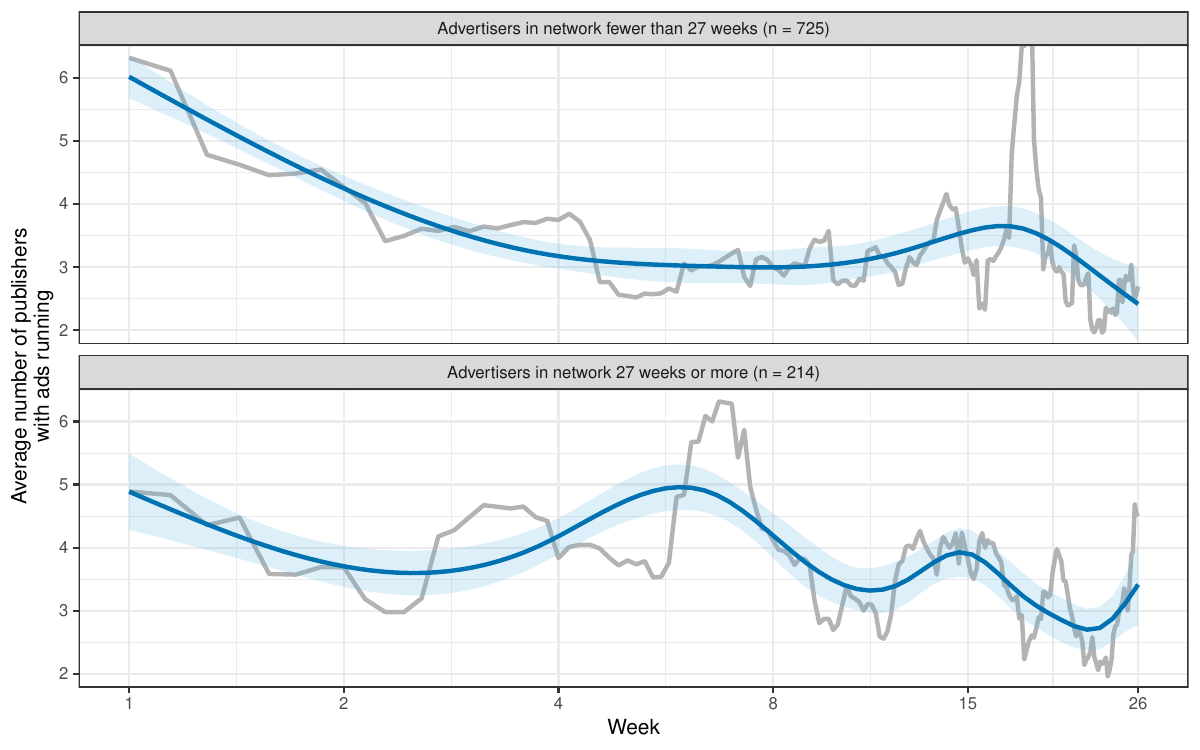}

\caption{Average Daily Number of Sites Running Ads by Advertisers' Continued
Use of the Network. The first 26 weeks of ads served by the focal
cluster of 165 sites are shown for 939 advertisers who placed ads
with two or more of these publishers. The top panel shows data from
advertisers who placed their last ad within the first 26 weeks, the
bottom panel advertisers who remained in the network longer past week
26. The line (and smoothed average) depicts the average daily number
of publishers serving ads for advertisers actively running ads. The
timing of ads is normalized relative to when the advertiser joined
the ad network. For example, if advertiser A first advertises in week
$t$ and advertiser B first advertises in week $t+k$, then week 1
in the figure depicts A's week $t$ decisions and B's week $t+k$
decisions. The plot is truncated to show weeks 1--26, and logarithmically
scaled to emphasize changes during the initial few weeks of advertising.}
\label{fig:sites_by_day}
\end{figure}

Figure \ref{fig:sites_by_day} depicts changes in the number of sites
concurrently used by advertisers running ads. For each advertiser
that placed at least one ad with one of the 165 liberal blogs and
news sites, we calculate the number of sites on which the advertiser
actively ran ads each day. To account for different advertiser cohorts
commencing their advertising on different dates, start dates are normalized
to the first ad placement for these time-series. For each day, we
calculate the average number of sites actively used by each advertiser,
conditional on that advertiser running at least one ad (because subscriptions
are typically 7 days or more, observations prior to 7 days are truncated).
At the end of the first week, advertisers have ads running, on average,
at about 6 sites. But over the next month, the number of sites used
drops by almost half. Over time, many advertisers cease advertising
entirely. This raises the possibility that the pattern observed in
Figure \ref{fig:sites_by_day} arises due to selection, and not due
to changes in advertisers' decisions. To alleviate this concern, Figure
\ref{fig:sites_by_day} contains two panels containing data from advertisers
whose last purchase occurred within the first 26 weeks (top panel)
and advertisers who remain active in the network beyond the period
depicted in the figure. The pattern evident in both panels is predicted
by learning, as advertisers begin with a larger set of sites and eventually
stop placing ads at less effective sites.\footnote{\label{fn:Shrinking-budgets-are}Shrinking budgets are another potential
explanation for a reduction in the number of sites used by advertisers
over time, but unlike learning, shrinking budgets do not explain the
broader patterns in the data. First, shrinking budgets alone cannot
explain why advertisers are more (less) likely to repeat buy at publisher
sites with lower (higher) costs per click. Second, lower ad budgets
cannot explain the tendency of advertisers in the data to explore
new sites over time, as opposed to simply cutting sites (by week 26
since joining the ad network, nearly 1/3 of all ad purchases are on
new sites). Third, diminished budgets cannot explain why many advertisers
spend more over time, as 35\% of the 853 advertisers buying ads on
two or more occasions spent more in the second half of their tenure
using the network.}

\begin{figure}[!t]
\includegraphics[width=1\columnwidth]{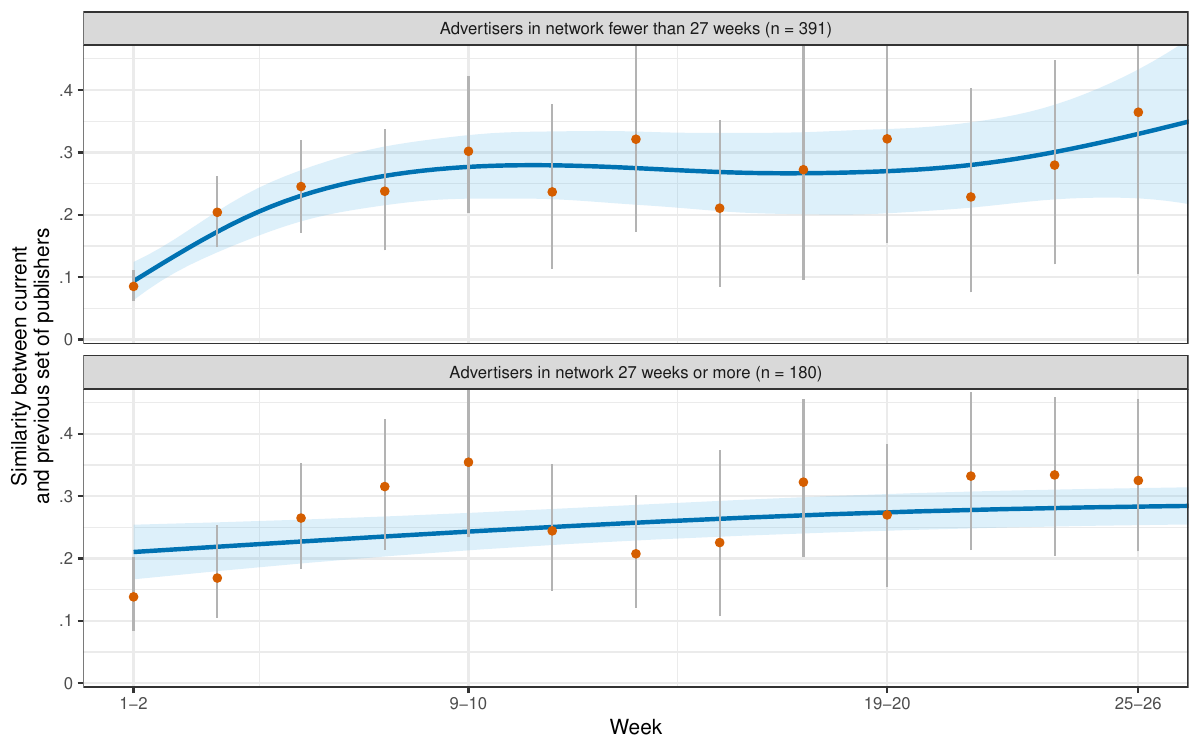}\caption{Persistence in Set of Publishers Chosen by Advertisers' Continued
Use of the Network. The first 26 weeks of ads purchased from the focal
cluster of 165 sites are shown for 571 advertisers who placed ads
with two or more publishers of these publishers in two or more two-week
periods. The top panel shows data from advertisers who placed their
last ad within the first 26 weeks, the bottom panel advertisers who
remained in the network longer past week 26. The timing of ad buys
is normalized relative to when the advertiser joined the ad network.
For example, if advertiser A first advertises in week $t$ and advertiser
B first advertises in week $t+k$, then week 1 in the figure depicts
A's week $t$ decisions and B's week $t+k$ decisions. Similarity
is calculated as the Jaccard coefficient between i) the set of publishers
used in a two-week period, and ii) the set of publishers from the
most recent, previous two-week period with an ad buy. Points (vertical
lines) indicates the average (bootstrap 95\% CI's) similarity in the
set of sites used among advertisers purchasing ads.\label{fig:Persistence-in-Site}}
\end{figure}

Figure \ref{fig:Persistence-in-Site} depicts how the set of sites
at which advertisers place ads changes over time. Similar to Figure
\ref{fig:sites_by_day}, the data are separated into two panels to
account for any potential selection effects. Similarity is calculated
as the Jaccard coefficient between two sets of publishers, over rolling,
bi-weekly periods. The first set contains all publishers used by an
advertiser in the focal bi-week. The second set contains all publishers
used by an advertiser in the previous bi-week (exclusive of periods
with no purchases). This coefficient, defined as the cardinality of
the intersection of the two sets divided by the cardinality of the
union of the two sets, ranges from 0 to 1, with a higher number meaning
that the set of sites in the current and preceding buy show greater
overlap. Over time, the set of sites an advertiser places ads at is
more likely to resemble the previous set, as the Jaccard coefficient
triples from 0.1 to 0.3. Note that it does not necessarily follow
that the Jaccard index must increase as set sizes decrease, meaning
that Figure \ref{fig:sites_by_day} and Figure \ref{fig:Persistence-in-Site}
are not isomorphic.

The analyses depicted in Figures \ref{fig:sites_by_day} and \ref{fig:Persistence-in-Site}
imply that advertisers decisions about ad placements change over time.
Specifically, the data indicate a convergence towards smaller, more
stable sets of publishers. If advertisers' ex-ante placements were
ex-post optimal, one would not expect to see these patterns.

\subsubsection{Resolving Advertiser Uncertainty Through Experience}

\begin{figure}[t]
\includegraphics[width=1\columnwidth]{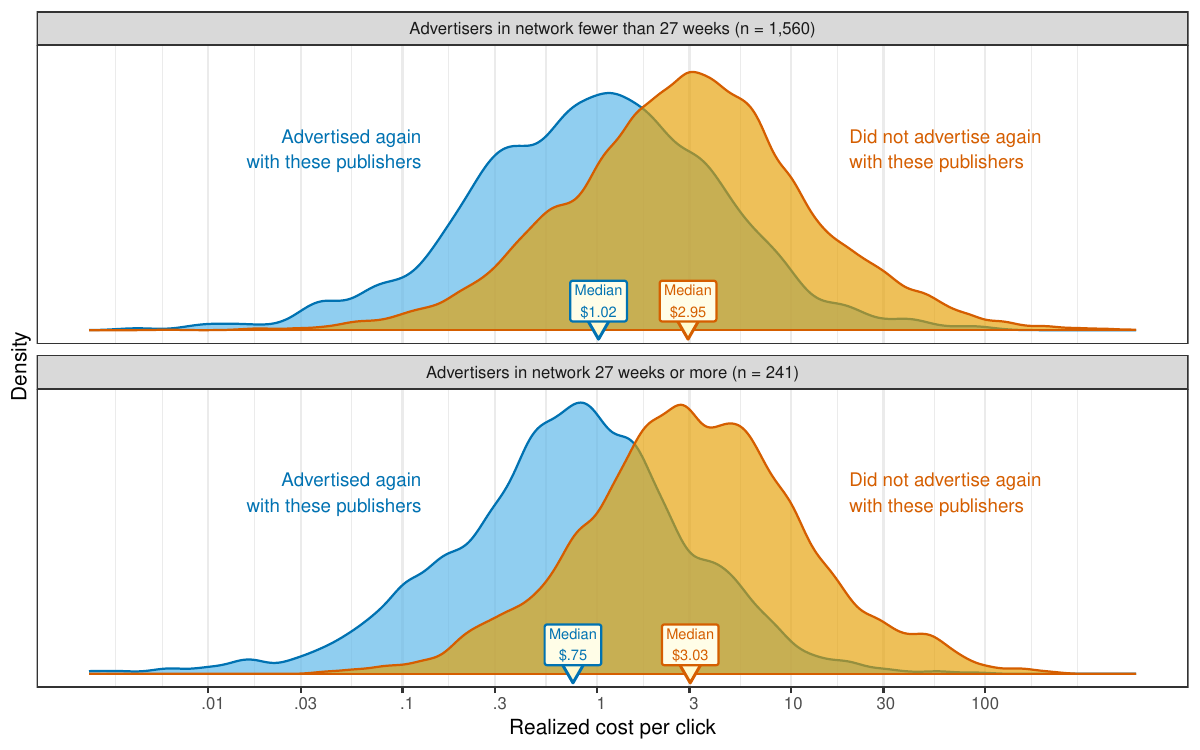}

\caption{Distribution of Realized Cost per Click, Grouped by Action After First
Ad Run is Completed, and Advertisers' Continued Use of the Network.
Based on 1,801 advertisers who ran ads with any publisher in the focal
cluster of 165 sites, showing only ads placed by these advertisers
at these sites. Cost per click is cumulative through last observed
ad placement. Medians are indicated.}
\label{fig:cpc_density}
\end{figure}

Another marker of learning, as indicated in our third conjecture,
would be advertisers resolving uncertainty about advertising outcomes
from data generated by their own campaigns. To visualize this phenomenon,
Figure \ref{fig:cpc_density} shows the distribution of \emph{realized}
cost per click (CPC) for all ads placed at the subset of 165 liberal
blogs and news sites by 1,801 advertisers. Because ads are placed
on a fixed-price basis, CPC is an outcome that is only observed \emph{after}
the ad has run. Hence, after comparing the number of clicks and impressions
an ad received against the amount of spent on ad placements, the advertiser
can calculate the effective CPC for each site they used. If the advertiser
never used a particular site again, then the realized cost per click
is depicted in the distribution to the right of both panels in Figure
\ref{fig:cpc_density} (labeled ``Did not advertise again with these
publishers''). If the advertiser did place additional ads at a site,
then the realized CPC is included in the distribution to the left.
The difference between these two distributions is consistent with
advertisers abandoning sites where the returns to advertising (as
reflected in the ex-post number of clicks received relative to the
ex-post number of impressions) are insufficiently high to justify
the cost.\footnote{\label{fn:We-conduct-a}A similar analysis is conducted at the advertiser
level to control for unobserved heterogeneity. Specifically, we compute,
for each advertiser the mean difference in realized cost per click
between sites chosen again and sites not chosen again (for advertisers
where both types of choices are observed). The analysis yields a median
difference across advertisers of \$2.20 (versus implied differences
of \$1.93 and \$2.28 in Figure \ref{fig:cpc_density}).}

In sum, there is heterogeneity in ad performance across sites. Against
this backdrop, we see the number of sites chosen by advertisers decreasing,
but the set of publishers within those decreasing sets exhibiting
greater consistency over time. Moreover, advertisers appear to be
more likely to advertise again with sites that yielded relatively
better CTRs relative to the cost of advertising. Collectively, these
behaviors would be consistent with advertiser learning behavior. In
the next section, we outline the specific approach used to model these
learning behaviors.

\section{Model and Estimation\label{sec:Model-and-Estimation}}

This section details the model of advertisers' site choice, and how
learning affects these choices. It concludes by detailing our estimation
approach.

\subsection{Advertising Payoffs\label{subsec:Payoff-from-Advertising}}

Each week $w$, advertiser $a$ considers placing an ad at each of
the top 20 sites, $s$, where the advertiser does not already have
an ad running. When choosing to place an ad at site $s$, the advertiser
considers a menu of potential subscription lengths (i.e., how many
days the ad is to run), $x$, and prices, $p$, that are charged by
sites. For example, an advertiser considering site $s$ might have
the following options: $\{$7 days for \$100, 14 days for \$180, 30
days for \$350, not advertising at site $s$$\}$. \label{rn:participation}The
option not to advertise is always available, and is represented as
a subscription of length $x=0$ days, obtained at a price of $p=0$.

The expected payoff generated by the ad running for $x$ days at site
$s$ starting at week $w$ depends on the expected number of site
visitors who will be shown the ad. Our formulation is intended to
be consistent with the display advertising literature, which typically
construes advertisers as valuing ad impressions rather than clicks
\citep{Choi_et_al_2020}. The average daily number of visitors exposed
to ads at site $s$ is denoted $t_{s}$ and is common knowledge to
all sites and advertisers. Because all site visitors are served ads
for all active subscriptions, an ad running for $x$ days at site
$s$ yields $t_{s}x$ total expected impressions. Advertiser $a$'s
expected payoff from an ad running for $x$ days at site $s$ is given
as follows:
\begin{gather}
\pi_{asw}\left(x\right)=\delta_{asw}\,\zeta_{a}^{-1}\,\log\left(1+t_{s}x\zeta_{a}\right)-p_{sw}(x)\label{eq:advertiser_payoff}
\end{gather}
where the first additive term represents the advertiser's valuation
from the expected advertising outcome, and the second term ($p_{sw}(x)$)
represents the price the publisher charges to the advertiser.\footnote{\label{fn:Because-we-observe}We follow \citet{Berry1992} and \citet{Bajari2007}
and incorporate ad costs directly in the payoff function without a
price coefficient. The corresponding interpretation of the first term
in Equation (\ref{eq:advertiser_payoff}) as a dollar valuation is
consistent with the display ad literature in exchange markets, where
bid payments are expressed in monetary terms, and bid valuations are
interpreted in a dollar metric \citep{Ahmadi2023,Alcobendas2021,bompaire2021causal,Tunuguntlan_Hoban_2020,Waisman2019}.
To the degree a price coefficient were to exist and be normalized,
it would be absorbed into the advertiser fixed-effect contained within
$\delta_{asw}$ (described in Equation (\ref{eq:conditional_match})),
which is multiplicatively separable in the advertiser value function
in the first part of Equation (\ref{eq:advertiser_payoff}).} Choosing not to advertise yields a payoff of $\pi_{asw}\left(0\right)=0$.
This function for advertiser valuation is a special case of the generalized,
translated, constant elasticity of substitution utility function discussed
in \citet{Bhat2008} and \citet{Lee2014}. The parameter $\zeta_{a}>0$
determines the rate at which the advertiser satiates on additional
impressions \citep{dube2005empirical}. \label{rev:satiation}The
satiation parameter implies a tradeoff between lower prices and more
impressions. While longer subscriptions result in more impressions
at a lower price, the marginal value of those impressions is diminishing.
To the extent the marginal value of impressions is lower than the
marginal decrease in prices, advertisers prefer shorter subscriptions.
The term $\delta_{asw}>0$ determines the overall scale of payoffs,
and in particular, the change in marginal payoff at the point of zero
impressions.\footnote{\label{fn:Extending-the-approach}Extending the approach from a per
subscription to a per impression selling model is straightforward.
In this case, the right hand side of equation \ref{eq:advertiser_payoff}
is replaced with $\delta_{asw}\,\zeta_{a}^{-1}\,\log\left(1+x\zeta_{a}\right)-p_{sw}x$
where $p_{sw}$ is the per impression price and $x$ is the number
of impressions purchased.}

The term $\delta_{asw}$ is subscripted by both advertiser and site,
reflecting the empirical regularity that the same number of impressions
served by two similar sites can generate different payoffs for the
same advertiser \citep{Perlich2012}. This term can be interpreted
as translating the log number of expected impressions to advertiser
monetary value. For example an ad that generates a higher rate of
sales conversions than another would lead to more advertiser revenue.
\label{rev:learn_about_match} The $\delta_{asw}$ term represents
the \emph{match }between the advertiser's ad content and the site's
audience. A higher match value leads to higher expected returns, and
thus a higher likelihood of advertising at site $s$. A priori, the
value of $\delta_{asw}$ is uncertain to the advertiser, but it can
be learned over time, as described in Section \ref{subsec:Advertiser-Site-Match-and}.
Were the advertiser to be overly optimistic about its match with site
$s$, it would initially advertise too much at that site. Were the
advertiser to be overly pessimistic, it would advertise too little.
Hence, there is value in learning efficiently about match.

\subsection{Advertiser-Site Match and Learning\label{subsec:Advertiser-Site-Match-and}}

We next discuss how advertisers learn about their match value with
each site. Clicks and impressions are assumed to provide an unbiased
signal about match, under the presumption that a higher CTR reflects
greater interest in the advertised good. Hence, we first link match
($\delta_{asw}$) to CTRs ($c_{as}$), and then show how changes in
beliefs about CTRs (i.e., learning) translate into changes in beliefs
about match.

\subsubsection{Linking Clicks to Match}

The proportion of site $s$'s audience who click on advertiser $a$'s
ads---that is, the true CTR---is a noisy measure of the general
effectiveness of advertiser $a$'s ad impressions when served to site
$s$'s audience. We denote the true CTR as $c_{as}$, and note that
this rate is unknown to both the advertiser and the site. Conditional
on this unknown quantity, the true match between advertiser and site
is a function of their true CTR, as given by
\begin{gather}
\delta_{asw}\big|c_{as}=\frac{c_{as}}{\gamma_{a}}\exp\left(\xi_{a}+\eta_{s}+\phi_{\tau[a,s,w]}+\psi_{m[w]}+\epsilon_{asw}\right).\label{eq:conditional_match}
\end{gather}
\label{rev:value-of-CTR}Equation \ref{eq:conditional_match} implies
that CTRs enter the value function indirectly in Equation \ref{eq:advertiser_payoff}
via the match term $\delta_{asw},$ which translates expected impressions
to advertiser monetary value. Hence, the model implies that advertisers
value impressions directly, not CTRs \citep{Choi_et_al_2020}. That
is, higher site CTRs imply that advertisers are more likely to value
those sites' \emph{impressions}. For a given number of impressions,
a higher site CTR corresponds with receiving more clicks. As clicks
are frequently antecedent to sales, higher site CTRs typically correspond
with greater monetization, and thus higher advertiser-site match.
Higher CTRs can also reflect greater interest or engagement with the
advertised product, and thus signal higher efficacy per impression
for brand awareness.

\label{rev:fixed-effects}Other advertiser and site aspects are embedded
in the match term, $\delta_{asw}$. \label{rn:first-the-term-xi}On
the advertiser side, some firms have higher margins or conversion
rates. These effects would be captured via an advertiser-specific
match component because greater margins or conversion rates increase
profits and revenues. As we do not observe sales, margins, or conversions,
we control for them using advertiser fixed effects. Specifically,
the term $\xi_{a}$ is an advertiser-specific fixed effect that is
common across sites.\footnote{Note that equation \ref{eq:advertiser_payoff} rescales the price
coefficient to one. As a result, any advertiser-specific price effects
are embedded in $\exp(\xi_{a})$.} On the publisher side, sites often post information about their audience
demographics. Consider a site with a wealthy audience that is more
likely to convert, leading to greater revenues for all advertisers
placing ads at that site. This wealth-audience effect suggests a higher
site-specific match component for all advertisers. The parameter $\eta_{s}$
reflects such differences in the value of site impressions that are
shared by all advertisers.\footnote{\label{fn:Equation--accommodates}Equations (\ref{eq:advertiser_payoff})
and (\ref{eq:conditional_match}) accommodate a restricted degree
of heterogeneity in satiation from additional impressions at different
sites. This is because the term in Equation (\ref{eq:advertiser_payoff})
representing the value from advertising is proportional to $c_{as}\exp(\xi_{a}+\eta_{s})\,\zeta_{a}^{-1}\,\log\left(1+t_{s}x\zeta_{a}\right)$.
At the point of no advertising, the marginal value of the first impression
is proportional to $c_{as}\exp\left(\xi_{a}+\eta_{s}\right)$; and
at the point of an expected $t_{s}x$ impressions, the marginal value
of an additional impression is proportional to $c_{as}\exp\left(\xi_{a}+\eta_{s}\right)\left(1+t_{s}x\zeta_{a}\right)^{-1}$.
Moreover, as advertisers learn about their true CTRs ($c_{as}$),
the extent of satiation is expected to change. Following the broader
literature on learning models in marketing and economics, the satiation
parameter is not assumed to be fully unrestricted over advertisers
and time, $\zeta_{asw}$. Relaxing this restriction could prove a
useful extension to the learning literature.}\label{rn:phi-tau}Although, more generally, any ad-site-week observable
covariates could be incorporated into the conditional match expression
for $\delta_{asw}$, we only observe $\phi_{\tau[a,s,w]}$, a fixed
effect that is selected according to how long ago advertiser $a$
first placed an ad at site $s$.\footnote{\label{fn:There-are-separate}There are separate fixed effects for
each of the following groups of weeks since first placing an ad at
a given site (the fixed effect for $\tau=0$ is normalized to 0):
$\tau\in\left\{ 0,1\text{\textendash}4,5\text{\textendash}8,9\text{\textendash}12,13\text{\textendash}16,17\text{\textendash}20,21\text{\textendash}24,25\text{\textendash}33,34\text{\textendash}41,42+\right\} $.
As $\phi_{1-4}$ includes the first four weeks since an ad first appeared,
it might be correlated with lagged advertising, and thus endogenous
(all other $\phi$ are invariant to lag purchase). The correlation
with lag advertising is $0.22$ and the correlation with clicks and
impressions are $-0.09$ and $-0.10$, suggesting the inclusion of
this control variable has little potential to bias the structural
parameters. Another interpretation of $\phi$ is that it controls
for wear-out in CTRs when ad content satiates, as might happen if
creatives never change, in which case $\delta_{asw}\big|c_{as}\propto c_{as}/\widetilde{\gamma}_{a}$,
with $\tilde{\gamma_{a}}\equiv\gamma_{a}/\exp(\phi_{\tau[a,s,w]})$
interpreted as a satiation-adjusted threshold.} The purpose of this fixed effect is to reflect dynamics in the value
or efficacy of advertising that are unrelated to learning about match
(e.g., wear-in or wear-out; \citealp{Little1969}).\footnote{A log-log regression of the cumulative number of creatives used over
time at a site on the number of cumulative subscriptions purchased,
controlling for site and advertiser fixed effects, yields a coefficient
of 1.11 ($t=199.4,$$se=0.006)$ implying a 1.11\% increase in the
number creatives used for each 1\% increase in subscriptions purchased
at a site. Hence copy wear-out from seeing the same creative over
time is not likely to be substantial.\textcolor{red}{{} \label{fn:copy-wearout}}} The term $\psi_{m[w]}$ is a month fixed effect, included to account
for any seasonal dynamics affecting the entire ad network. Finally,
$\epsilon_{asw}$ is an idiosyncratic demand shifter at the advertiser-site-week
level that is observed by the advertiser and not the econometrician.

\subsubsection{Learning About Click-Through Rates\label{subsec:Learning-About-Clicks}}

The expression $c_{as}\big/\gamma_{a}$ indicates the ratio of advertiser
$a$'s true (but a priori unknown) CTR at site $s$ to the parameter
$\gamma_{a}$. The parameter $\gamma_{a}$, which is known to the
advertiser but not the econometrician, serves as both i) advertiser
$a$'s prior expectation for its unknown CTR at any new site, and
ii) the baseline for judging how successful its advertising has been.
For example, if the true CTR at site $s$ turns out to be greater
than what the advertiser previously expected, then $c_{as}\big/\gamma_{a}$
will be greater than $1$, reflecting better than anticipated returns
to advertising at site $s$. To the extent initial beliefs about click
through rates are too high (low) at a given site, one would expect
advertisers' initial probabilities of advertising to be higher (lower).

We assume prior beliefs about the true CTR, $c_{as}$, denoted $\tilde{c}_{as}$,
follow a beta distribution, owing to the beta's flexibility in representing
distributions bounded between 0 and 1. Specifically, 
\begin{gather}
\tilde{c}_{as}\,\big|\,\gamma_{a}\,\sim\mathrm{Beta}\left(1,\left(1-\gamma_{a}\right)\big/\gamma_{a}\right),\label{eq:prior_ctr}
\end{gather}
which implies $\mathbb{E}\left[\tilde{c}_{as}|\gamma_{a}\right]=\gamma_{a}$,
and places the bulk of probability density between 0 and $4\gamma$.
Hence, $\tilde{c}_{as}$, like most CTRs, is a priori skewed towards
0. The parameter $\gamma_{a}$ reflects advertiser $a$'s prior belief
about its CTR at a previously unused site.

Conditional on the true CTR, $c_{as},$ the likelihood for the cumulative
number of clicks as of week $w$, $n_{asw}^{C}$, follows a binomial
distribution with likelihood
\begin{gather}
n_{asw}^{C}\,\big|\,n_{asw}^{I},c_{as}\,\sim\mathrm{Binomial}\left(n_{asw}^{I},c_{as}\right),\label{eq:views_and_clicks}
\end{gather}
where $n_{asw}^{I}$ is the cumulative number of advertising impressions
served. The number of clicks received, $n_{asw}^{C}$, depends on
the true CTR at site $s$, $c_{as}$, hence impressions and clicks
provide information about the true, but unknown CTR. \label{rn:beta-distribution}Accordingly,
we specify a Bayesian updating process on advertiser beliefs about
the CTR, leading to a $\mathrm{Beta}\left(1+n_{asw}^{C},(1-\gamma_{a})/\gamma_{a}+n_{asw}^{I}-n_{asw}^{C}\right)$
posterior CTR belief distribution. The mean of this updated posterior
distribution is
\begin{gather}
\mathbb{E}\left[\tilde{c}_{as}\,\big|\,n_{asw}^{I},n_{asw}^{C},\gamma_{a}\right]=\gamma_{a}\frac{1+n_{asw}^{C}}{1+\gamma_{a}n_{asw}^{I}},\label{eq:updated_ctr}
\end{gather}
and the expected ratio of CTR to $\gamma_{a}$ is $\mathbb{E}\left[\tilde{c}_{as}/\gamma_{a}\,\big|\,n_{asw}^{I},n_{asw}^{C},\gamma_{a}\right]=(1+n_{asw}^{C})/(1+\gamma_{a}n_{asw}^{I})$.

\begin{figure}[t]
\noindent\includegraphics[width=1\textwidth]{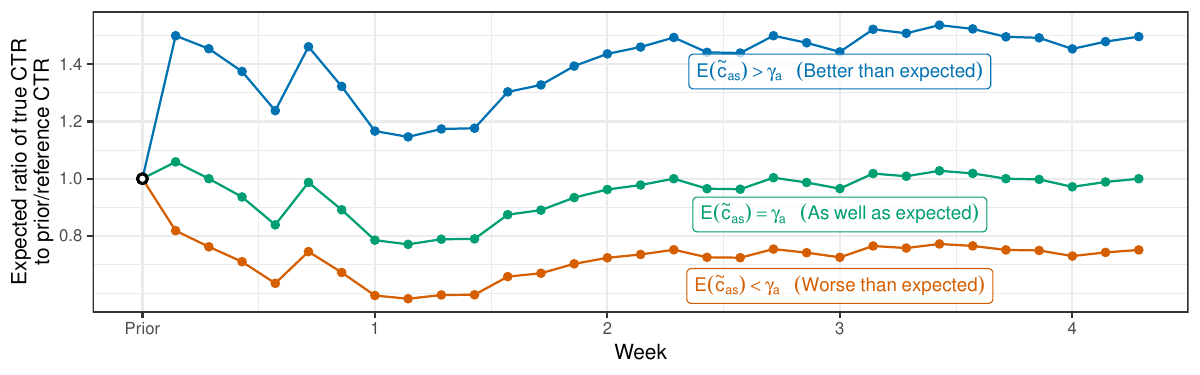}

\caption{Simulated Posterior CTR Beliefs. The evolution of posterior beliefs
for $\mathbb{E}\left[\tilde{c}_{as}/\gamma_{a}\,\big|\,n_{asw}^{I},n_{asw}^{C},\gamma_{a}\right]$
is simulated under a true click through rate of $0.0075$. The advertiser
prior is either pessimistic (blue plot, $\gamma_{a}=0.005$), accurate
(green plot, $\gamma_{a}=0.0075$) or optimistic (red plot, $\gamma_{a}=0.01$).}
 \label{fig:Simulated-Posterior-CTR}
\end{figure}

As previously noted, Equation (\ref{eq:updated_ctr}) implies that
the advertiser initially expects a CTR of $\mathbb{E}\left[\tilde{c}_{as}\big|\gamma_{a}\right]=\gamma_{a}$
prior to advertising (when $n_{asw}^{C}=0$ and $n_{asw}^{I}=0$).
As $n_{asw}^{C}\rightarrow\infty$ and $n_{asw}^{I}\rightarrow\infty$
, the ratio $n_{asw}^{C}/n_{asw}^{I}\rightarrow c_{as}$, and thus
$\mathbb{E}\left[\tilde{c}_{as}\big|\gamma_{a}\right]\rightarrow c_{as}$.
Hence, initial beliefs about the CTR begin as $\gamma_{a}$, but eventually
converge with enough advertising to the true CTR, $c_{as}$. If initial
beliefs are correct, the expected value of the ratio $\tilde{c}_{as}\big/\gamma_{a}$
remains at one and never changes (in expectation). If initial beliefs
are low (high), this ratio increases above (decreases below) one.
For example, if the true CTR is less than $\gamma_{a}$, then the
number of clicks received, $n_{asw}^{C}$, will grow at a slower rate
than the a priori expected number of clicks received, $\gamma_{a}n_{asw}^{I}$.
Hence $\gamma_{a}$ in Equation (\ref{eq:updated_ctr}) will be multiplied
by a ratio that converges to a value between 0 and 1. If initial beliefs
are too high, the advertiser spends too much on advertising in early
periods, with the added expense not recovered by sales. Conversely,
if initial beliefs are too low, the advertiser advertises too little,
or not at all. Figure \ref{fig:Simulated-Posterior-CTR} shows a hypothetical
learning process for the ratio of CTR to $\gamma_{a}$ for optimistic,
accurate, and pessimistic prior beliefs, and illustrates how these
beliefs change over time.

\subsubsection{Learning About Match}

The expression for true match values in Equation (\ref{eq:conditional_match})
is conditioned on the true CTR, $c_{as}$. Taking the expectation
of $\delta_{asw}$ in Equation (\ref{eq:conditional_match}) with
respect to the posterior distribution of $\tilde{c}_{as}$ , the advertiser's
updated expected match with site $s$, $\tilde{\delta}_{asw}$, is
given by
\begin{align}
\mathbb{E}\left[\tilde{\delta}_{asw}\,\big|\,n_{asw}^{I},n_{asw}^{C},\epsilon_{asw},\theta\right] & =\check{\mu}_{asw}\cdot\exp(\epsilon_{asw})\nonumber \\
 & =\frac{1+n_{asw}^{C}}{1+\gamma_{a}n_{asw}^{I}}\exp\left(\xi_{a}+\eta_{s}+\phi_{\tau[a,s,w]}+\psi_{m[w]}\right)\cdot\exp(\epsilon_{asw}),\label{eq:update_match_longer}
\end{align}
where $\theta$ represents the model parameters to be estimated. Defining
$\check{\mu}_{asw}\equiv\mathbb{E}\left[\tilde{\delta}_{asw}\,\big|\,n_{asw}^{I},n_{asw}^{C}\right]/\exp(\epsilon_{asw})$
will prove useful when deriving the likelihood function. This expression
for expected match has a simple interpretation: Prior to advertising,
the advertiser would have expected $\gamma_{a}n_{asw}^{I}$ clicks
after the network served $n_{asw}^{I}$ impressions. Instead, it received
$n_{asw}^{C}$ clicks. The ratio of these two quantities determines
whether the advertiser now expects higher or lower match, relative
to when it placed the first ad. Combining Equation (\ref{eq:update_match_longer})
with Equation (\ref{eq:advertiser_payoff}), where true match is replaced
by expected match, yields an expression for expected advertiser payoffs
conditional on past impressions ($n_{asw}^{I}$) and clicks ($n_{asw}^{C}$).
Note that advertisers learn about CTRs at specific sites, thus learning
is about match; vertical quality is represented by $\eta_{s}$.

\subsection{Identification\label{subsec:Identification}}

This section outlines how the parameters in the advertiser demand
model are identified by different sources of variation in the data.
First, following \citet{Lee2014}, a two-parameter version of the
advertiser payoff function in Equation (\ref{eq:advertiser_payoff}),
namely $\pi(x)=\check{\mu}_{a}\exp(\epsilon_{asw})\zeta_{a}^{-1}\log\left(1+xt_{s}\zeta_{a}\right)-p_{sw}\left(x\right)$,
is identified by repeated observations of ad subscription purchases
lasting $x$ days at a price of $p(x)$. \citet{Lee2014} show that
$\check{\mu}_{a}$ is primarily identified by the proportion of non-advertising
choices, as a higher match parameter $\check{\mu}_{a}$ increases
the overall attractiveness of advertising. The satiation parameter
$\zeta_{a}$ is primarily identified by variation in observations
with $x>0$, in combination with non-linearity in observed prices
as a function of $x$ (longer subscriptions cost less per day than
shorter subscriptions). A preponderance of shorter (longer) subscriptions
in the data implies a larger (smaller) value of $\zeta_{a}$, as the
higher the value of the satiation parameter $\zeta_{a}$, the sharper
the decrease in marginal returns to more impressions.

Second, by observing choices from many advertisers at multiple sites
over time, $\check{\mu}$ can be factored multiplicatively into advertiser,
site, and month fixed-effects, as given by Equation \ref{eq:conditional_match}.
Moreover, the fixed effects $\phi_{\tau[asw]}$, which are selected
according to the number of weeks since an advertiser placed an ad
at a site after the first purchase, are identified by differences
in demand (common to all advertisers) along this advertiser-site time
dimension.

Third, differences in demand at the advertiser-site level in the face
of differences in the observed cumulative advertiser-site CTR allow
separate identification of a time-varying match component, which we
posit takes the form of a Bayesian learning model. In contrast with
standard quality learning models in the literature \citep{Ching2013},
wherein the number of signals is observed but their values are latent,
our empirical setting allows us to observe both the number and value
of signals in the data (in the form of impressions and clicks). One
can, therefore, observe how advertisers' subscription purchase patterns
change after obtaining a specific number of clicks and impressions,
and this variation identifies $\gamma_{a}$.

Of note, Equations (\ref{eq:advertiser_payoff}) and (\ref{eq:conditional_match})
include three parameters defined at the advertiser level. These are
separately identified based on differences in advertiser demand along
three dimensions of the data. The parameters for satiation ($\zeta_{a}$)
and initial CTR prior ($\gamma_{a}$) are separately identified because
$\zeta_{a}$ applies in all periods, whereas $\gamma_{a}$ (which
cancels out of the payoff function when $n^{I}=0$) only applies to
choices after using a site for the first time. A similar argument
allows separate identification of $\gamma_{a}$ and the advertiser
fixed effect ($\xi_{a}$).\footnote{\label{fn:price-effects}As mentioned, price effects are set to one
because advertiser value is in a dollar metric; price effects are
thus incorporated into $\exp(\xi_{a})$, which can be loosely interpreted
as an advertisers relative preference for quality over price. Price
effects are identified by within-publisher price variation across
an offering of different subscription lengths $x$, along with the
functional form of the impressions function in Equation (\ref{eq:advertiser_payoff}).
Intuitively, if advertisers tend to choose cheaper subscriptions within
a publisher controlling for match and impressions, price sensitivity
is greater.} Finally, separate identification of $\xi_{a}$ from $\zeta_{a}$
is due to variation in choices $x$ at different prices, as previously
described in the context of a two-parameter payoff function.

\subsection{Estimation\label{subsec:Estimation}}

The likelihood formulation uses the method described in \citet{Lee2014}.
The central idea behind this approach is to derive a set of inequality
constraints on $\epsilon$ that rationalize the observed set of choices.
For example, if an advertiser buys a 7 day ad run at a given price,
the advertiser must expect a higher payoff compared to a shorter or
longer subscription at an alternative price. Many advertiser choices
are observed over many periods, yielding a large number of inequality
constraints from which one can derive a likelihood.

Advertiser site choices are observed at the weekly level. Hence, we
consider the idiosyncratic demand shifters, $\epsilon_{asw}$, in
Equation (\ref{eq:update_match_longer}), and their implications for
advertiser payoffs given by Equation (\ref{eq:advertiser_payoff}).
Replacing the true match with expected match, the expected payoff
equation becomes
\begin{align}
V_{asw}^{\theta}(x,\epsilon_{asw}) & \equiv\mathbb{E}\left[\pi_{asw}(x)|\theta,\epsilon_{asw},n_{asw}^{I},n_{asw}^{C}\right]\nonumber \\
 & =\check{\mu}_{asw}\exp(\epsilon_{asw})\,\zeta_{a}^{-1}\,\log\left(1+t_{s}x\zeta_{a}\right)-p_{sw}(x),\label{eq:V_x}
\end{align}
where the set of parameters to estimate is $\theta=\{\gamma,\xi,\eta,\phi,\psi,\zeta\}.$
Of special interest in this set is $\gamma$, the vector of advertisers'
prior beliefs about the efficacy of their advertising, which is informative
about whether their naivety induces them to advertise too much or
too little.

Conditional on a set of parameters $\theta$, the likelihood of $x_{asw}$
is given by the probability density of the unobserved $\epsilon_{asw}$
after integrating it over the region of $\epsilon$ that can rationalize
the observed choice of $x_{asw}$. This region is defined by upper
and lower bounds, which are themselves determined by a pair of inequality
constraints. Let $\shortuparrow\!\!x$ denote an alternative subscription
that is longer than the subscription of length $x$ that was purchased,
and let $\shortuparrow\!\!p$ denote its (higher) price. Similarly,
let $\shortdownarrow\!\!x$ denote an alternative, shorter subscription,
and $\shortdownarrow\!\!p$ its (lower) price. Moreover, assume for
the moment that both shorter and longer alternatives to $x_{asw}$
were offered by site $s$.

The lower bound for the region of $\epsilon$ that can rationalize
$x_{asw}$ is obtained from the observation that the advertiser did
not buy the shorter subscription, $\shortdownarrow\!\!x$. Hence,
$V_{asw}^{\theta}(x,\epsilon_{asw})>V_{asw}^{\theta}(\shortdownarrow\!\!x,\epsilon_{asw})$.
Substituting Equation (\ref{eq:V_x}) into this inequality and isolating
$\epsilon_{asw}$ leads to the following lower bound for $\epsilon_{asw}$
\citep{Lee2014}:
\begin{gather}
\epsilon_{asw}>\ell b^{\theta}(x_{asw},p_{sw}),\qquad\ell b^{\theta}(x,p)\equiv\log(p\,-\shortdownarrow\!\!p)-\log\left[\check{\mu}_{asw}\zeta_{a}^{-1}\log\left(\frac{t_{s}\,x\,\zeta_{a}+1}{t_{s}\!\shortdownarrow\!\!x\,\zeta_{a}+1}\right)\right]\label{eq:lb}
\end{gather}
Similarly, because the advertiser did not buy the longer subscription,
$V_{asw}^{\theta}(x,\epsilon_{asw})>V^{\theta}(\shortuparrow\!\!x,\epsilon_{asw})$,
and thus
\begin{gather}
\epsilon_{asw}<ub^{\theta}(x_{asw},p_{sw}),\qquad ub^{\theta}(x,p)\equiv\log(\shortuparrow\!\!p-p)-\log\left[\check{\mu}_{asw}\zeta_{a}^{-1}\log\left(\frac{t_{s}\!\shortuparrow\!\!x\,\zeta_{a}+1}{t_{s}\,x\,\zeta_{a}+1}\right)\right].\label{eq:ub}
\end{gather}
\label{rn:participation-again}Finally, if $x_{asw}=0$, meaning the
advertiser did not buy a subscription, then there is no lower bound,
and thus $\epsilon_{asw}>-\infty$. Similarly, if a longer alternative
to $x_{asw}$ was not offered, then there is no known upper bound,
and thus $\epsilon_{asw}<\infty$. Defining $\ell b$ or $ub$ as
negative or positive infinity when $x_{asw}$ lies at one of these
boundaries, the resulting likelihood for each observation is obtained
by integrating over the joint density of $\epsilon$, denoted $f\left(\epsilon_{asw}\right)$,
over the regions indicated by Equations (\ref{eq:lb}) and (\ref{eq:ub}):
\begin{gather}
L(x,p|\theta)=\int_{\ell b^{\theta}(x,p)}^{ub^{\theta}(x,p)}f(\epsilon)d\epsilon.\label{eq:likelihood}
\end{gather}
The intuition behind this likelihood is that the observed choice probability
is maximized if the interval of $\epsilon$ that can rationalize the
observed choice is wide, and minimized if the interval that can rationalize
the observed choice is narrow. As noted by \citet{Lee2014}, this
discrete likelihood approach admits the possibility that intermediate
options between the observed choice and the closest available alternatives
might have been preferred, had they been offered. It therefore does
not assume that the observed choice $x_{asw}$ was optimal, but rather
that it was simply better than the closest alternatives.\footnote{\label{fn:Conditioned-upon-first}Conditioned upon first observing
an advertiser purchase on the network, all subsequent non-purchase
decisions ($x=0$) are included in the advertiser's choice set. Prior
to the advertiser's first purchase on any site, non-choices do not
enter the likelihood. Not including the zero choices before the first
purchase implies that either the advertiser was unaware of the sites
prior to their first purchase, or they purchased a subscription in
the week they joined the network. The timing of the advertisers' first
purchases on the network are assumed to be exogenous \citep{Balseiro2017,Balseiro2019,Bimpikis2020,Wu2015}.}

To complete the likelihood, we assume $f(\epsilon)$ is a normal pdf
with mean 0 and variance $\sigma^{2}$, and that the $\epsilon$s
are independent. Prior distributions for the fixed effects $\eta$,
$\xi$, $\phi$, and $\psi$ are independent Student-$t$ with $4$
degrees of freedom, mean $0$, and unit variance. The prior distribution
for $\sigma$ is exponential, and the penalized complexity approach
of \citet{Simpson2017} is used to choose the exponential rate parameter.
This entails choosing a rate such that $\Pr\left[\sigma>U\right]=\alpha$.
We set $U=1$ and $\alpha=0.1$, so that $\Pr\left[\sigma>1\right]=0.1$.
The resulting prior distribution is $\mathrm{Exponential}\left(-\log\left(\alpha\right)/U\right)$,
leading to $\sigma\sim\mathrm{Exponential}\left(\log10\right)$. The
prior distributions for $\zeta_{a}$ and $\gamma_{a}$ are defined
hierarchically to allow pooling across advertisers, and both parameters'
prior distributions are derived by transforming exponential variates.
The conditional prior for $\zeta_{a}$ is exponential, shifted by
$0.01$ to improve numerical stability during estimation, with mean
$0.01+\bar{\zeta}$. $\bar{\zeta}$ is exponential with prior mean
$10/\log10$, so that $\Pr\left[\bar{\zeta}>10\right]=0.1$. Accordingly,
the marginal prior mean for $\zeta_{a}$ is $0.01+10/\log10$. The
conditional prior for $\gamma_{a}$ resembles a unimodal beta distribution
with most of the mass near zero, but is derived from $\gamma_{a}\,\big|\,\bar{\gamma}=g_{a}/\left(\bar{\gamma}^{-1}+g_{a}\right)$,
with $g_{a}\sim\mathrm{Exponential}\left(1\right)$. The prior distribution
for $\bar{\gamma}$ is exponential with rate $-\log\left(0.1\right)/0.002$,
so that $\Pr\left[\bar{\gamma}>0.002\right]=0.1$. The resulting marginal
prior mean for $\gamma_{a}$ is $0.00087$, with $\Pr\left[\gamma_{a}>0.002\right]\approx0.12$
(for reference, the average CTR is $0.00045$). We sample from the
model's posterior distribution using Hamiltonian Monte Carlo (HMC),
as implemented in the cmdstanr package for R \citep{StanTeam2017}.

\subsection{Subscription Prices\label{subsec:Subscription-Prices}}

Subscription prices, $p_{sw}(x)$, vary by week, publisher, and subscription
length.\footnote{Subscription lengths $x$ are observed only when purchased in a given
week. We impute unobserved subscription lengths in a given week for
a given site using a nearest neighbor approach. If a subscription
of length $x$ was offered within 4 weeks before or after week $w$,
we assume it was also offered in week $w$.} Although these prices are generally stable over time within publisher-subscription
length, the series can sometimes exhibit perturbations around their
means. While their effect on estimation is limited, these perturbations
induce counterfactual analyses in some instances that yield results
with unusual valuations (for example, in a week where the longer ad
subscription is priced more than the shorter one). \label{rn:price-eq-fixed-effects}To
alleviate this problem, we use OLS to regress the log of subscription
prices on week fixed effects (to account for seasonality or other
time-varying trends), site fixed effects (to account for differences
in audience attractiveness or site popularity), and the log of subscription
length, 
\begin{gather}
\mathbb{E}\left[\log\left(p_{sw}(x)\right)\right]=\mu_{w}+\mu_{s}+\nu\log(x).\label{eq:pricing}
\end{gather}
The $R^{2}$ of the log pricing regression is 0.94, evidencing high
fit. When estimating the likelihood from Equations (\ref{eq:lb})
and (\ref{eq:ub}), price is set to be
\begin{gather}
p_{sw}(x)=\exp\left(\frac{\hat{\varsigma}^{2}}{2}+\hat{\mu}_{w}+\hat{\mu}_{s}+\hat{\nu}\log(x)\right)
\end{gather}
where $\hat{\varsigma}^{2}$ is the residual variance from the pricing
regression for $\log\left(p_{sw}(x)\right)$.\footnote{\label{fn:To-ascertain-the}Because prices are inferred from observed
sales, one cannot rule out the possibility of negotiated instead of
posted prices. To ascertain the potential for negotiated prices, we
checked whether prices paid for the same inventory from the same publisher
differ across advertisers in the same week. The coefficient of variation
(bias corrected) for price for the same ad inventory sold to different
advertisers in the same week is 12.4\%, meaning the price variation
is small. In addition, in 2/3 of all instances for which we observe
multiple advertisers at a site, they pay exactly the same price. In
other words, some price negotiation may exist, but its degree is slight,
so prices are set to be constant across advertisers within a given
week at a given publisher.}

\section{Results\label{sec:Results}}

\subsection{Model Fit}

\begin{table}[!t]
\begin{centering}
\smaller%
\begin{tabular}{ld{-4.1}d{-4.1}d{-4.1}}
\toprule 
 & \multicolumn{1}{c}{Model 1} & \multicolumn{1}{c}{Model 2} & \multicolumn{1}{c}{Model 3}\tabularnewline
\midrule
Time-Varying & \multicolumn{1}{c}{} & \multicolumn{1}{c}{} & \multicolumn{1}{c}{}\tabularnewline
\quad{}Fixed Effects & \multicolumn{1}{c}{No} & \multicolumn{1}{c}{Yes} & \multicolumn{1}{c}{Yes}\tabularnewline
Learning & \multicolumn{1}{c}{No} & \multicolumn{1}{c}{No} & \multicolumn{1}{c}{Yes}\tabularnewline
$p_{{\scriptscriptstyle LOO}}$ & 95.8 & 113.4 & 148.2\tabularnewline
$ELPD_{{\scriptscriptstyle LOO}}$ & -3672.7 & -3525.2 & -3393.9\tabularnewline
 & (138.8) & (135.9) & (132.4)\tabularnewline
Incremental &  & -131.3 & -147.5\tabularnewline
\quad{}Improvement &  & (18.2) & (18.2)\tabularnewline
\bottomrule
\end{tabular}
\par\end{centering}
\caption{Model Fit and Comparison. $ELPD_{{\scriptscriptstyle LOO}}$ is the
Bayesian leave-out-one estimate of log pointwise predictive density,
an estimate of model fit; and $p_{{\scriptscriptstyle LOO}}$ is the
effective number of parameters, an estimate of model complexity \citep{Vehtari2016}.
Incremental improvements compare Model 2 to Model 1, and Model 3 to
Model 2. Values in parentheses show standard errors of point estimates.
\label{tab:Model-Fit}}
\end{table}

In addition to estimating the advertiser learning model described
in Section \ref{sec:Model-and-Estimation}, we estimate three simpler
models. These omit learning (setting the ratio $c_{as}/\gamma_{a}=1$
for all choices) and/or the time-varying fixed effects (setting $\phi_{\tau}$
and $\psi_{m}$ to zero), and help to ascertain the predictive value
of modeling advertiser learning. The fit of each model is presented
in Table \ref{tab:Model-Fit}. Results suggest that modeling advertiser
learning leads to a substantial improvement in fit. The improvement
in fit from modeling advertiser learning (a decrease in $ELPD_{{\scriptscriptstyle LOO}}$
of $147.5$) is comparable in magnitude to the improvement in fit
from including time-varying fixed effects (a decrease in $ELPD_{{\scriptscriptstyle LOO}}$
of $131.3$).

\begin{figure}[!t]
\begin{centering}
\includegraphics[width=1\textwidth]{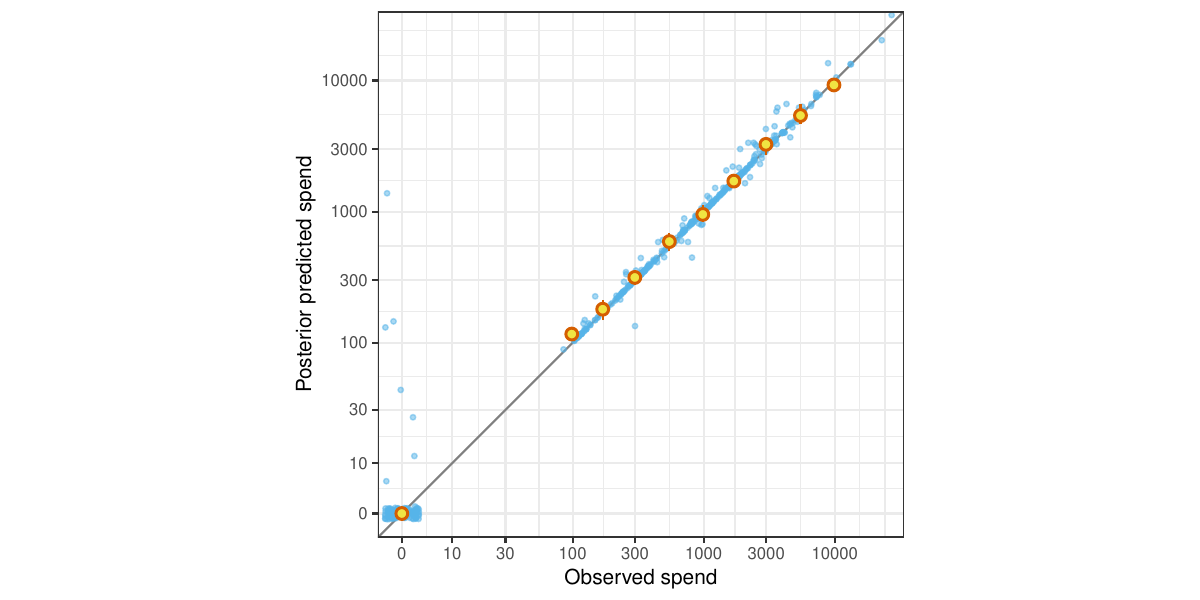}
\par\end{centering}
\centering{}\caption{Posterior Predictive Fit for Advertising Spend. Expected total spend
for all advertiser-site pairs (averaged over 500 simulations) is shown
with binned medians and inner 50th percentiles. Points corresponding
with advertiser-site pairs with zero observed spend are jittered.
The observed spend of zero does not represent the inaugural campaign
(where spend is necessarily positive), but instead weeks in which
no spend is observed in the data. The distribution of positive forecasted
spend amounts when observed spend is zero reflects left truncated
prediction errors.\label{fig:Posterior-Predictive-Fit}}
\end{figure}

The posterior predictive fit for advertiser spending is depicted in
Figure \ref{fig:Posterior-Predictive-Fit}, where each point represents
the total amount an advertiser spent with a publisher over 27 weeks
(the procedure for simulating from the posterior distribution is described
in Section \ref{subsec:Simulation-Procedure-Baseline}). Overall,
the model predicts spending well, including cases of no spending.

\subsection{Parameter Estimates}

\begin{figure}
\begin{centering}
\includegraphics[scale=0.8]{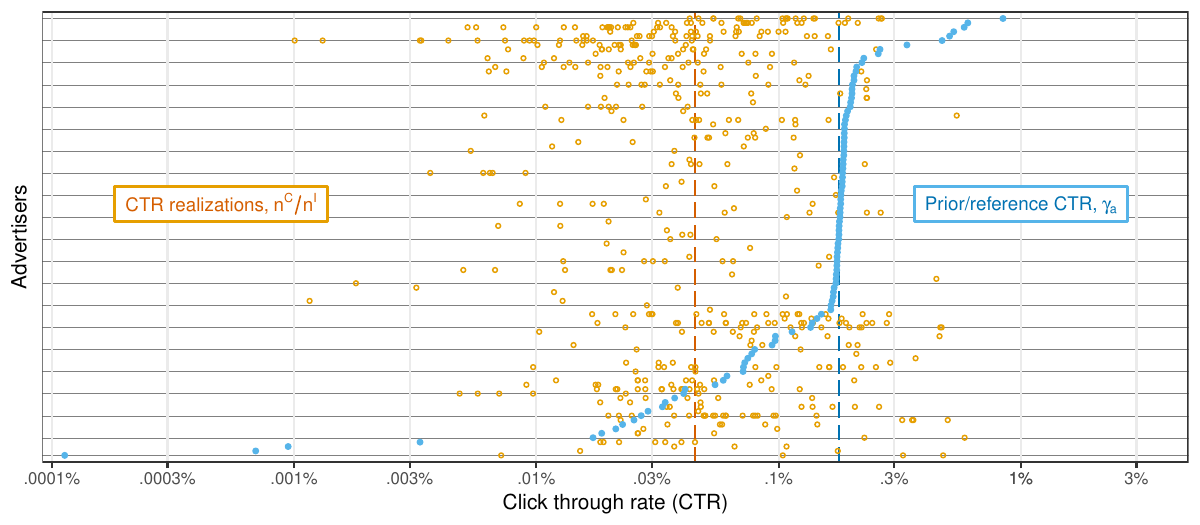}
\par\end{centering}
\centering{}\caption{Realized and Prior/Reference Click Through Rates. Each row represents
an advertiser and shows the posterior mean of $\gamma_{a}$ and their
observed CTRs for sites at which they placed ads. The median CTR is
0.045\%, and the median prior/reference CTR is 0.177\%; both are depicted
as vertical dashed lines. \label{fig:Realized-and-prior}}
\end{figure}
Recall, the key parameter estimates pertain to the $\gamma_{a}$'s,
which represent advertisers' prior beliefs about CTRs, and a central
question is whether these beliefs are optimistic or pessimistic. Given
that Section \ref{subsec:Descriptive-Analyses} shows that the number
of sites used by advertisers tends to decrease over time and that
advertisers tend to switch away from those initial choices, we hazard
that advertisers tend to be over-optimistic about site match. Hence,
the parameter $\gamma_{a}$, which reflects prior beliefs about match
(in terms of CTR), should reflect this (see Section \ref{subsec:Learning-About-Clicks}).
Estimates indicate that for nearly all advertisers, the $\gamma_{a}$s
are indeed higher than their empirical CTRs. Figure \ref{fig:Realized-and-prior}
plots the estimates for $\gamma_{a}$ against the realized CTRs for
all observed advertiser-site pairs. Note that the median prior belief
for click through rates is 0.177\%, whereas the median true CTR is
about one quarter that amount, 0.045\%. This suggests that advertisers
are, on average, initially over-optimistic. There is considerable
heterogeneity in initial beliefs, as well as in how accurate those
beliefs are, as some are correct, some are optimistic, and some are
pessimistic.

\begin{figure}
\begin{centering}
\includegraphics[width=1\textwidth]{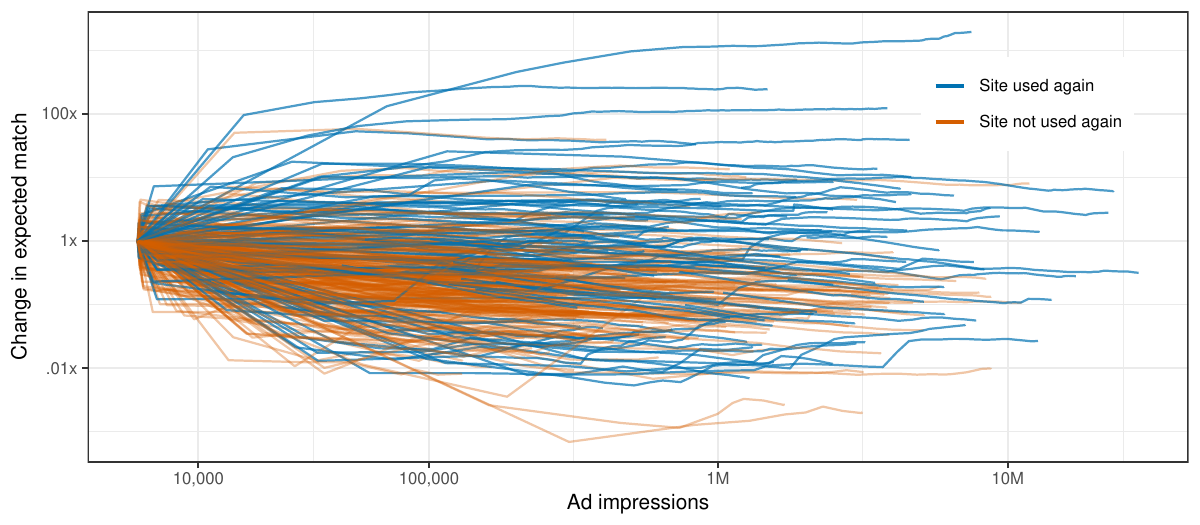}
\par\end{centering}
\caption{Advertiser Match Beliefs Over Time. Depicts outcomes for the estimation
sample, with each line representing an advertiser-site combination.
Change in expected match is the ratio of an advertiser's CTR beliefs
to their initial prior, $\tilde{c}_{as}/\gamma_{a}$, which updates
based on impressions and clicks observed each day. The horizontal
axis presents the cumulative number of impressions (rather than time)
to allow comparison across advertiser-site pairs, which differ in
impressions served per day and the timing of ad purchases. \label{fig:Advertiser-Match-Beliefs-Over-Time-1}}
\end{figure}

Figure \ref{fig:Advertiser-Match-Beliefs-Over-Time-1} depicts the
changes in advertiser CTR beliefs over time, scaled by the prior belief
($\tilde{c}_{as}/\gamma_{a}$). Prior to advertising at a site, this
ratio is equal to one. With more information, the ratio approaches
$c_{as}/\gamma_{a}$. For example, if the true CTR is twice the initial
belief, then this ratio will approach two. Values less than one imply
advertisers were initially optimistic.

Because most of the belief paths in Figure \ref{fig:Advertiser-Match-Beliefs-Over-Time-1}
decrease from their starting point of 1, advertisers appear largely
optimistic about the likelihood of clicks they will obtain on a site
when they first advertise. Accordingly, advertisers often spend too
much, but sometimes spend too little. Moreover, beliefs take several
thousand impressions before they begin to stabilize, meaning an advertiser
could misspend a considerable sum on advertising before learning its
true site match. Whether optimistic or pessimistic, advertisers' best
advertising choices---were they to know their true match with each
site up front---would frequently differ from their actual advertising
choices. Advertisers thus face losses relative to being fully informed.

One might expect that publishers and the platform have little incentive
to address over-optimistic advertisers. Were one to correct advertisers'
prior beliefs about their match with an initial set of sites, one
might expect ad spend across those sites and the platform as a whole
to be lower. However, this expectation ignores the possibility that
an advertiser might switch to a better matched publisher, potentially
increasing total advertising on the publisher sites and the platform,
thereby increasing revenues. Section \ref{sec:The-Role-of-Information}
quantifies these trade-offs and considers a potential solution that
can benefit advertisers, sites, and the platform.

\subsection{Robustness Checks\label{subsec:Robustness-Checks}}

Several issues present with our specification, including the potential
for forward-looking behavior by advertisers, supply-side considerations,
and other concerns. Below, we detail these issues and offer some analyses
to better rationalize our modeling choices. We highlight the findings
of various robustness checks in this section and provide further detail
in Online Appendices \ref{online:Test-and-Learn}--\ref{online:Additional-Analyses}.
Model performance and insights are robust across these checks.

\subsubsection{Test and Learn Behavior\label{subsec:Test-and-Learn}}

Test and learn behavior entails advertisers choosing to ``test''
sites with lower expected match, but higher match uncertainty. Resolving
this uncertainty creates future value in the event the advertiser
discovers a better than expected match. The demand model assumes advertisers
are not forward-looking when making choices. One reason for this assumption
is that advertisers are often unsophisticated. Recently, test and
learn algorithms have been automated in ad exchange markets, showing
great gains in ad performance \citep{Schwartz2017,tunuguntla_2022};
such gains would not be possible if advertisers were already using
a test and learn strategy \citep{Feit_Berman_2019}. However, to the
degree test and learn behavior is present, a model that assumes otherwise
would yield over-optimistic priors ($\gamma_{a}$), because lower
demand arising from lower match uncertainty would be rationalized
as a lower-than-expected CTR.

To explore the question of whether advertisers are forward-looking,
we consider advertiser switching across ad subscription lengths within
a publisher. Forward-looking advertisers, all else equal, should tend
to first use shorter subscriptions and, if the short trial works at
the publisher, switch to longer subscriptions (a form of ``dipping
one's toes in the water before taking a swim''). No evidence exists
for this behavior. In 1,240 cases, advertisers switched from shorter
to longer subscriptions, in 1,266 cases the next ad subscription was
shorter, and in 3,007 the next ad subscription was the same length.

In addition, to more formally address this concern, the ad subscription
model in Section \ref{subsec:Payoff-from-Advertising} is extended
to incorporate a UCB component for test and learn \citep{auer2002finite,Wang_et_al_2025,zhuo2022exploit}.
We estimate models using two different UCB approximations to the value
of exploration; results indicate that neither changes model fit, as
indicated by the $ELPD_{{\scriptscriptstyle LOO}}$, and prior belief
estimates are essentially identical to those from the main specification.
This suggests the model is robust to the assumption of myopic advertisers.
Details of this analysis are reported in Online Appendix \ref{online:Test-and-Learn}.

\subsubsection{Supply-side Considerations\label{subsec:Supply-side-Considerations}}

The model and counterfactuals do not consider how aggregate advertiser
demand might affect publisher prices or CTRs. This section briefly
overviews these considerations next. Detailed analyses are in Online
Appendix \ref{online:Supply-Side-Considerations-Appendix}.

\paragraph{Pricing.\label{rev:endogeneity}}

Equation (\ref{eq:advertiser_payoff}) raises the potential for price
endogeneity bias. Such a bias could manifest if site-week omitted
factors i) correlate with the advertiser's value for ads, and ii)
are not captured by the observables in the estimation equation (these
observables include week and site fixed effects to control for site
popularity and seasonality). If price endogeneity exists, one would
expect prices to change in response to demand, affecting both inference
(lowering the implied price effect) and counterfactual analyses. To
explore the issue, we conduct a variance decomposition of the original
(log-transformed) subscription prices at the top 20 sites during the
estimation period. Variation in prices is decomposed in terms of the
log of subscription durations (days), publisher, and week---per Equation
(\ref{eq:pricing})---as well as a publisher-week interaction and
current and lagged variables reflecting advertiser demand. Only the
variables in Equation (\ref{eq:pricing}) have significant $F$ statistics
in the variance decomposition, collectively explaining $89.34\%$
of variation in prices. The publisher-week interaction uses 400 degrees
of freedom to explain just 3.18\% of price variation, and the variables
related to advertiser demand explain just $0.51\%$. We find no evidence
to suggest that prices are sensitive to contemporaneous or lagged
advertiser demand, reducing the potential for endogeneity bias. In
spite of these analyses, to the extent that unobserved variables do
induce any remaining endogeneity, counterfactual welfare results would
reflect a lower bound on the welfare gains, as imputed price sensitivity
would increase after controlling price endogeneity, amplifying the
welfare effects from changes in ad spend.

\paragraph{Crowding.\label{par:Crowding}}

Another concern pertains to crowding. When more advertisers appear
concurrently on a site, it becomes possible that CTRs could decrease.
If that were the case, it would be necessary to control for how the
number of advertisers on a site affects clicks. However, in direct
advertising contexts, such crowding effects can be small because ads
appear over a long duration and are rotated. For example, the rate
card in Figure \ref{fig:2024-Star-News} indicates that in-column
side bars ads are rotated with each impression, with an estimated
50,000 impressions monthly. This means that, even if several ads are
running simultaneously, there is still ample opportunity for repeat
visitors to see an ad in a prominent position on the page. To test
this conjecture more formally, site-week clicks are regressed on the
number of advertisers, controlling for site fixed effects and average
advertiser clicks. Results indicate a small and insignificant effect
(and of the opposite sign of that predicted by crowding) of the number
of advertisers on clicks. This result suggests that crowding effects
are not considerable in this context.

\subsubsection{Additional Considerations\label{subsec:Additional-Considerations}}

We briefly consider several additional issues, including the sample
selection criterion, budget constraints, heterogeneity in advertiser
priors across publishers, and the assumed prior variance of CTR beliefs.
A detailed analysis of these considerations is presented in Online
Appendix \ref{online:Additional-Analyses}.

\paragraph{Sample Selection.}

Results are based on the largest 20 publishers and 100 random advertisers.
We limit the number of publishers to the 20 largest because we believe
this is a reasonable choice set for a typical advertiser. As the number
of publishers increases, counterfactual increases in match values
due to improving advertisers' priors should be higher, because the
opportunity for advertisers to sort into a better match is greater.
Results indicate this to be the case when the number of publishers
included in the sample increases by 25\%, provided advertisers' choice
sets include these additional publishers, making the reported overall
gains conservative estimates.

\paragraph*{Budget Constraints.}

To the extent that budget constraints exist, advertiser spending will
be attenuated in any counterfactual analysis. Budget constraints are
not observed, and there is no evidence advertisers are constrained.
Moreover, it is unclear why advertisers would not raise funds to pay
for positive investments. Because budget constraints are not observed,
analyses pertaining to this issue are speculative. However, under
the premise that advertiser budgets are constrained to some percentage
increase over the maximum spend in the data, it is possible to assess
how counterfactual insights would change in response to different
budget constraints. As one might expect, the tighter the budget constraint,
the less spending and welfare can increase under the counterfactual.

\paragraph{Heterogeneity in Advertiser Priors Across Publishers and Variance
in Learning Rates.\label{rev:heterogeneous_priors}}

We conduct two additional robustness checks related to advertiser
learning. First, we accommodate the potential for heterogeneity in
advertisers' prior beliefs across sites. In spite of adding a large
number of parameters, allowing heterogeneity in prior beliefs about
site-specific CTRs has a negligible impact on model fit and an immaterial
effect on prior belief estimates. Second, we estimate the prior variance
(and thus the learning rate) of beliefs about CTRs. This degrades
model fit and has no material effect on inferences about prior beliefs.

\paragraph{Stability of CTRs.}

The model presumes advertiser-site CTRs, $c_{as}$, are invariant
overt time. However, if ad content does not change, consumers may
become satiated with the content; or if content does change, consumers
may find it more or less engaging. Reviewing empirical CTRs by week
for each advertiser-site pair, we find no evidence in the data supporting
the idea that values of $c_{as}$ vary over time.

\section{The Role of Information Provision on Advertiser and Publisher Welfare\label{sec:The-Role-of-Information}}

The results in Section \ref{sec:Results} indicate that many advertisers
are over-optimistic about advertising efficacy and therefore tend
to overspend. Holding all else fixed, redressing this overoptimism
could cause advertisers to spend less, leaving advertisers better
off, but publishers and the platform worse off. On the other hand,
advertisers could migrate some of their ad spend to other sites that
are a better match, increasing their welfare and possibly their total
spend, thus increasing publisher and network revenue. Hence, whether
endowing advertisers with better information increases or reduces
overall welfare is an empirical question. This section explores how
informing advertiser priors, $\tilde{c}_{as}$, affects advertisers'
valuations and spending levels, and considers the consequences for
publisher and platform welfare. It first outlines approaches to affording
better advertiser priors and then considers the welfare implications
of these approaches.

\subsection{Information Provision}

We consider two mechanisms for counterfactual information provision:
i) whether or not the advertiser is endowed with full information
about the clicks they obtain from their chosen sites, and ii) whether
or not advertisers have access to pooled data from other advertisers.
We simulate advertising choices and compare advertising spend under
each of these mechanisms to a common baseline. Appendix \ref{appx:Simulation-Procedure}
describes the procedure used to simulate counterfactual data in the
baseline and other scenarios.

\subsubsection{Simulation Procedure and Baseline Scenario\label{subsec:Simulation-Procedure-Baseline}}

The baseline and counterfactual scenarios differ in terms of i) the
advertisers' initial information about their expected CTRs, and thus
expected match utilities in week 1; and ii) the advertisers' choice
sets, which are closely related to the information available for simulating
clicks.

\paragraph{Advertisers' initial information.}

In the baseline scenario, $B$, advertisers have the same initial
information as in the estimation sample. Thus, for sites at which
the advertiser placed ads prior to week 1, their initial expected
ratio of CTR to $\gamma_{a}$ in the baseline scenario is $\mathbb{E}[\tilde{c}_{as0}/\gamma_{a}|n_{as0}^{I},n_{as0}^{C},\gamma_{a}]=(1+n_{as0}^{C})/(1+\gamma_{a}n_{as0}^{I})$,
with $n_{as0}^{I}>0$; for all other sites, $\mathbb{E}\left[\tilde{c}_{as0}/\gamma_{a}|\gamma_{a}\right]=1$.
The counterfactual scenarios augment these initial information states
with additional CTR information.

\paragraph{Advertisers' choice sets.}

In the baseline scenario, advertisers' choice sets are restricted
to the sites with observed purchases in the estimation sample, as
this allows us to simulate clicks using an approximation to $c_{as}$
based on all advertising outcomes in the full data set. In the counterfactual
simulations, we use these values of $c_{as}$ to simulate clicks whenever
they are available from the observed advertising data. In some counterfactual
scenarios, we expand the advertisers' choice sets to include all sites
in the estimation sample, and accordingly simulate clicks for these
sites based on imputed values of $c_{as}$, which are obtained via
a procedure described in Section \ref{subsec:Pooling}.

\subsubsection{Full Information Within Advertiser\label{subsec:Oracle}}

The first counterfactual scenario considers endowing advertisers with
full information about the sites for which we observe ad buys in the
data. It is labeled the ``full information'' counterfactual and
denoted $C_{F}$. In this counterfactual scenario, we set $n_{as0}^{I}=10^{12}$
and $n_{as0}^{C}=\left\lfloor 10^{12}c_{as}\right\rfloor $, so that
$\mathbb{E}[\tilde{c}_{as0}/\gamma_{a}|n_{as0}^{I},n_{as0}^{C},\gamma_{a},C_{F}]\approx c_{as}/\gamma_{a}$.\footnote{We do not use the value of $c_{as}/\gamma_{a}$ directly because i)
it leads to numerical discrepancies between this ratio and simulated
clicks and impressions related to floating point precision for large
numbers, and ii) it completely eliminates learning. By setting $n^{I}=10^{12}$
advertisers' priors are strongly updated, but there is still sufficient
opportunity for learning and adaptation.} The value of $c_{as}$ is the same value used in the baseline scenario,
and advertisers' choice sets are restricted to observed advertiser-site
pairs, as in the baseline. Hence the only difference between $C_{F}$
and $B$ lies in advertisers' prior information. This counterfactual
yields advertiser spending and revenue as if the advertiser had already
known the response they would eventually receive from ads placed on
the site (i.e., an oracle prior).

While this counterfactual can yield insights into the advertisers'
losses in the baseline scenario, relative to a hypothetical endowment
with complete information about advertising responses at chosen sites,
it does not consider how advertiser outcomes might change at sites
where no spending occurred. As noted previously, this limitation is
a consequence of only observing CTRs for sites that advertisers actually
used in the data. Moreover, this approach to providing advertisers
with better information is infeasible because full information about
site CTRs is not revealed until after sites are chosen. By then, it
is too late for advertisers to change course. Hence, we consider an
alternative that involves sharing information across advertisers.

\subsubsection{Pooling Information Across Advertisers\label{subsec:Pooling}}

Given the ad network observes CTRs across all advertisers over time
(as well as their ad creatives), it is well positioned to better inform
advertisers about initial match by pooling information across advertisers.
This counterfactual scenario updates advertisers' prior information
by considering the performance of similar advertisers and ads on the
platform, and extrapolating this information to sites that advertisers
did not purchase from in the data. Reflecting advertisers' augmented
information states, we also expand their choice sets to allow ad purchases
at any site. When simulating clicks, we use the observed (i.e., full
information) CTRs when available, and the imputed CTRs otherwise.
Hence, for sites with observed purchases, there is a discrepancy between
advertisers' expectations, which are based on imputed CTRs, and realizations
of clicks, which are based on observed CTRs.

For the purposes of this counterfactual simulation, we expect that
ads with similar content generate similar CTRs when placed with a
given publisher. More conceptually, we expect advertisers with similar
ads to have similar levels of match with sites' audiences, and thus
generate similar match signals in terms of CTR. Inferences from this
scenario depend on information that is readily available to platforms,
hence this approach to pooling is feasible for platforms to implement.

\paragraph{Estimating advertiser similarities.}

Our approach entails i) deriving a measure of the similarity between
any two advertisers, and ii) pooling outcomes among similar advertisers
to predict ad performance. This raises the practical question of how
to measure similarity between advertisers. As is common on many ad
platforms, there is no advertiser demographic information in the data.
However, in addition to recording the number of impressions and clicks,
the platform also retains the images used by advertisers in their
display ad campaigns, and these images contain useful information
about the advertisers.

\label{rn:tf-idf-by-creative}To estimate similarity between advertisers,
we first retrieve a set of concept tags from Google's Cloud Vision
API\footnote{\url{https://cloud.google.com/vision} \href{https://web.archive.org/web/20240229224934/https://cloud.google.com/vision}{(Archive.org)}}
describing the content of all ad images placed by advertisers at any
sites in the focal cluster of 165 liberal blogs. We count the frequency
for every advertiser-tag pair, and construct a term frequency/inverse
document frequency (TF-IDF) matrix reflecting how prevalent and distinctive
each tag is in characterizing each advertiser. A square matrix of
pairwise cosine similarities between advertisers is then computed
from the TF-IDF measures. Finally, we use the similarities within
each row of this matrix to impute click through rates for all advertiser-site
pairs in the estimation sample, including those for which no ads were
placed. Appendix \ref{appx:GoogleAPI} details these specific steps.

\paragraph{Imputing CTRs.}

Given a vector of similarities, $r_{a}$, between advertiser $a$
and all other advertisers, as well as the set of advertisers, $\mathcal{A}_{s}$,
for which we can approximate $c_{as}$ for site $s$, we impute advertiser
$a$'s CTR at site $s$ as
\begin{gather}
\bar{c}_{as}=\frac{\sum_{j\in\mathcal{A}_{s}\backslash a}r_{aj}c_{js}}{\sum_{j\in\mathcal{A}_{s}\backslash a}r_{aj}}.\label{eq:ctr-imputed}
\end{gather}
In this counterfactual scenario, which we refer to as ``pooling''
and denote $C_{P}$, we set $n_{as0}^{I}=10^{12}$ and $n_{as0}^{C}=\left\lfloor 10^{12}\bar{c}_{as}\right\rfloor $,
so that $\mathbb{E}[\tilde{c}_{as0}/\gamma_{a}|n_{as0}^{I},n_{as0}^{C},\gamma_{a},C_{P}]\approx\bar{c}_{as}/\gamma_{a}$.

\begin{figure}[!t]
\begin{centering}
\includegraphics[width=1\textwidth]{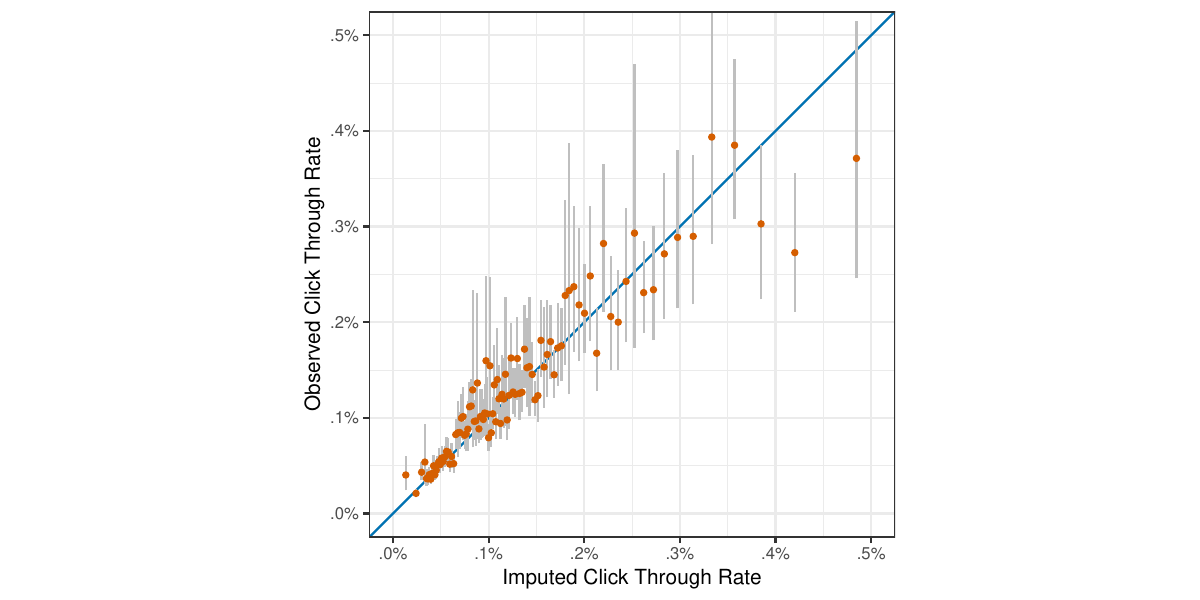}
\par\end{centering}
\centering{}\caption{Click Through Rate Imputation\label{fig:Click-Through-Rate-Prediction}.
The horizontal axis shows imputed CTRs for advertiser-publisher pairs
observed in the estimation sample, grouped into bins of equal numbers
of observations and truncated above the 99th percentile. The vertical
axis shows the distribution of actual CTRs for each bin, with points
indicating means and vertical lines indicating bootstrapped 95\% CIs
for the means. A 45 degree diagonal line is shown; observations closer
to this line indicate more accurate imputations.}
\end{figure}

For the subset of advertiser-site pairs with observed ad buys, the
imputed CTRs, $\bar{c}_{as}$, can be compared against the observed
CTRs, $c_{as}$. Figure \ref{fig:Click-Through-Rate-Prediction} shows
this comparison, grouping imputed CTRs into bins of roughly equal
size, and plotting the average observed CTRs corresponding with these
advertiser-site pairs. The observations in the plots largely hew to
a 45 degree line, meaning that the information from like advertisers
can be used by the platform to predict initial click through rates.\footnote{\label{fn:As-a-robustness}A linear regression of observed CTRs on
imputed CTRs yields an intercept of $.00014$ (SE = $.000037$), a
slope of $.957$ ($.0244$), and an $R^{2}$ of $.144$. The computation
borrows observed CTRs from future outcomes in the data as an approximation
to the knowledge the platform would have accumulated based on past
experiences with campaigns across a large number of advertisers. A
robustness check calculates expected CTRs using only impressions and
clicks obtained prior to the estimation period, obtaining an intercept
of $0.00032$ ($0.00009$), a slope of $1.02971$ ($0.06060$), and
an $R^{2}$ of $0.140$. Thus, using past data only, imputed CTRs
vary little in expectation, but are noisier, as one might expect.
Given both approaches yield unbiased estimates of CTRs, CTRs are imputed
based on the full data, as they are less noisy for the purposes of
counterfactual evaluation.}

\subsubsection{Full Information Combined with Pooling}

We also consider a third counterfactual that combines $C_{F}$ and
$C_{P}$. In this scenario, advertisers are endowed with full information
about their CTRs at sites with observed purchases, as in $C_{F}$;
as well as imputed CTRs at all other sites, as in $C_{P}$. We refer
to this as the ``combined'' scenario and denote it $C_{F+P}$. Comparing
$C_{F+P}$ to $C_{P}$ provides information about the value of better
match information (i.e., oracle versus approximate information about
CTRs). Comparing $C_{F+P}$ to $C_{F}$ provides information about
the value of expanding advertisers' choice sets.

\subsubsection{Combining Information Within and Across Advertisers}

\begin{table}[!b]
\centering{}\smaller%
\begin{tabular}{cccc}
\toprule 
 &  & \multicolumn{2}{c}{Own Information}\tabularnewline
 &  & No Information & Full Information\tabularnewline
\midrule 
\multirow{2}{*}{Shared Information} & No Sharing & $B$ & $C_{F}$\tabularnewline
 & Pooled Information & $C_{P}$ & $C_{F+P}$\tabularnewline
\bottomrule
\end{tabular}\caption{Counterfactual Scenarios\label{tab:CFs}. The full information condition
replaces the first periods' advertiser CTR priors with the last periods'
oracle CTR posteriors. The shared information condition replaces the
first period advertisers' CTR priors with a similarity weighted average
of other advertisers' CTRs.}
\end{table}

Integrating the respective approaches outlined in Sections \ref{subsec:Oracle}
and \ref{subsec:Pooling} implies a fully crossed two-by-two counterfactual
design, as indicated in Table \ref{tab:CFs}. Cell $B$ represents
the base case, wherein the advertiser has no additional information
about other sites or its imminent clicks. In the no sharing cells
($B$ and $C_{F})$, advertisers are constrained to choose only from
the set of sites at which they advertised in the estimation data.
Cell $B$ involves naïve priors, uses observed CTRs to simulate clicks,
and serves as the base case for measuring welfare gains from enhancing
prior information. Cell $C_{F}$ affords advertisers ex-ante full
information on the clicks they obtained in the full data (i.e., their
ex-post advertising outcomes), erasing their lack of prior information.
Contrasting $C_{F}$ with $C_{B}$ yields theoretical insights into
the cost advertisers bear owing to their lack of a priori knowledge
about the sites on which they advertised.

Cell $C_{P}$ pools advertiser information using the approach outlined
in Section \ref{subsec:Pooling}. Contrasting cell $C_{P}$ to $B$
yields the advertiser value of pooling information via the direct
ad network platform (or any other information pooling entity), and
expanding the advertisers' choice set to those sites where the advertisers
did not place ads in the data. Finally, cell $C_{F+P}$ relative to
$B$ contrasts having information both from other advertisers (via
pooling) and within advertiser (via full information). Other comparisons
exist. For example, contrasting differences in $C_{P}$ and $B$ with
differences in $C_{F}$ and $B$ yields a sense for which information
sets generate greater improvements in advertiser outcomes.

\subsection{Advertiser Welfare}

\begin{table}
\centering{}\smaller%
\begin{tabular}{cccccc}
\toprule 
 &  & \multicolumn{4}{c}{Own Information}\tabularnewline
\cmidrule{3-6}
 &  & \multicolumn{2}{c}{No Information} & \multicolumn{2}{c}{Full Information}\tabularnewline
 &  & Spend & Welfare & Spend & Welfare\tabularnewline
\midrule 
\multirow{2}{*}{Shared Information} & No Sharing & \$0 & \$0 & $-\$70$ ($-8.5$\%) & $\$686$ ($5.0\%$)\tabularnewline
 & Pooled Information & $\$946$ ($98.3\%$) & $\$3,621$ ($18.3\%$) & $\$823$ ($76.9\%$) & $\$3,753$ ($26.2\%$)\tabularnewline
\bottomrule
\end{tabular}\caption{Advertiser Counterfactual Changes in Spend and Welfare\label{tab:Advertiser-Counterfactual-Outcomes}.
Each cell reports the median advertiser's posterior mean counterfactual
change in ad spend and welfare (relative to baseline), with the associated
percentage increase reported in parentheses.}
\end{table}

\begin{figure}[b]
\begin{centering}
\includegraphics[width=1\textwidth]{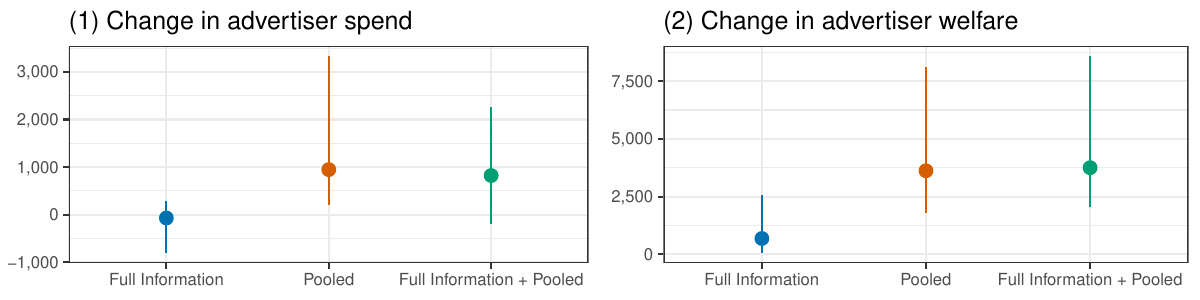}
\par\end{centering}
\caption{Advertiser Counterfactual Changes in Spend and Welfare\label{fig:Counterfactual-Advertiser-Welfare}.
Points and vertical lines indicate medians and inner 50th percentiles
of advertisers' posterior mean counterfactual changes. Panel (1) presents
changes in ad spend relative to baseline across the three counterfactual
conditions. Panel (2) reports changes in advertiser welfare.}
\end{figure}

Table \ref{tab:Advertiser-Counterfactual-Outcomes} reports the results
of the advertiser counterfactual outcomes in terms of relative gains
compared to the baseline. Advertiser outcomes are computed using two
metrics, change in ad spend, and change in welfare. Figure \ref{fig:Counterfactual-Advertiser-Welfare}
reports, for each metric, the median and inner 50th percentile of
advertisers' posterior mean counterfactual changes (relative to baseline)
in each counterfactual cell in Table \ref{tab:CFs}. Beginning with
the full information comparison ($C_{F}$ vs $B$), results indicate
that the median ad spend decreases by $\$70$ (a decrease of $8.5\%$
relative to baseline). In $C_{F}$, the preponderance of advertiser
priors tend to be updated negatively across the selected sites (i.e.,
sites with advertising in the data), hence advertisers spend less,
or stop spending entirely at these sites. As advertisers' entire ad
budgets can be shifted away from the lower match publishers in the
set they ultimately selected, the median welfare increases by $\$686$
($5.0\%$).

\begin{figure}[t]
\begin{centering}
\includegraphics[scale=0.8]{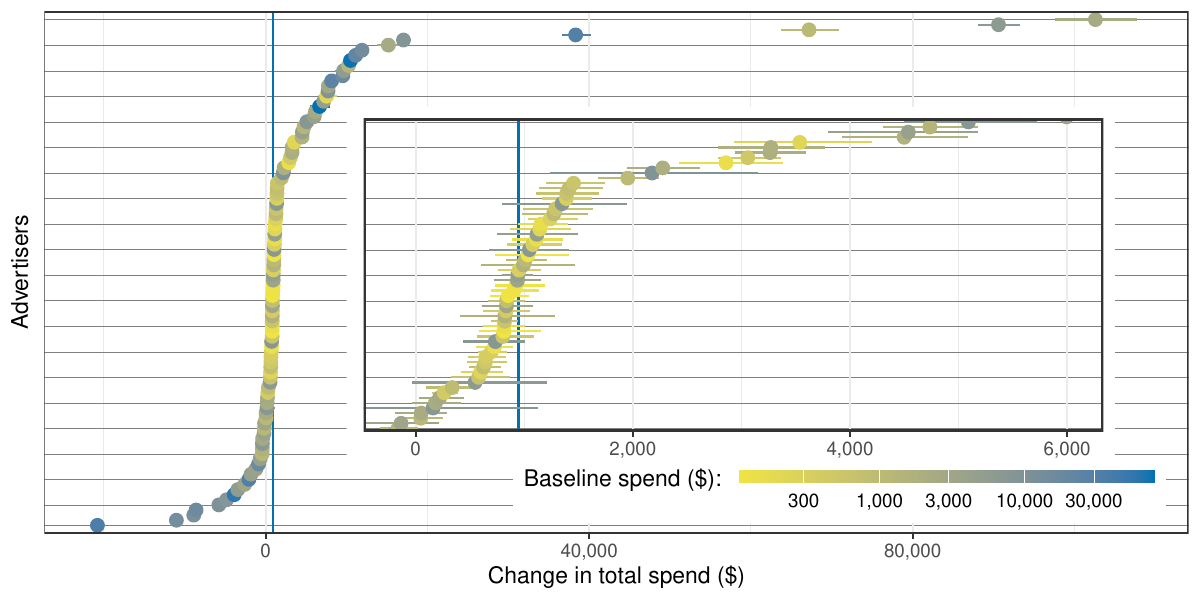}
\par\end{centering}
\centering{}\caption{Change in Spend by Advertiser Under the Pooling Counterfactual. Inset
shows the inner 60\% of advertisers at higher resolution. Points indicate
posterior means, horizontal bars are bootstrap 95\% CIs for the means.
Advertisers are sorted from lowest to highest mean change in spend.
Vertical line indicates the median outcome. \label{fig:Change-in-Spend-Advertiser}}
\end{figure}

Considering next the implications of pooled information ($C_{P}$
vs $B$), we observe a $\$946$ ($98.3\%$) counterfactual increase
in median spend. Unlike the $C_{F}$ case, advertisers can shift spend
to any publisher, not just those used in the data and baseline case.
Hence pooling information enables advertisers to find better matches
with new publishers, leading to an increase in ad spending in $C_{P}$
compared to $B$. The median advertiser value increases $\$3,621$
($18.3\%$) under $C_{P}$. The final case, $C_{F+P}$, combines both
within-advertiser oracle and pooled information. The combined case's
results largely hew to the pooled information counterfactual, because
the oracle priors apply only to a limited number of sites. Hence,
the solutions across $C_{F+P}$ and $C_{P}$ are similar. The median
advertising spend increases $\$823$ ($76.9\%$) and the median advertiser
welfare increases $\$3,753$ ($26.2\%$). As the information requirements
from the $C_{P}$ case are closest to what could be implemented in
practice, subsequent discussion focuses on the pooled information
scenario, $C_{P}$.

Figure \ref{fig:Change-in-Spend-Advertiser} reports the mean change
in spend in the pooling counterfactual, $C_{P}$, for each of the
100 advertisers. The figure suggests considerable heterogeneity in
spending changes, presumably owing to considerable differences across
advertisers in both information states and match values. Advertisers
with higher levels of spending tend to show the greatest magnitude
changes in counterfactual spend.

\begin{figure}
\centering{}\includegraphics[width=1\textwidth]{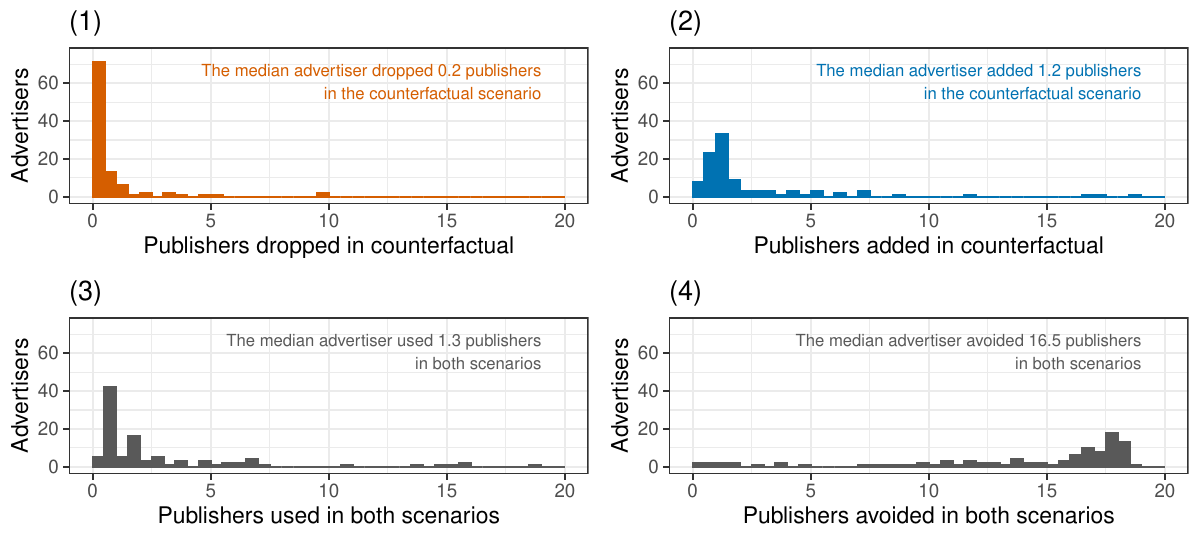}\caption{Change in Publisher Usage by Advertisers Under the Pooling Counterfactual\label{fig:publisher-usage-by-advertiser}.
Each vertical bar represents the number of advertisers that evidenced
a given level change in publisher usage from the baseline to the counterfactual
condition. For example, the first bar in panel (1) indicates that
71 advertisers dropped between 0 and 0.5 publishers on average in
the counterfactual simulations; the second bar in panel (1) indicates
that 13 advertisers dropped between 0.5 and 1 publisher.}
\end{figure}

Figure \ref{fig:publisher-usage-by-advertiser} depicts changes in
the number of publishers used by advertisers under the pooling counterfactual,
$C_{P}$. If the high and incorrect priors dominate, then most advertisers
will decrease the number of publishers used. If the assortive matching
effect dominates, then advertisers will sort into using more sites.
In this regard, the four quadrants of Figure \ref{fig:publisher-usage-by-advertiser}
decompose the aggregate changes in the number of publishers used into
partial effects, representing the number of publishers dropped or
added in the counterfactual (panels 1 and 2), and the number used
or avoided in both scenarios (panels 3 and 4). The lower right quadrant
(panel 4) indicates that most sites not chosen by advertisers under
$B$ continue not to be chosen under $C_{P}$. For example, on average,
18 of the 100 advertisers did not use between 17.5 and 18 of the 20
publishers in both the baseline and counterfactual settings. Comparing
the histogram in the upper left cell (panel 1) to the one in the upper
right (panel 2) indicates that i) more new publishers were added by
advertisers than were removed, as advertisers found good matches;
and ii) pooling information induces considerable change in advertiser
behavior. Overall, this figure suggests that, rather than concentrating
on a single publisher or merely eliminating sites where prior beliefs
were too high, advertisers seem to be sorting into better matches
(that is, finding sites that generate higher advertising value).

\subsection{Publisher and Platform Welfare\label{subsec:supply-side}}

\begin{table}
\centering{}\smaller%
\begin{tabular}{cccc}
\toprule 
 &  & \multicolumn{2}{c}{Own Information}\tabularnewline
 &  & \multicolumn{1}{c}{No Information} & \multicolumn{1}{c}{Full Information}\tabularnewline
\midrule 
\multirow{2}{*}{Shared Information} & No Sharing & \$0 & $-\$2,810$ ($-13.5\%$)\tabularnewline
 & Pooled Information & \$13,558 ($77.7\%$) & $\$10,403$ ($59.9\%$)\tabularnewline
\bottomrule
\end{tabular}\caption{Publisher Counterfactual Revenue Outcomes\label{tab:Publisher-Counterfactual-Outcomes}}
\end{table}

\begin{figure}[b]
\centering{}\includegraphics[scale=0.8]{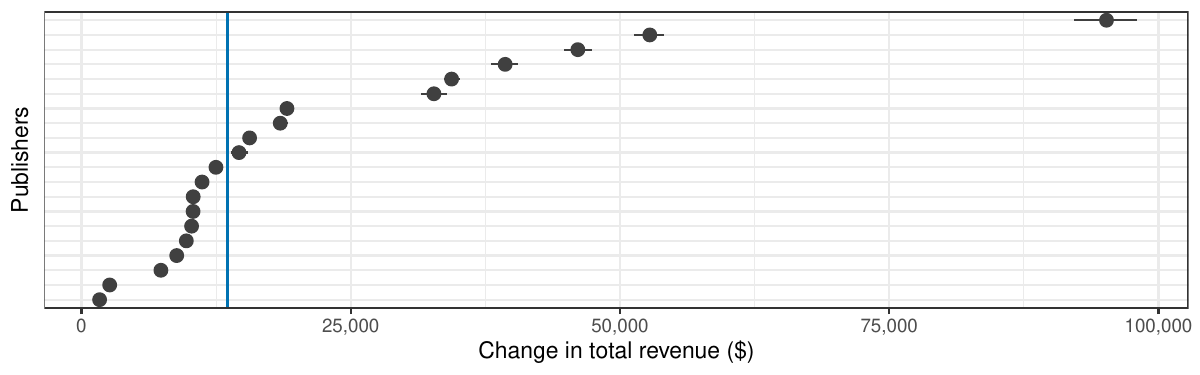}\caption{Change in Publisher Revenue by Site Under the Pooling Counterfactual\label{fig:Change-in-Publisher-Revenue}.
Points indicate posterior means, horizontal bars are bootstrap 95\%
CIs for the means. Publishers are sorted from lowest to highest mean
change in revenue. Vertical line indicates the median outcome.}
\end{figure}

Table \ref{tab:Publisher-Counterfactual-Outcomes} reports, and Figure
\ref{fig:Change-in-Publisher-Revenue} depicts how revenues change
across the 20 publishers under the pooling counterfactual. The network
collects a flat percentage of publisher revenues, thus these changes
are proportional to the network's gains and losses.

Two effects of information provision---that is, endowing advertisers
with better priors---are possible. First, as advertisers are generally
over-optimistic about the sites they chose in the estimation sample,
demand at publisher sites could decrease and revenues could fall.
Comparing $C_{F}$ to $B$ indicates that 17 of the 20 publishers
lose revenue and that the median revenue loss is $\$2,810$ ($-13.5\%$),
because the optimistic advertiser prior effect dominates under $C_{F}$.
Advertisers realize that they were too optimistic, but because they
are not provided information about better potential matches, they
cut (but do not reallocate) spending. In such a scenario, it is likely
that the network and publishers would not seek to inform advertisers
that their priors are too high, as the network and publisher sites
would lose money. However, comparing $C_{P}$ to $B$ indicates that
information pooling causes all publishers to gain revenue, with a
median increase of \$13,558 ($77.7\%$). As suggested by Figure \ref{fig:Change-in-Publisher-Revenue},
advertisers can better sort into a match across publisher sites, finding
new options that generate greater value than the set of publishers
selected in the baseline setting. The revenue from $C_{F+P}$ is lower
than $C_{P}$ because the negative effect on spending from correcting
advertisers' optimistic priors slightly offsets the positive effect
from better matching. Under both $C_{P}$ and $C_{F+P}$, the sorting
effect dominates the prior effect, and the pooled information mechanism
generates not only positive welfare outcomes for advertisers, but
also for publishers.

\begin{figure}[t]
\centering{}\includegraphics[scale=0.8]{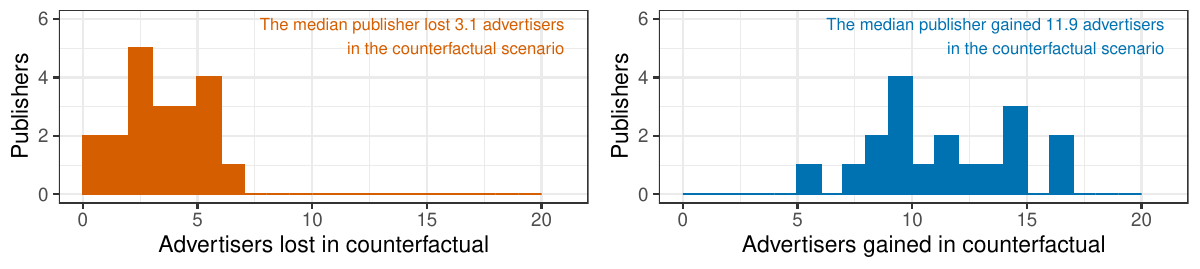}\caption{Distribution of Advertisers Gained and Lost by Publishers Under the
Pooling Counterfactual\label{fig:Change-in-Advertiser-Composition}.
Each vertical bar represents the number of publishers that evidenced
a given level change in advertisers across the baseline and counterfactual
condition. For example, the first bar in the left hand panel indicates
that two publishers lost between 0 and 1 advertisers.}
\end{figure}

Focusing again on the pooling counterfactual, $C_{P}$, we see that
publishers not only gain revenue, but they also tend to serve more
advertisers. Figure \ref{fig:Change-in-Advertiser-Composition} reports
the distribution (across publishers) of the number of advertisers
lost (left panel) and gained (right panel) in $C_{P}$ versus $B$.
Comparing the two distributions, it is evident that more advertisers
are gained than lost, as the gains from enhancing match (enabling
advertisers to find higher value publishers) more than offset the
losses from rectifying advertisers' over-optimistic priors.

Collectively, the counterfactual results suggest that the ad network
would share information to improve advertisers' match information,
because all publishers gain revenue under the pooling counterfactual
and the network's change in revenue is proportional to these gains.
Moreover, the information needed to impute advertiser CTRs is readily
available to the network so there would be few barriers to implementing
such a system. Yet two considerations present. First, the network
might be reticent to provide advertisers with imputed CTRs, as they
might be construed as guarantees on ad performance. As an alternative,
the network might choose to suggest a set of high-match sites. Second,
although the largest publishers (i.e., the top 20) in the data set
all see increases in revenue, the same might not be true for the smallest
publishers. However, given the low baseline of revenues at these sites,
there is a floor on how much these sites can be hurt by lower advertiser
spend, but no ceiling on how much they can benefit from better sorting.

\subsection{Counterfactual Summary\label{subsec:Counterfactual-Summary}}

The total increase in revenue across 20 publishers is $\$453,167$;
the median advertiser's welfare gain is $\$3,621$. We therefore estimate
the combined welfare gains across the 100 advertisers and 20 publishers
over the six-month window in the estimation data to be on the order
of $\$800\textrm{K}$ under the pooling counterfactual. The median
advertiser's spending rises $\$946$. Extrapolating the spending increase
from the random sample of 100 advertisers to the population of 8,000
advertisers suggests the platform's revenue could grow its total revenue
on the order of $\$7.5\mathrm{M}$ (an analogous extrapolation across
publishers is misleading, as our sample focuses on the 20 largest).
To the extent a similar pooled approach could be used across multiple
ad networks, or other ad channels such as retail media, gains would
be larger.

As a caveat, it should be noted that our counterfactual analysis constitutes
a partial equilibrium. Publisher sites, faced with increased demand,
could raise subscription prices. Although, as Appendix \ref{online-subsec:Pricing}
indicates, little such evidence for price response is observed in
the data. To the extent this does occur, advertiser gains could be
somewhat diminished and publisher gains could be somewhat decreased.
However, most gains in our analysis arise from better sorting, and
there is little concentration in advertisers across sites (see Online
Appendix \ref{online-subsec:Crowding}). As suggested by Figure \ref{fig:Change-in-Advertiser-Composition},
sites tend to face a change in advertiser composition owing to better
sorting, rather than having all advertisers concentrate into a single
site. To the extent sites are imperfect substitutes, this lack of
concentration should offset the tendency of any dominant site to substantially
raise prices. In sum, changes in advertising information could have
downstream consequences for firms' pricing strategies that offset
some of their gains by increasing or decreasing price competition;
these issues are beyond the scope of our research.

\section{Conclusions\label{sec:Conclusions}}

This paper considers the implications of advertiser learning in direct
display advertising markets. In contrast to exchange markets, direct
advertising is bought in bulk (many thousands of impressions at a
time) rather than sequentially, which limits advertisers' opportunities
to use test and learn strategies to efficiently identify publishers
whose audiences are a good match for their ads \citep{Tunuguntlan_Hoban_2020}.
In direct markets, advertisers' initial information states are economically
consequential because, lacking the ability to test and learn incrementally,
they are effectively guessing which publishers to use. Hence, there
is a considerable potential to choose sites incorrectly and, as a
corollary, to enhance advertiser (and possibly publisher) welfare
by endowing advertisers with better information.

To ascertain whether advertisers have incorrect beliefs and assess
the potential consequences thereof, we collect direct sales data from
an ad exchange. The data describe ad purchases from the exchange's
inception, providing an ideal setting to explore the nature of advertiser
learning as advertisers enter the network. Patterns in these data
suggest evidence of advertiser learning. Advertisers initially run
ads at relatively many sites and, after additional exploration, ultimately
settle on a smaller number where they presumably find greater customer
response. Using a direct utility model of advertiser choices of ad
subscriptions across sites, in which match is allowed to be learned
from past advertising outcomes, we document that advertisers' initial
priors are quite misinformed. Advertisers overestimate initial CTRs
by a factor of four, as median beliefs are 0.177\%, compared to a
median CTR of 0.045\%. In short, advertisers are initially far too
optimistic about the publishers they have chosen for advertising.

We consider various mechanisms to redress this problem and ascertain
the impact of advertiser information on advertiser welfare and publisher
revenues. We consider two such mechanisms. The first uses advertiser
outcomes at the end of the data to inform their priors at the start
of the data. While this suggests how much advertiser outcomes could
potentially improve on the sites they actually explored, the approach
cannot inform us about how well advertisers would fare on sites they
did not explore. Moreover, the mechanism is infeasible, as it is impossible
to know ad outcomes ex-ante. Hence, we consider a second mechanism
wherein the platform can pool its information across advertiser on
ad outcomes, drawing on similar advertisers who previously placed
ads on the considered sites. We operationalize this information sharing
mechanism using Google's Cloud Vision API to measure similarities
between ads and advertisers, and then imputing CTRs for advertisers
at new publisher sites. The approach is predictive of advertising
performance.

Endowing advertisers with pooled information leads to a median advertiser
increase in ad spend of $98.3\%$ ($\$946$), and a median advertiser
welfare gain of $18.3\%$ ($\$3,621$). On the publisher side, there
exist two opposing forces: Rectifying optimistic priors should decrease
ad spend and publisher revenue, while better sorting between advertisers
and publishers should increase revenue. The latter effect dominates,
generating a median publisher gain of $77.7\%$ ($\$13,558$). All
publishers gain in revenue. In other words, redressing advertiser
uncertainty in direct ad networks leads to large welfare gains for
publishers and advertisers alike. Extrapolating these gains across
all 8,000 advertisers in our considered network suggests platform
revenue gains up to $\$7.5\mathrm{M}$ over six months. The information
pooling approach, if ported to other direct advertising networks,
would yield even greater gains. Of note, direct ad purchasing also
exists in non-digital channels as well, such as television, radio,
and out of home. It stands to reason that advertisers also face a
learning problem in those contexts, but the number of publisher alternatives
in digital dwarfs these other channels, and likely exacerbates the
learning problem and potential gains from redressing it.

Advertiser learning will become a more substantial research topic
and practical opportunity in the coming years. Because advertisers
are often ill-informed, poor advertiser decisions can induce advertisers
to cease advertising, creating an incentive to shift decision rights
from the advertiser to publishers and platforms. Examples of this
shift already include platform-recommended ad delivery optimization,
audience selection, and bidding, as well as generative AI to create
ad content. In all these instances, platforms and publishers pool
information across advertisers to enhance advertisers' overall outcomes.
As the ad ecosystem becomes more complex, the opportunities to pool
information grow. In the face of cookie deprecation, these opportunities
are especially relevant for first-party sites, including platforms
and publishers such as those this paper considers. Better utilization
of information, such as ascertaining whether advertisers learn site
match at the level of individual creatives, rather than across all
creatives, could generate additional welfare gains for advertisers,
platforms, and publishers.\label{rev:learn_at_site-creative}

A related question of interest is why advertisers are ill-informed.
In our context, their priors are far too optimistic. Possible reasons
include advertisers' use of market research that lacks causal variation,
and potentially overstated claims of ad effects by publishers and
ad networks.\footnote{The authors thank Nils Wernerfelt for sharing these observations.}

Our analysis focuses on the advertiser demand side. An interesting
potential extension is to consider how publisher ad pricing responds
to changes in advertiser demand due to informing priors. To the extent
that advertiser demand increases, publishers can raise ad prices and
increase their share of total welfare gains at the cost of advertisers.
That said, efficient sorting can reduce the concentration of advertisers
on some sites and raise it on others, making the overall effect ambiguous
and highly variable across publishers. Given that publishers face
a large competitive set in the face of dynamic demand as a result
of learning, the supply side problem would be a challenging, but useful
extension in the analysis of direct display advertising markets. Another
issue of interest is how to balance inventory and pricing across both
direct and exchange markets \citep{Baleiro_et_al_2014}, a challenge
that is also exacerbated by advertiser learning. In sum, we hope this
research sparks more interest on these and other topics in the large
and economically consequential direct advertising market.

\bibliographystyle{informs2014}
\phantomsection\addcontentsline{toc}{section}{\refname}\bibliography{advertiser_learning}

\begin{thebibliography}{52}
\providecommand{\natexlab}[1]{#1}
\providecommand{\url}[1]{\texttt{#1}}
\providecommand{\urlprefix}{}

\bibitem[{Ahmadi et~al.(2023)Ahmadi, Abou~Nabout, Skiera, Maleki, \protect\BIBand{} Fladenhofer}]{Ahmadi2023}
Ahmadi I, Abou~Nabout N, Skiera B, Maleki E, Fladenhofer J (2023) Overwhelming targeting options: Selecting audience segments for online advertising. \emph{International Journal of Research in Marketing} 41(1):24--40, ISSN 0167-8116.

\bibitem[{Alcobendas \protect\BIBand{} Zeithammer(2021)}]{Alcobendas2021}
Alcobendas M, Zeithammer R (2021) Adjustment of bidding strategies after a switch to first-price rules. \emph{SSRN Electronic Journal} ISSN 1556-5068.

\bibitem[{Allouah \protect\BIBand{} Besbes(2017)}]{Allouah_Besbes_2017}
Allouah A, Besbes O (2017) Auctions in the online display advertising chain: A case for independent campaign management. \emph{SSRN Electronic Journal} ISSN 1556-5068.

\bibitem[{Athey \protect\BIBand{} Ellison(2011)}]{Athey2011}
Athey S, Ellison G (2011) Position auctions with consumer search. \emph{The Quarterly Journal of Economics} 126(3):1213--1270, ISSN 1531-4650.

\bibitem[{Auer et~al.(2002)Auer, Cesa-Bianchi, \protect\BIBand{} Fischer}]{auer2002finite}
Auer P, Cesa-Bianchi N, Fischer P (2002) Finite-time analysis of the multiarmed bandit problem. \emph{Machine learning} 47:235--256.

\bibitem[{Bajari et~al.(2007)Bajari, Benkard, \protect\BIBand{} Levin}]{Bajari2007}
Bajari P, Benkard CL, Levin J (2007) Estimating dynamic models of imperfect competition. \emph{Econometrica} 75(5):1331--1370, ISSN 1468-0262.

\bibitem[{Balseiro et~al.(2015)Balseiro, Besbes, \protect\BIBand{} Weintraub}]{Balseiro2015}
Balseiro SR, Besbes O, Weintraub GY (2015) Repeated auctions with budgets in ad exchanges: Approximations and design. \emph{Management Science} 61(4):864--884, ISSN 1526-5501.

\bibitem[{Balseiro \protect\BIBand{} Candogan(2017)}]{Balseiro2017}
Balseiro SR, Candogan O (2017) Optimal contracts for intermediaries in online advertising. \emph{Operations Research} 65(4):878--896, ISSN 1526-5463.

\bibitem[{Balseiro et~al.(2014)Balseiro, Feldman, Mirrokni, \protect\BIBand{} Muthukrishnan}]{Baleiro_et_al_2014}
Balseiro SR, Feldman J, Mirrokni V, Muthukrishnan S (2014) Yield optimization of display advertising with ad exchange. \emph{Management Science} 60(12):2886--2907.

\bibitem[{Balseiro \protect\BIBand{} Gur(2019)}]{Balseiro2019}
Balseiro SR, Gur Y (2019) Learning in repeated auctions with budgets: Regret minimization and equilibrium. \emph{Management Science} 65(9):3952--3968, ISSN 1526-5501.

\bibitem[{Benkard(2000)}]{Benkard2000}
Benkard CL (2000) Learning and forgetting: The dynamics of aircraft production. \emph{American Economic Review} 90(4):1034--1054, ISSN 0002-8282.

\bibitem[{Berry(1992)}]{Berry1992}
Berry ST (1992) Estimation of a model of entry in the airline industry. \emph{Econometrica} 60(4):889, ISSN 0012-9682.

\bibitem[{Bhat(2008)}]{Bhat2008}
Bhat CR (2008) The multiple discrete-continuous extreme value (mdcev) model: Role of utility function parameters, identification considerations, and model extensions. \emph{Transportation Research Part B: Methodological} 42(3):274--303, ISSN 0191-2615.

\bibitem[{Bimpikis et~al.(2020)Bimpikis, Elmaghraby, Moon, \protect\BIBand{} Zhang}]{Bimpikis2020}
Bimpikis K, Elmaghraby WJ, Moon K, Zhang W (2020) Managing market thickness in online business-to-business markets. \emph{Management Science} 66(12):5783--5822, ISSN 1526-5501.

\bibitem[{Bompaire et~al.(2021)Bompaire, Gilotte, \protect\BIBand{} Heymann}]{bompaire2021causal}
Bompaire M, Gilotte A, Heymann B (2021) Causal models for real time bidding with repeated user interactions. \emph{Proceedings of the 27th ACM SIGKDD Conference on Knowledge Discovery \& Data Mining}, 75--85.

\bibitem[{Cai et~al.(2017)Cai, Ren, Zhang, Malialis, Wang, Yu, \protect\BIBand{} Guo}]{Cai2017}
Cai H, Ren K, Zhang W, Malialis K, Wang J, Yu Y, Guo D (2017) Real-time bidding by reinforcement learning in display advertising. \emph{Proceedings of the Tenth ACM International Conference on Web Search \& Data Mining}, 661--670, WSDM 2017 (New York, NY, USA: Association for Computing Machinery), ISBN 9781450346757.

\bibitem[{Ching et~al.(2013)Ching, Erdem, \protect\BIBand{} Keane}]{Ching2013}
Ching AT, Erdem T, Keane MP (2013) Learning models: An assessment of progress, challenges, and new developments. \emph{Marketing Science} 32(6):913--938, ISSN 1526-548X.

\bibitem[{Choi et~al.(2020)Choi, Mela, Balseiro, \protect\BIBand{} Leary}]{Choi_et_al_2020}
Choi H, Mela CF, Balseiro SR, Leary A (2020) Online display advertising markets: A literature review and future directions. \emph{Information Systems Research} 31(2):556--575, ISSN 1526-5536.

\bibitem[{Choi \protect\BIBand{} Sayedi(2019)}]{Choi2019}
Choi WJ, Sayedi A (2019) Learning in online advertising. \emph{Marketing Science} 38(4):584--608, ISSN 1526-548X.

\bibitem[{Dub{\'e} et~al.(2005)Dub{\'e}, Hitsch, \protect\BIBand{} Manchanda}]{dube2005empirical}
Dub{\'e} JP, Hitsch GJ, Manchanda P (2005) An empirical model of advertising dynamics. \emph{Quantitative Marketing and Economics} 3(2):107--144, ISSN 1573-711X.

\bibitem[{Feit \protect\BIBand{} Berman(2019)}]{Feit_Berman_2019}
Feit EM, Berman R (2019) Test \& roll: Profit-maximizing a/b tests. \emph{Marketing Science} 38(6):1038--1058, ISSN 1526-548X.

\bibitem[{Gope \protect\BIBand{} Jain(2017)}]{Gope2017}
Gope J, Jain SK (2017) A survey on solving cold start problem in recommender systems. \emph{2017 International Conference on Computing, Communication and Automation (ICCCA)}, 133--138 (IEEE).

\bibitem[{Guan \protect\BIBand{} Jiang(2018)}]{guan2018nonparametric}
Guan M, Jiang H (2018) Nonparametric stochastic contextual bandits. volume~32 (Association for the Advancement of Artificial Intelligence (AAAI)), ISSN 2159-5399.

\bibitem[{Hendel \protect\BIBand{} Spiegel(2014)}]{Hendel2014}
Hendel I, Spiegel Y (2014) Small steps for workers, a giant leap for productivity. \emph{American Economic Journal: Applied Economics} 6(1):73--90, ISSN 1945-7790.

\bibitem[{Lam et~al.(2008)Lam, Vu, Le, \protect\BIBand{} Duong}]{Lam2008}
Lam XN, Vu T, Le TD, Duong AD (2008) Addressing cold-start problem in recommendation systems. \emph{Proceedings of the 2nd International Conference on Ubiquitous Information Management and Communication}, 208--211, ICUIMC08 (ACM).

\bibitem[{Lee \protect\BIBand{} Allenby(2014)}]{Lee2014}
Lee S, Allenby GM (2014) Modeling indivisible demand. \emph{Marketing Science} 33(3):364--381, ISSN 1526-548X.

\bibitem[{Levitt et~al.(2013)Levitt, List, \protect\BIBand{} Syverson}]{Levitt2013}
Levitt SD, List JA, Syverson C (2013) Toward an understanding of learning by doing: Evidence from an automobile assembly plant. \emph{Journal of Political Economy} 121(4):643--681, ISSN 1537-534X.

\bibitem[{Lika et~al.(2014)Lika, Kolomvatsos, \protect\BIBand{} Hadjiefthymiades}]{Lika2014}
Lika B, Kolomvatsos K, Hadjiefthymiades S (2014) Facing the cold start problem in recommender systems. \emph{Expert Systems with Applications} 41(4):2065--2073, ISSN 0957-4174.

\bibitem[{Little \protect\BIBand{} Lodish(1969)}]{Little1969}
Little JD, Lodish LM (1969) A media planning calculus. \emph{Operations Research} 17(1):1--35, ISSN 1526-5463.

\bibitem[{Liu \protect\BIBand{} Toubia(2018)}]{Liu2018}
Liu J, Toubia O (2018) A semantic approach for estimating consumer content preferences from online search queries. \emph{Marketing Science} 37(6):930--952, ISSN 1526-548X.

\bibitem[{Perlich et~al.(2012)Perlich, Dalessandro, Hook, Stitelman, Raeder, \protect\BIBand{} Provost}]{Perlich2012}
Perlich C, Dalessandro B, Hook R, Stitelman O, Raeder T, Provost F (2012) Bid optimizing and inventory scoring in targeted online advertising. \emph{Proceedings of the 18th ACM SIGKDD International Conference on Knowledge Discovery \& Data Mining}, 804--812, KDD 12 (ACM).

\bibitem[{Ren et~al.(2018)Ren, Zhang, Chang, Rong, Yu, \protect\BIBand{} Wang}]{Ren2018}
Ren K, Zhang W, Chang K, Rong Y, Yu Y, Wang J (2018) Bidding machine: Learning to bid for directly optimizing profits in display advertising. \emph{IEEE Transactions on Knowledge and Data Engineering} 30(4):645--659, ISSN 1041-4347.

\bibitem[{Schein et~al.(2002)Schein, Popescul, Ungar, \protect\BIBand{} Pennock}]{Schein2002}
Schein AI, Popescul A, Ungar LH, Pennock DM (2002) Methods and metrics for cold-start recommendations. \emph{Proceedings of the 25th Annual International ACM SIGIR Conference on Research and Development in Information Retrieval}, 253--260, SIGIR02 (New York, NY, USA: Association for Computing Machinery), ISBN 1581135610.

\bibitem[{Schwartz et~al.(2017)Schwartz, Bradlow, \protect\BIBand{} Fader}]{Schwartz2017}
Schwartz EM, Bradlow ET, Fader PS (2017) Customer acquisition via display advertising using multi-armed bandit experiments. \emph{Marketing Science} 36(4):500--522, ISSN 1526-548X.

\bibitem[{Scott(2010)}]{Scott2010}
Scott SL (2010) A modern bayesian look at the multi-armed bandit. \emph{Applied Stochastic Models in Business and Industry} 26(6):639--658, ISSN 1526-4025.

\bibitem[{Silge \protect\BIBand{} Robinson(2016)}]{Silge2016}
Silge J, Robinson D (2016) tidytext: Text mining and analysis using tidy data principles in r. \emph{The Journal of Open Source Software} 1(3):37, ISSN 2475-9066.

\bibitem[{Simpson et~al.(2017)Simpson, Rue, Riebler, Martins, \protect\BIBand{} S{\o}rbye}]{Simpson2017}
Simpson D, Rue H, Riebler A, Martins TG, S{\o}rbye SH (2017) Penalising model component complexity: A principled, practical approach to constructing priors. \emph{Statistical Science} 32(1), ISSN 0883-4237.

\bibitem[{{Stan Development Team}(2017)}]{StanTeam2017}
{Stan Development Team} (2017) Rstan: the r interface to stan. r package version 2.16.2. \urlprefix\url{http://mc-stan.org}.

\bibitem[{Tadelis et~al.(2023)Tadelis, Hooton, Manjeer, Deisenroth, Wernerfelt, Dadson, \protect\BIBand{} Greenbaum}]{Tadelis_et_al_2023}
Tadelis S, Hooton C, Manjeer U, Deisenroth D, Wernerfelt N, Dadson N, Greenbaum L (2023) Learning, sophistication, and the returns to advertising: Implications for differences in firm performance. Working Paper 31201, National Bureau of Economic Research.

\bibitem[{Theussl \protect\BIBand{} Hornik(2023)}]{Theussl2023}
Theussl S, Hornik K (2023) \emph{Rglpk: R/GNU Linear Programming Kit Interface}. \urlprefix\url{https://CRAN.R-project.org/package=Rglpk}, r package version 0.6-5.

\bibitem[{Toubia \protect\BIBand{} Netzer(2017)}]{Toubia2017}
Toubia O, Netzer O (2017) Idea generation, creativity, and prototypicality. \emph{Marketing Science} 36(1):1--20, ISSN 1526-548X.

\bibitem[{Tunuguntla(2022)}]{tunuguntla_2022}
Tunuguntla S (2022) \emph{Display ad measurement using observational data: A reinforcement learning approach}. Ph.D. thesis, University of Wisconsin-Madison.

\bibitem[{Tunuguntla \protect\BIBand{} Hoban(2020)}]{Tunuguntlan_Hoban_2020}
Tunuguntla S, Hoban PR (2020) A near-optimal bidding strategy for real-time display advertising auctions. \emph{Journal of Marketing Research} 58(1):1--21, ISSN 1547-7193.

\bibitem[{Varian(2007)}]{Varian2007}
Varian HR (2007) Position auctions. \emph{International Journal of Industrial Organization} 25(6):1163--1178, ISSN 0167-7187.

\bibitem[{Vehtari et~al.(2016)Vehtari, Gelman, \protect\BIBand{} Gabry}]{Vehtari2016}
Vehtari A, Gelman A, Gabry J (2016) Practical bayesian model evaluation using leave-one-out cross-validation and waic. \emph{Statistics and Computing} 27(5):1413--1432, ISSN 1573-1375.

\bibitem[{Waisman et~al.(2019)Waisman, Nair, \protect\BIBand{} Carrion}]{Waisman2019}
Waisman C, Nair HS, Carrion C (2019) Online causal inference for advertising in real-time bidding auctions.

\bibitem[{Wang et~al.(2025)Wang, Tao, \protect\BIBand{} Zhang}]{Wang_et_al_2025}
Wang Y, Tao L, Zhang XX (2025) Recommending for a multi-sided marketplace: A multi-objective hierarchical approach. \emph{Marketing Science} 44(1):1--29, ISSN 1526-548X.

\bibitem[{Wu(2015)}]{Wu2015}
Wu C (2015) Matching value and market design in online advertising networks: An empirical analysis. \emph{Marketing Science} 34(6):906--921, ISSN 1526-548X.

\bibitem[{Xu et~al.(2022)Xu, Deng, \protect\BIBand{} Mela}]{Xu2022}
Xu B, Deng Y, Mela CF (2022) A scalable recommendation engine for new users and items. \emph{SSRN Electronic Journal} ISSN 1556-5068.

\bibitem[{Yao et~al.(2023)Yao, Kong, Lu, Bai, Yang, \protect\BIBand{} Xiong}]{Yao2023}
Yao Z, Kong D, Lu M, Bai X, Yang J, Xiong H (2023) Multi-view multi-task campaign embedding for cold-start conversion rate forecasting. \emph{IEEE Transactions on Big Data} 9(1):280--293, ISSN 2372-2096.

\bibitem[{Zhou et~al.(2020)Zhou, Li, \protect\BIBand{} Gu}]{Zhou2019}
Zhou D, Li L, Gu Q (2020) Neural contextual bandits with ucb-based exploration. III HD, Singh A, eds., \emph{Proceedings of the 37th International Conference on Machine Learning}, volume 119 of \emph{Proceedings of Machine Learning Research}, 11492--11502 (PMLR), \urlprefix\url{https://proceedings.mlr.press/v119/zhou20a.html}.

\bibitem[{Zhuo(2023)}]{zhuo2022exploit}
Zhuo R (2023) Exploit or explore? {An} empirical study of resource allocation in scientific labs. Technical report, Working Paper.

\end{thebibliography}

\clearpage{}

\appendix
\begin{center}
\normalfont\LARGE Appendices
\end{center}
\titleformat{\section}{\normalfont\Large\bfseries}{Appendix \thesection}{1em}{}

\section{Simulation Procedure\label{appx:Simulation-Procedure}}

All counterfactual scenarios share a common simulation procedure based
on 500 draws from the posterior distribution of the model parameters,
$\theta$. For each draw of $\theta$, $\epsilon$ is sampled from
the posterior predictive distribution of $\epsilon|\theta$, and the
same $\epsilon$ vector is used for each baseline/counterfactual pair
(each counterfactual is paired with a unique baseline simulation).
Most simulated choices by advertiser $a$ at site $s$ in week $w$
have a corresponding observed choice; for these simulated choices,
$\epsilon_{asw}$ is sampled from a distribution truncated by the
upper and lower bounds defined in Equations (\ref{eq:lb}) and (\ref{eq:ub}),
conditional on $\theta$. If there is not a corresponding choice in
the data (e.g., a subscription lasting $x>7$ days starting in week
$w$ at site $s$ prevents a choice at site $s$ in week $w+1$)---then
$\epsilon_{asw}$ is sampled from an unrestricted distribution. We
average simulated choices over draws of $\theta$ to obtain posterior
predictive means for the outcomes of interest.

Using the set of available ad durations, $x$, and prices, $p$, from
the estimation sample, purchases for advertiser $a$ are simulated
starting with week $w=1$ (or the first week the advertiser was observed
in the network, if the advertiser joined during the estimation period).
For each site $s$, advertiser $a$ chooses an optimal number of days
of advertising, $x$. If $x=0$, advertiser $a$ does not place an
ad at site $s$ in week $w$. Otherwise, $x>0$ and $n^{I}=x\cdot t_{s}$
impressions are delivered, yielding $n^{C}\big|n^{I},c_{as}\,\sim\mathrm{Binomial}(n^{I},c_{as})$
clicks. If $0\leq x\leq7$, then the next advertising choice for site
$s$ occurs in week $w+1$; otherwise the next choice occurs in a
later week, depending on the value of $x$ . The procedure repeats
for each site, and then advances to the next week. Choices in subsequent
weeks are informed by Bayesian updating on $\tilde{c}_{as}$ based
on the cumulative number of impressions, $n_{asw}^{I}$, and clicks,
$n_{asw}^{C}$, obtained from previous ad buys.

The outcomes of interest from these simulations are i) the amount
each advertiser spends in total or at each site, which is obtained
directly from the simulations; and ii) a measure of advertiser welfare.
Regarding the latter: because advertisers choose sites based on net
expected payoff (Equation (\ref{eq:advertiser_payoff})), expected
net payoffs are transformed into a ``true'' net payoff by replacing
the simulated value of $\mathbb{E}\left[\tilde{c}_{asw}|n_{asw}^{I},n_{asw}^{C}\right]$
with $\mathbb{E}\left[\tilde{c}_{asw}|n^{I}=10^{12},n^{C}=\left\lfloor 10^{12}c_{as}\right\rfloor \right]$.
We use $\mathbb{E}\left[\tilde{c}_{asw}|n^{I}=10^{12},n^{C}=\left\lfloor 10^{12}c_{as}\right\rfloor \right]$
rather than $c_{asw}$ because the procedure for counterfactually
manipulating advertisers' prior beliefs entails altering their state
space ($n^{I}$ and $n^{C}$) in a similar manner. Using $c_{asw}$
instead would introduce simulation error due to differences in floating
point precision for large numbers.\footnote{Owing to path dependencies in the simulation, it is possible for choices
in the baseline and counterfactual with common values of $a$, $s$,
and $w$, to depend on different time-varying fixed effects, $\phi_{\tau[a,s,w]}$.
In some cases, this leads to large differences in welfare that are
not meaningfully related to the most important differences between
the baseline and counterfactual scenarios. Hence, in these cases,
the baseline and counterfactual $\phi$s are replaced with their simple
average.}

All simulations incorporate the advertiser participation constraint.
Economic profits must be positive (i.e., $\pi_{asw}\left(x\right)>0$),
else advertisers would not advertise. For example, should match become
sufficiently low in the face of new information, an advertiser will
cease to advertise.

\section{Measuring Advertising Similarity\label{appx:GoogleAPI}}

\begin{figure}
\centering{}\includegraphics{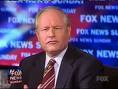}\caption{\label{fig:Example-of-an-ad}Example of an Ad Image in the Dataset.}
\end{figure}

To tag ad images, we use Google's Cloud Vision API, a pre-trained
machine learning model developed by Google that ingests images as
input and returns concept tags as output. Using this product, we query
the set of 10 tags that best characterize each ad image in our dataset.\footnote{A robustness check uses 20 tags (when available from the API) and
results are qualitatively the same.} Figure \ref{fig:Example-of-an-ad} depicts an example ad from the
dataset, and Google's Cloud Vision API retrieves the following tags
for this image: television program, television presenter, tie, news,
font, display device, event, cable television, electric blue and public
speaking.

Using the ad image tags returned by Google's Cloud Vision API, we
collect all tags describing all creatives for each advertiser, count
the frequency for every advertiser-tag pair, and construct a TF-IDF
matrix of advertisers by tags \citep{Silge2016}.\footnote{For a marketing application of TF-IDF, see \citet{Toubia2017}.}
For the purpose of computing advertiser similarity, distinctive characterizations
of an image are most informative. Hence, TF-IDF is an attractive metric,
as it upweights ad tags that are relatively unique, and down-weights
commonly appearing tags.

\begin{figure}[h]
\centering{}\includegraphics[width=1\textwidth]{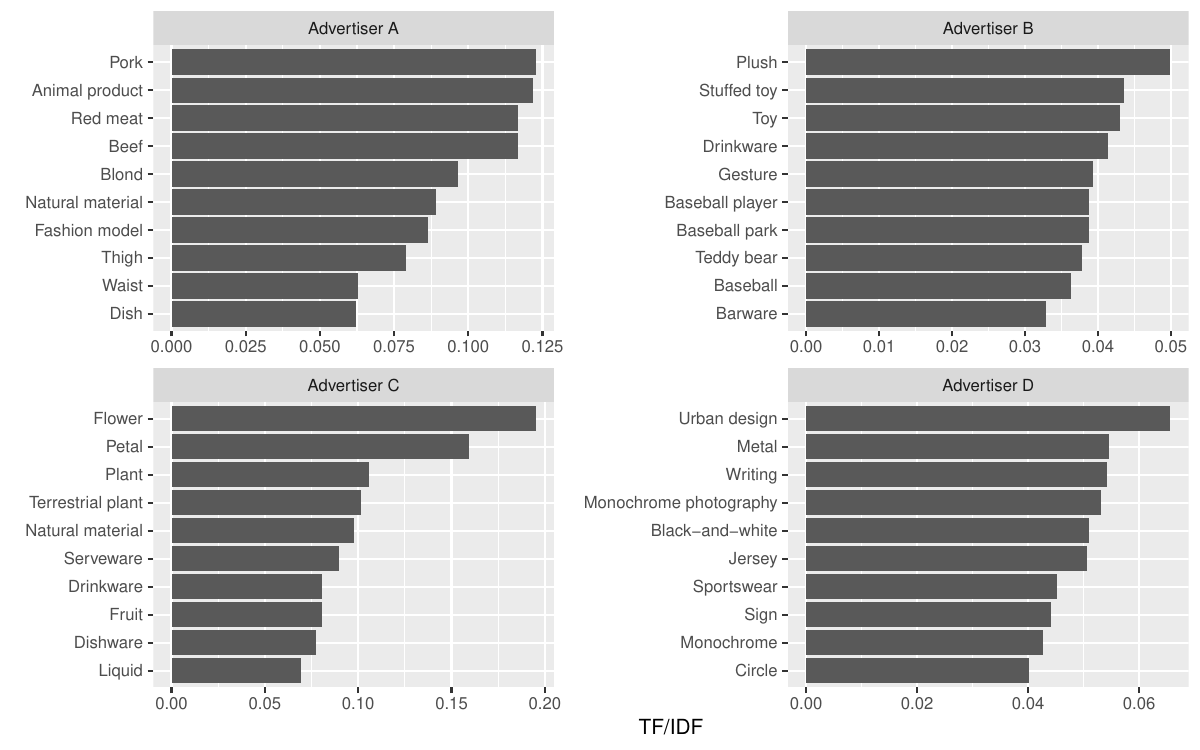}\caption{Top 10 Tags For The Top 4 Advertisers. Advertiser A placed the most
ads in the data set, Advertiser D the fourth most. Depicts the ten
tags with the highest TF-IDF for each advertiser. \label{fig:Top-10-TF-IDF}}
\end{figure}

Figure \ref{fig:Top-10-TF-IDF} reports the top 10 TF-IDF tags for
the four advertisers appearing most frequently in the data. The most
frequent advertiser (Advertiser A) is primarily characterized by tags
related to food. Similarly, the third most frequent advertiser (Advertiser
C) purveys products related to plants. In this sense, the TF-IDF describes
advertisers based on the ads shown in their campaigns.

To measure the similarity between advertisers based on their ad images
tags, we compute the cosine similarity of their TF-IDF vectors. The
cosine similarity between two vectors $a$ and $b$ is $f(a,b)=\frac{a\cdot b}{||a||\,||b||}=\frac{\sum_{k}a_{k}b_{k}}{\sqrt{\sum_{k}a_{k}^{2}}\sqrt{\sum_{k}b_{k}^{2}}}$
and ranges from 0 to 1: it is 0 when the ad vectors are orthogonal,
and 1 when the ad vectors are identical.\footnote{See \citet{Liu2018} for an application of cosine similarity in marketing.}
Stacking the vectors of advertisers' cosine similarities with other
advertisers yields a square and symmetric matrix $R$ with 1 in the
main diagonal and entries $r_{aj}$ containing the cosine similarity
between advertiser $a$ and advertiser $j$. The ad copy similarity
matrix $R$ is then used to generate a weighted prediction of CTRs
for advertisers on sites where they have not previously advertised,
per Equation (\ref{eq:ctr-imputed}), under the assumption that similar
items have similar click through rates on a site.

\clearpage
\setstretch{1.45}
\mbox{}\protect\vspace{-3ex}
\begin{center}\normalfont\LARGE Online Appendix\end{center}\vspace{-3ex}
\titleformat{\section}{\normalfont\Large\bfseries}{Online Appendix \thesection}{1em}{}
\setcounter{section}{0}
\setcounter{table}{0}
\setcounter{figure}{0}

\section{Test and Learn\label{online:Test-and-Learn}}

Recently, \citet{zhuo2022exploit} developed an upper confidence bound
(UCB) approach to address test and learn in dynamic models of choice,
and \citet{Wang_et_al_2025} developed a UCB approach based on uncertainty
in a Bayesian posterior distribution. The UCB approach has been shown
to achieve nearly optimal regret relative to the use of exact continuation
values \citep{auer2002finite,guan2018nonparametric,Zhou2019}. This
section first describes the test and learn problem, then discusses
the implementation using the approaches of \citet{zhuo2022exploit}
and \citet{Wang_et_al_2025}.

\subsection{Specification}

Consider the choice of subscription length of $x$ days made by advertiser
$a$ at site $s$ in week $w$. Because choices of the optimal $x$
at each site $s$ are independent, the discussion that follows suppresses
both the $a$ and $s$ subscripts. Denoting $\bar{c}\equiv n^{C}/n^{I}$,
the period (myopic) objective function defined by Equations (\ref{eq:update_match_longer})
and (\ref{eq:V_x}) can be written as 
\begin{gather*}
V(x;\epsilon,n^{I},\bar{c},w)=\frac{1+\bar{c}n^{I}}{1+\gamma n^{I}}\exp\left(\xi+\eta+\phi_{\tau[w]}+\psi_{m[w]}\right)\cdot\exp(\epsilon)\zeta^{-1}\,\log\left(1+tx\zeta\right)-p_{w}(x).
\end{gather*}
The Bellman equation for the corresponding dynamic problem would be
\begin{gather*}
W(\epsilon,n^{I},\bar{c},w)=\sup_{x\in\mathcal{X}_{w}}V(x;\epsilon,n^{I},\bar{c},w)+\beta\int W(\epsilon^{\prime},n^{I\prime},\bar{c}^{\prime},w+1)dF\left(\epsilon^{\prime},n^{I\prime},\bar{c}^{\prime}|x,\epsilon,n^{I},\bar{c},w\right)
\end{gather*}
where $\beta$ is the weekly discount rate, and $\mathcal{X}_{w}$
is the set of available subscription lengths in week $w$ (if a previously
purchased subscription is still running in week $w$, then $\mathcal{X}_{w}=\left\{ 0\right\} $).
Because the $\epsilon$s are i.i.d. across weeks and Bayes rule implies
$\mathbb{E}\left[\bar{c}^{\prime}|n^{I\prime},\bar{c}\right]=\bar{c}$,
the Bellman equation can be simplified to 
\begin{gather*}
W(\epsilon,n^{I},\bar{c},w)=\sup_{x\in\mathcal{X}_{w}}V(x;\epsilon,n^{I},\bar{c},w)+\beta\int\mathbb{E}_{\epsilon^{\prime}}\left[W(\epsilon^{\prime},n^{I\prime},\bar{c},w+1)\right]dF\left(n^{I\prime}|x,n^{I}\right),
\end{gather*}
which shows the link between forward-looking behavior and the choice
of $x$: For larger values of $x$ (i.e. more days of advertising),
$n^{I}$ increases (in expectation by $tx$, where $t$ is expected
daily traffic), increasing the available information about match.
Hence, if there is a discrepancy between choices under myopic and
forward-looking behavior, it is likely due to a trade-off between
a higher value of $x$ that produces a less favorable $V(x)$ in the
current period, but which maximizes the Bellman equation due to its
positive influence on $n^{I}$.\footnote{The evolution of $w$ is also part of the Bellman recursion, as it
determines the values of the time-varying fixed effects $\phi_{\tau[w]}$
and $\psi_{m[w]}$. We do not consider these mean 0, time-varying
control variables to be paramount for understanding dynamic behavior,
and focus on the cumulative number of impressions.} The corresponding choice-specific value function for this Bellman
equation is
\begin{gather*}
v(x;\epsilon,n^{I},\bar{c},w)=V(x;\epsilon,n^{I},\bar{c},w)+\beta\int\mathbb{E}_{\epsilon^{\prime}}\left[W(\epsilon^{\prime},n^{I\prime},\bar{c},w+1)\right]dF\left(n^{I\prime}|x,n^{I}\right)
\end{gather*}

\subsection{Test and Learn using \citet{zhuo2022exploit}'s UCB Approach \label{online:Test-and-Learn-Zhuo}}

\subsubsection{Approach}

Applying \citeauthor{zhuo2022exploit}'s approach to approximating
the UCB continuation value yields 
\begin{gather*}
\beta\int\mathbb{E}_{\epsilon^{\prime}}\left[W(\epsilon^{\prime},n^{I\prime},\bar{c},w+1)\right]dF\left(n^{I\prime}|x,n^{I}\right)=\mathsf{1}\left(x>0\right)\frac{\lambda}{\sqrt{j}}
\end{gather*}
where $\mathsf{1}(x>0)$ indicates that a subscription is purchased,
and thus new impressions will arrive, $\lambda>0$ is a parameter
to be estimated, and $j$ represents the number of signals received.
In our case, ad impressions are signals, hence the choice-specific
value function becomes\footnote{Following \citet{zhuo2022exploit}, $j=1$ when no signals are observed
(i.e., $n^{I}=0$). In our context, this would be essentially no information
because $j=1$ is the equivalent of receiving one impression (that
is, a very small fraction of the typical 821K impressions in a one-week
subscription). Setting $j=0$ is impracticable because it would imply
an infinite continuation value.} 
\begin{gather*}
v(x;\epsilon,n^{I},\bar{c},w)=V(x;\epsilon,n^{I},\bar{c},w)+\mathsf{1}\left(x>0\right)\frac{\lambda}{\sqrt{1+n^{I}}}.
\end{gather*}
The intuition behind this approach is that sites with no or little
prior advertising receive an increase in continuation value when $n^{I}$
is small, encouraging forward-looking advertisers to explore new sites.
The parameter $\lambda$ indicates the relative importance of the
continuation value; a large estimated value of $\lambda$ suggests
the data are consistent with forward-looking behavior. Variation in
CTRs identify the match parameter, whereas the test and learn parameter
is identified from variation in accumulated impressions, leading to
a different likelihood of subscription purchase.

\subsubsection{Estimation}

Conditional on the state variable for the cumulative number of impressions,
$n^{I}$, the approximate UCB continuation value takes on one of two
possible values. If $x=0$, then the continuation value is $0$, and
if $x>0$, then the continuation value is $\lambda/\sqrt{1+n^{I}}$.
The implications of this for estimation are the following. 

First, when a value of $x=0$ is observed in the data, the upper bound
on $\epsilon$ given in Equation (9) changes to 
\begin{gather*}
ub^{\theta}(x,p)\equiv\log\left(\shortuparrow\!\!p-\frac{\lambda}{\sqrt{1+n^{I}}}-p\right)-\log\left[\check{\mu}_{asw}\zeta_{a}^{-1}\log\left(\frac{t_{s}\!\shortuparrow\!\!x\,\zeta_{a}+1}{t_{s}\,x\,\zeta_{a}+1}\right)\right],
\end{gather*}
and because $p=0$ and $x=0$ in this case, the above can be further
simplified as
\begin{gather*}
ub^{\theta}(x,p)\equiv\log\left(\shortuparrow\!\!p-\frac{\lambda}{\sqrt{1+n^{I}}}\right)-\log\left[\check{\mu}_{asw}\zeta_{a}^{-1}\log\left(\frac{t_{s}\!\shortuparrow\!\!x\,\zeta_{a}+1}{1}\right)\right].
\end{gather*}
One can interpret $-\frac{\lambda}{\sqrt{1+n^{I}}}$ as an amount
by which the price of the shortest subscription is effectively decreased
(relative to the choice not to advertise) under forward-looking behavior.
When $n^{I}=$0 (i.e. an advertiser has yet to try a site), this price
decrease is equal to the monetary value of the time-discounted, net
benefits from future advertising at this site due to having better
information. By effectively lowering the price of the forgone ad subscription
when observing $x=0$, the UCB term lowers, compared to the myopic
case, the upper bound on the range of $\epsilon$s that can rationalize
$x=0$. As the advertiser grows more experienced advertising at the
site, $n^{I}$ will be higher, the value of future information will
be diminished (because the value of reducing uncertainty is diminished),
and the effective decrease in the price of advertising will eventually
go to zero. Second, a similar logic applies when the observed $x>0$
is the smallest available, non-zero subscription length (i.e., $x=\inf\mathcal{X}_{w}\setminus\left\{ 0\right\} $).
In this case (noting that $\downarrow p=0$ and $\downarrow x=0$),
the lower bound on $\epsilon$ given in Equation (8) changes to
\begin{gather*}
\ell b^{\theta}(x,p)\equiv\log\left(p\,-\frac{\lambda}{\sqrt{1+n^{I}}}\right)-\log\left[\check{\mu}_{asw}\zeta_{a}^{-1}\log\left(\frac{t_{s}\,x\,\zeta_{a}+1}{1}\right)\right].
\end{gather*}
The interpretation of the UCB term as a price reduction applies here
as well. Given $x>0$ was observed instead of $x=0$, and in light
of the information gains from advertising, the lower bound of the
$\epsilon$ that rationalizes $x>0$ can be even lower than in the
myopic case. Third, in all other cases, the UCB term differences out
of the expressions for the upper and lower bounds, so these bounds
are unchanged compared to the myopic case.

\subsubsection{Results}

Adopting this approach, several findings emerge. First, the inclusion
of test and learn behaviors into our model does not improve its overall
predictive performance, with $ELPD_{LOO}$ decreasing by $2.4$ with
a standard error of this difference equal to $0.8$. Posterior estimates
of $\gamma$ (prior beliefs on CTRs) are essentially identical to
the null model of no forward-looking behavior, with the p-value from
a two-sided Kolmogorov-Smirnov test equal to $0.994$. A WLS regression
($\hat{\gamma}_{a}^{\mathrm{UCB}}\sim1+\hat{\gamma}_{a}^{\mathrm{}}$)
of the 100 advertiser prior belief estimates in the UCB model, $\hat{\gamma}_{a}^{\mathrm{UCB}}$,
on the 100 prior beliefs in the non-UCB model, $\hat{\gamma}_{a}$
(where weights are the inverse squared standard errors of the UCB
prior belief estimates, $\left(\hat{\sigma}^{UCB}\right)^{-2}$),
yields an intercept of $0.000000001666$ ($0.000000197$) and slope
of $0.9849$ ($0.004384$) with an $R^{2}$ of $0.998$. Third, the
$\lambda$ parameter in the UCB approximation to the value function
is estimated to be $\hat{\lambda}=2.07$ with a posterior standard
deviation of $2.07$. Owing to the dollar metric for the advertiser
utility function, the UCB approximation to the value function can
be interpreted as a \$2.07 (\$2.07) incentive to use test and learn
on new sites (against an average subscription price of \$1,226). Thus,
advertisers in our data place relatively little value on exploration
(i.e., their willingness to pay for information is roughly 0.2\% of
the purchase price).

\subsection{Test and Learn using \citet{Wang_et_al_2025}'s UCB Approach \label{online:Test-and-Learn-Wang}}

\subsubsection{Approach}

The \citet{Wang_et_al_2025} UCB approach models the value of \textquotedblleft exploration\textquotedblright{}
of an option as being proportional to the posterior standard deviation
of the value of that option, yielding
\begin{gather*}
\beta\int\mathbb{E}_{\epsilon^{\prime}}\left[W(\epsilon^{\prime},n^{I\prime},\bar{c},w+1)\right]dF\left(n^{I\prime}|x,n^{I}\right)=\lambda\sqrt{\mathrm{Var}\left[\pi(x)|\epsilon,n^{I},\bar{c},w\right]}
\end{gather*}
where again $\lambda$ is a parameter to be estimated. Advertisers
are uncertain about their match with sites, hence the uncertainty
that drives the value of exploration is ultimately derived from uncertainty
about site-specific CTRs. To simplify the derivation of this uncertainty
term, define $\delta\equiv\check{\delta}c$, where $\delta$ is the
match term defined in Equation (\ref{eq:conditional_match}), and
$z(x)\equiv\zeta^{-1}\log\left(1+tx\zeta\right)$ is the (unscaled)
value of $tx$ ad impressions, so that advertiser payoff in Equation
(\ref{eq:advertiser_payoff}) can be written as $\pi(x)=c\check{\delta}z(x)-p(x)$.
The posterior variance of $\pi(x)$ is 
\begin{align*}
\mathrm{Var}\left[\pi(x)|\epsilon,n^{I},\bar{c},w\right] & =\int_{0}^{1}(\pi(x)-\mathrm{E}[\pi(x)|\epsilon,n^{I},\bar{c},w)])^{2}p(\tilde{c}|n^{I},\bar{c}))d\tilde{c}\\
 & =\int_{0}^{1}\left[\tilde{c}\check{\delta}z(x)-p(x)-\mathrm{E}[\tilde{c}|n^{I},\bar{c}]\check{\delta}z(x)+p(x)\right]^{2}p(\tilde{c}|n^{I},\bar{c}))d\tilde{c}
\end{align*}
where $p(\tilde{c}|n^{I},\bar{c})$ is the PDF of the $\mathrm{Beta}\left(1+n^{I}\bar{c},\frac{1-\gamma}{\gamma}+n^{I}(1-\bar{c})\right)$
distribution. Rearranging terms, we arrive at
\begin{align*}
\mathrm{Var}\left[\pi(x)|\epsilon,n^{I},\bar{c},w\right] & =\left(\check{\delta}z(x)\right)^{2}\int_{0}^{1}\left[\tilde{c}-\mathrm{E}[\tilde{c}|n^{I},\bar{c}]\right]^{2}p(\tilde{c}|n^{I},\bar{c}))d\tilde{c}\\
 & =\left(\check{\delta}z(x)\right)^{2}\mathrm{Var}[\tilde{c}|n^{I},\bar{c}]\\
 & =\left(\check{\delta}z(x)\right)^{2}\frac{\gamma^{2}\left(1+\bar{c}n^{I}\right)\left(1-\gamma-\gamma\bar{c}n^{I}+\gamma n^{I}\right)}{\left(1+\gamma n^{I}\right)^{2}(1+\gamma(1+n^{I}))}\\
 & =\left(\frac{1+\bar{c}n^{I}}{1+\gamma n^{I}}\check{\delta}z(x)\gamma\right)^{2}\frac{\left(1-\gamma-\gamma\bar{c}n^{I}+\gamma n^{I}\right)}{\left(1+n^{I}\bar{c}\right)(1+\gamma(1+n^{I}))}
\end{align*}
The resulting choice-specific value function is thus \allowdisplaybreaks
\begin{align*}
v(x;\epsilon,n^{I},\bar{c},w) & =V(x;\epsilon,n^{I},\bar{c},w)+\lambda\sqrt{\mathrm{Var}\left[\pi(x)|\epsilon,n^{I},\bar{c},w\right]}\\
 & =\frac{1+\bar{c}n^{I}}{1+\gamma n^{I}}\exp\left(\xi+\eta+\phi_{\tau[w]}+\psi_{m[w]}\right)\cdot\exp(\epsilon)\zeta^{-1}\,\log\left(1+tx\zeta\right)-p_{w}(x)+\\
 & \qquad\lambda\frac{1+\bar{c}n^{I}}{1+\gamma n^{I}}\check{\delta}z(x)\gamma\sqrt{\frac{\left(1-\gamma-\gamma\bar{c}n^{I}+\gamma n^{I}\right)}{\left(1+n^{I}\bar{c}\right)(1+\gamma(1+n^{I}))}}\\
 & =\left[\frac{1+\bar{c}n^{I}}{1+\gamma n^{I}}\exp\left(\xi+\eta+\phi_{\tau[w]}+\psi_{m[w]}\right)\cdot\exp(\epsilon)\zeta^{-1}\,\log\left(1+tx\zeta\right)\right]-p_{w}(x)+\\
 & \qquad\bigg[\lambda\frac{1+\bar{c}n^{I}}{1+\gamma n^{I}}\exp\left(\xi+\eta+\phi_{\tau[w]}+\psi_{m[w]}\right)\cdot\exp(\epsilon)\zeta^{-1}\,\log\left(1+tx\zeta\right)\cdot\\
 & \qquad\qquad\sqrt{\frac{\left(1-\gamma-\gamma\bar{c}n^{I}+\gamma n^{I}\right)}{\left(1+n^{I}\bar{c}\right)(1+\gamma(1+n^{I}))}}\bigg]\\
 & =\check{\mu}_{w}\exp(\epsilon)\zeta^{-1}\log(1+tx\zeta)\left[1+\lambda\sqrt{\frac{1-\gamma-\gamma\bar{c}n^{I}+\gamma n^{I}}{(1+\bar{c}n^{I})(1+\gamma(1+n^{I}))}}\right]-p(x)
\end{align*}
This expression differs from Equation (\ref{eq:V_x}) insofar as the
match term is multiplied by $1+\lambda\sqrt{\frac{1-\gamma-\gamma n^{C}+\gamma n^{I}}{(1+n^{C})(1+\gamma(1+n^{I}))}}$.
In the limit of no uncertainty about the component of match related
to CTR, the advertiser will make decisions consistent with Equation
(\ref{eq:V_x}). Prior to advertising, the match value of a site is
$1+\frac{1-\gamma}{1+\gamma}\approx2$ times higher than what is obtained
using Equation (\ref{eq:V_x}).

Because this UCB approach alters the match value of each option, the
lower and upper bounds on $\epsilon$ in Equations (\ref{eq:lb})
and (\ref{eq:ub}) become 
\begin{gather*}
\ell b^{\theta}(x,p)\equiv\log(p\,-\shortdownarrow\!\!p)-\log\left[\check{\mu}_{asw}\zeta_{a}^{-1}\log\left(\frac{t_{s}\,x\,\zeta_{a}+1}{t_{s}\!\shortdownarrow\!\!x\,\zeta_{a}+1}\right)\left(1+\varsigma_{asw}\right)\right]
\end{gather*}
and 
\begin{gather*}
ub^{\theta}(x,p)\equiv\log(\shortuparrow\!\!p-p)-\log\left[\check{\mu}_{asw}\zeta_{a}^{-1}\log\left(\frac{t_{s}\!\shortuparrow\!\!x\,\zeta_{a}+1}{t_{s}\,x\,\zeta_{a}+1}\right)\left(1+\varsigma_{asw}\right)\right]
\end{gather*}
where $\varsigma_{asw}\equiv\lambda\sqrt{\frac{1-\gamma_{a}-\gamma_{a}n_{asw}^{C}+\gamma_{a}n_{asw}^{I}}{(1+n_{asw}^{C})(1+\gamma_{a}(1+n_{asw}^{I}))}}$.
No other modifications to the likelihood are necessary.

\subsubsection{Results}

Results are qualitatively similar to those obtained using the \citet{zhuo2022exploit}
approach, with $ELPD_{LOO}$ decreased by $2.3$ and a standard error
of this difference equal to $0.8$. Estimates of $\gamma_{a}$ deviate
slightly further from the main model estimates, compared to the \citet{zhuo2022exploit}
approach, but the deviations are still very small. A K-S test yields
a p-value of $0.016$, and the analogous regression to the one previously
described yields an intercept of $0.00000021$ ($0.00000019$) and
slope of $0.954$ ($0.0044$). The average $\gamma_{a}$ estimate
differs by $-0.0076\%$ between the main model and this UCB approach.
The posterior mean of $\lambda$ is $0.0692$ ($0.0688$). We again
find no evidence that a specification approximating forward-looking
advertiser behavior fits the data better than a model that does not.

\section{Supply-Side Considerations\label{online:Supply-Side-Considerations-Appendix}}

\subsection{Pricing\label{online-subsec:Pricing}}

It is possible that there are unobserved factors jointly affecting
a publisher's prices and site demand in a given week. If this is the
case, then prices might not be counterfactually invariant to changes
in demand due to information pooling. To better understand if this
is an issue, we conduct a variance decomposition of the original (but
log-transformed) subscription prices at the top 20 sites using all
subscriptions purchased during the six-month estimation period. We
decompose log prices in terms of (log) duration of the subscription,
week, and publisher; as well as four variables representing site-week
demand: the number of advertisers running ads and their cumulative
(pooled) CTR, in both the current and previous week.

Table \ref{tab:Variance-Decomposition-of-Log-Prices} shows the contribution
to price variation from each component. Only the contributions for
duration, week, and publisher (the components in the main price regression
specification) have significant $F$ statistics (all with p-values
less than $2.0\times10^{-16}$). The publisher-week component explains
$3.18\%$ of the observed variation in prices, but uses 400 degrees
of freedom, and thus has $F_{400,722}=0.8253$, $p=0.9841$. All other
components combine to less than $1\%$ of the variation in prices.
As there is negligible variation in prices over time within publisher,
we do not find compelling evidence against the assumption that prices
are counterfactually invariant. In addition, as implied by the publisher-number
of advertisers component explaining only 0.14\% of the variation in
prices, there is no evidence that publisher prices are moderated by
the number of advertisers at the site. That is, there is no evidence
that larger publishers (in terms of advertisers) charge higher prices,
all else equal.

\begin{table}
\begin{centering}
\begin{tabular}{ld{4.0}d{0.4}d{2.1}}
\toprule 
Term & \multicolumn{1}{c}{DF} & \multicolumn{1}{c}{Sum of Squares} & \multicolumn{1}{c}{Variation Explained (\%)}\tabularnewline
\midrule 
Log Subscription Duration & 1 & 104.14 & 6.102\tabularnewline
Week & 27 & 92.94 & 5.446\tabularnewline
Publisher (Site) & 19 & 1327.68 & 77.792\tabularnewline
Publisher-Week & 400 & 54.34 & 3.184\tabularnewline
Publisher-No. Advertisers & 20 & 2.37 & 0.139\tabularnewline
Publisher-Lag No. Advertisers & 20 & 1.09 & 0.064\tabularnewline
Publisher-Cumulative CTR & 20 & 3.32 & 0.194\tabularnewline
Publisher-Lag Cumulative CTR & 20 & 2.00 & 0.117\tabularnewline
Residuals & 722 & 118.84 & 6.963\tabularnewline
\bottomrule
\end{tabular}
\par\end{centering}
\caption{Variance Decomposition of Subscription Prices for the Top 20 Sites
in the Estimation Period.\label{tab:Variance-Decomposition-of-Log-Prices}}
\end{table}

\subsection{Crowding\label{online-subsec:Crowding}}

\begin{figure}[!t]
\begin{centering}
\includegraphics[width=1\textwidth]{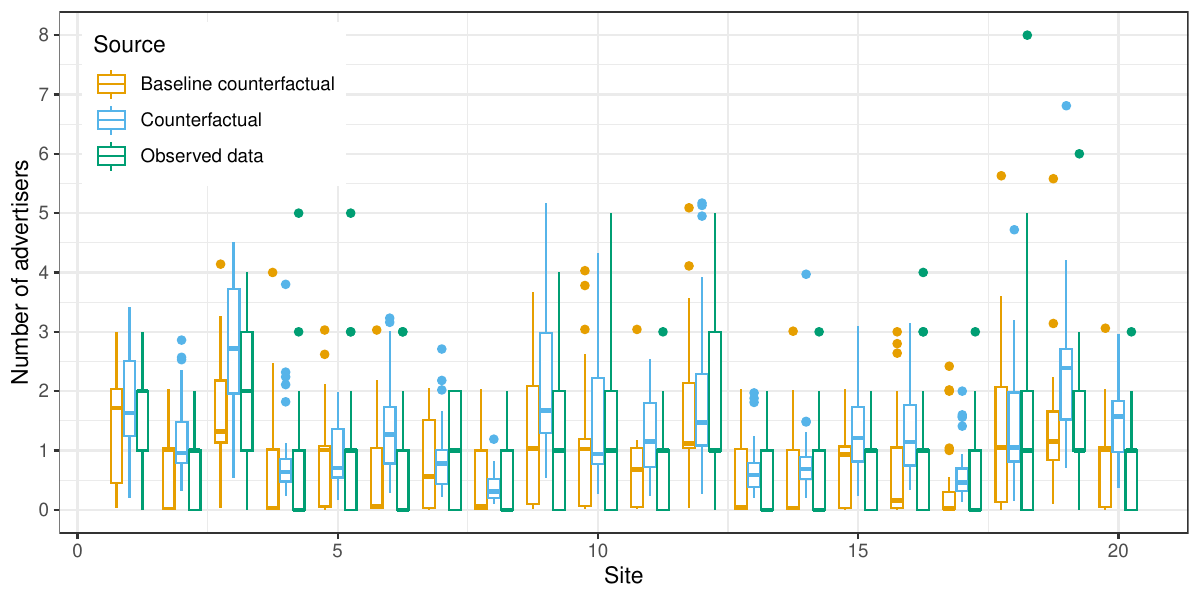}
\par\end{centering}
\caption{The Number of Advertisers On Each Site. The horizontal axis lists
the 20 publishers and the vertical axis is a count of the number of
advertisers on the publisher site. Box plots show the distribution
of the number of advertisers over the weeks in the data in the baseline
and observed counterfactuals, along with the observed data. Boxes
and horizontal lines show the 25th, 50th, and 75th percentiles. Lines
extend to 1.5 times the inner-quartile range, and observations outside
this range are shown as dots. \label{fig:The-Number-of-Advdertisers}}
\end{figure}

To address the effect of crowding, we conduct two analyses. First,
we report the number of instances across sites and weeks with multiple
subscriptions in both estimation and the counterfactual. As depicted
in Figure \ref{fig:The-Number-of-Advdertisers}, in most instances
only one or two advertisers are present at any given point in time
at a given publisher site. Hence, crowding is limited and not likely
to have a large effect on ad effectiveness as defined by clicks.

Second, we regress advertiser clicks by site-week on the number of
advertisers on the site, controlling for site fixed effects and advertiser-specific
click propensity. The latter is defined as the sum of the average
number of clicks over time received by the advertiser across sites
each week. More specifically, the regression is $q_{sw}\sim A_{sw}+\digamma_{s}+\bar{q}_{sw}$,
where $q_{sw}=\sum_{a}\left(n_{asw}^{C}-n_{asw-1}^{C}\right)$ is
the number of incremental clicks across all advertisers with ads running
at site $s$ in week $w$, $A_{sw}$ is the number of advertisers
with ads running at site $s$ in week $w$, and $\digamma_{s}$ is
a fixed effect for site $s$; $\bar{q}_{sw}$ is the average of $\bar{q}_{a}$
for advertisers with ads running at site $w$ in week $w$, and $\bar{q}_{a}$
is advertiser $a$'s average weekly incremental clicks per site. As
reported in Table \ref{tab:Crowding}, the effect of number of advertisers
on a site and the clicks that the advertiser receives is not significant.
Overall, little evidence of crowding exists in our context.

\begin{table}
\centering{}%
\begin{tabular}{ld{-3.4}d{3.2}d{2.2}d{0.2}}
\toprule 
DV: Weekly site-level clicks & \multicolumn{1}{c}{Estimate} & \multicolumn{1}{c}{Std. error} & \multicolumn{1}{c}{t-statistic} & \multicolumn{1}{c}{p-value}\tabularnewline
\midrule 
Intercept & -311.47 & 173.07 & -1.80 & 0.07\tabularnewline
Sum of average clicks among active advertisers & 1.05 & 0.06 & 16.81 & 0.00\tabularnewline
Number of active advertisers & 15.09 & 31.02 & 0.49 & 0.63\tabularnewline
Site fixed effects & \multicolumn{1}{c}{yes} &  &  & \tabularnewline
\midrule
$R^{2}$ & 0.7461 &  &  & \tabularnewline
Observations & \multicolumn{1}{c}{504} &  &  & \tabularnewline
\bottomrule
\end{tabular}\caption{The Effect of Advertiser Crowding at a Publisher Site on Advertiser
Clicks. The table reports a regression of advertiser clicks by site-week
on the number of advertisers, controlling for site fixed effects and
advertiser-specific average clicks across sites.\label{tab:Crowding}}
\end{table}

\section{Additional Analyses\label{online:Additional-Analyses}}

\subsection{Sample Selection\label{online-subsec:Sample-Selection}}

The estimation sample is restricted to the top 20 sites by revenue.
To ascertain the effect of increasing the number of publishers, we
expand the number of sites by 25\% from 20 to 25 and repeat the information
pooling counterfactual analyses in Section \ref{sec:The-Role-of-Information}.
The median increase in publisher revenue decreases from $\$13,558$
($77.7\%$) to $\$12,552$ ($122.3\%$). The median increase in publisher
revenue is smaller, but it represents a larger increase relative to
baseline revenue---this is consistent with the inclusion of smaller
publishers in this analysis. The median advertiser's increase in ad
spend increases from $\$946$ ($98.3\%$) to $\$1,828$ ($229.4\%$),
and the proportion of advertisers whose spending increases under pooling
increases from $79\%$ to $87\%$. These increases in ad spend are
consistent with the larger choice set, as advertisers have a larger
set of publishers to match with.

\subsection{Budget Constraints\label{online-subsec:Budget-Constraints}}

Because we simulate advertiser spending using draws from the (constrained)
posterior distribution of $\epsilon_{asw}$ (as discussed in Section
\ref{subsec:Estimation}), total spend in the baseline condition is
very close to the total observed spend. For example, 22 advertisers
spend exactly the same amount in the baseline condition as they do
in the estimation data, 86 spend between 95\% and 105\% of their observed
budget, and 92 spend between 90\% and 110\%. Moreover, because we
report changes in the median advertiser spend, results are not very
sensitive to the few advertisers predicted to spend beyond their observed
budgets in the baseline and/or counterfactual scenarios.

\begin{figure}[!t]
\begin{centering}
\includegraphics[width=1\textwidth]{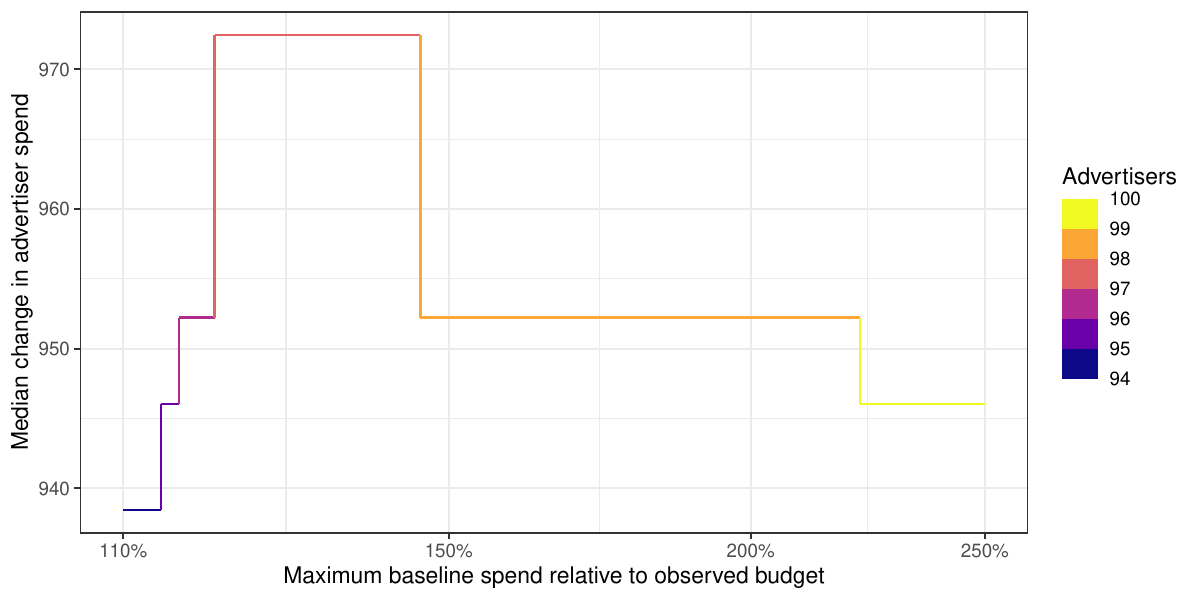}
\par\end{centering}
\caption{Median Change in Advertiser Spend under $C_{P}$ as a Function of
Baseline Spend and Observed Budget. \label{fig:cfl_spend_by_budget}}
\end{figure}

Taking observed budgets as a proxy for budget constraints, Figure
\ref{fig:cfl_spend_by_budget} shows how the median change in advertiser
spend under $C_{P}$ changes when excluding advertisers spending more
than $X\%$ of their observed budget in the baseline scenario. The
idea behind this analysis is to assess the robustness of our findings
to violations of imputed budget constraints by excluding advertisers
who exceed it in the baseline scenario. For example, when considering
only the 94 advertisers whose baseline spend was less than 110\% of
their observed budget, the median increase in spending is $\$938$;
and when considering the 98 advertisers whose baseline spend was less
than 120\% of their observed budget, the median increase in spending
is $\$972$. Thus, the key results in the paper do not rely on advertisers
who exceed the budget constraints implied by their observed spend.

We also perform simulations in which advertisers are constrained to
a particular budget each week, $b_{aw}$, which decrements with each
ad buy. For these simulations, rather than choosing subscription lengths
for each site in a given week independently, advertiser $a$ solves
an optimization problem constrained by the available budget: $\text{\ensuremath{\vec{x}_{aw}}}=\underset{\vec{x}}{\arg\max}\sum_{s}\pi_{asw}\left(x_{s}\right)$,
subject to $\sum_{s}p_{sw}\left(x_{s}\right)\leq b_{aw}$. The GNU
Linear Programming Kit, called via the \texttt{Rglpk} R package interface,
performs the optimization quickly and efficiently \citep{Theussl2023}.

Under the pooling counterfactual without budget constraints in the
main text (Table \ref{tab:Advertiser-Counterfactual-Outcomes}), $79\%$
of advertisers increase spend, with the median advertiser's spend
nearly doubling relative to baseline. When we constrain advertisers
to spend within a budget equal to their observed spend in the data
(in both the baseline and counterfactual), the median advertiser's
spend drops by $14.3\%$ and only $22\%$ of advertisers increase
their spend; moreover, the median change in welfare is $\$0$. Because
overall ad spend drops, the platform has no incentive to pool information.
Indeed, unless some advertisers are able to increase total spend,
platform revenue cannot improve under any counterfactual.

When advertiser budgets are allowed to increase to as much as $150\%$
of the observed budget (in either the baseline or counterfactual),
$46\%$ of advertisers increase their spend under pooling, with a
median advertiser decrease of just $2.4\%$. And as the budget constraint
rises to 300\% of observed spend, $67\%$ of advertisers increase
their spend, and the median advertiser's spend and welfare increase
over baseline by $19.8\%$ and $6.6\%$ respectively. Hence total
platform revenues increase when advertisers are allowed to spend as
much as 300\% of their observed budget and the platform has a positive
incentive to pool information. Notably, although the budget constraint
is set to 300\% of the observed budget, not all advertisers increase
their budgets, and among those that do increase their spend, the median
increase is only $53.1\%$ above total observed spend.

\subsection{Heterogeneity in Match and Prior Beliefs Within Advertiser and Across
Publisher\label{online-subsec:Heterogeneity-in-Priors}}

It is possible for an advertiser to have different beliefs about match
at the publisher level---that is, to have different prior beliefs
across sites about the distribution of factors affecting match. The
main specification accounts for this in Equation (\ref{eq:conditional_match})
through the multiplicative term $\exp(\eta_{s})$. Because there is
no parameter in Equation (\ref{eq:conditional_match}) that is indexed
by both advertiser and site, the relative increase in expected match
at site $s$ versus site $s^{\prime}$ is presumed the same for all
advertisers. Relaxing this assumption requires the alternative assumption
that advertisers are aware of and consider how match differs across
all potential available sites. Yet, it is not clear what type of information
about persistent match---which by definition cannot change after
observing advertising outcomes---would be available to advertisers
prior to placing their first ad with a publisher. In addition to these
informational concerns, there are also insurmountable problems related
to inference. Specifically, for sites where no advertising is observed,
an advertiser-site match term, say $\exp(\upsilon_{as})$, would be
only functionally identified, heavily biased toward $-\infty$, and
thus essentially identical across sites and advertisers (that most
advertiser-site pairs have identical fixed effects contradicts the
notion that private information would lead each advertiser to have
its own persistent valuation for each site). Finally, because it is
not possible to know which advertiser behaviors this term would capture,
it is difficult to defend their counterfactual invariance to the strength
of evidence platforms provide about the efficacy of unused sites.
As a practical matter, the alternative assumption of advertiser-site
specific priors for expected match is an informationally demanding
task for advertisers and, from an inference perspective, could induce
an incidental variables concern in our specification.

\begin{table}
\begin{centering}
\begin{tabular}{ld{4.0}d{0.4}d{2.1}}
\toprule 
Term & \multicolumn{1}{c}{DF} & \multicolumn{1}{c}{Sum of Squares} & \multicolumn{1}{c}{Variation Explained (\%)}\tabularnewline
\midrule 
Advertiser & 99 & 0.0111226 & 92.45\tabularnewline
Residuals & 1900 & 0.0009088 & 7.55\tabularnewline
\bottomrule
\end{tabular}
\par\end{centering}
\caption{Variance Decomposition of Posterior Means of Advertiser-Site Priors,
$\gamma_{as}$.}

\label{tab:anova-site-specific-gamma}
\end{table}

An alternative approach that places slightly less informational requirements
on advertisers is to explore whether and how the assumption of common
prior beliefs across sites affects inference. We therefore estimate
a model wherein advertisers have site-specific CTR priors: $\gamma_{as}=g_{as}/\left(\bar{\gamma}^{-1}+g_{as}\right)$,
with $g_{as}\sim\mathrm{Exponential}\left(1\right)$. In the main
model specification, there are $100$ advertisers, hence there are
$100$ parameters representing the $\gamma_{a}$s. In the augmented
model, because there are $20$ sites, there are $2000$ $\gamma_{as}$
parameters. The following results obtain: First, model fit is marginally
improved, with $ELPD_{LOO}$ decreasing by $20.9$ points (the SE
of the difference is $7.0$) due to adding 1900 additional parameters.
The two models recover similar prior beliefs, with a correlation in
advertiser-level point estimates of $.993$ (i.e., a correlation between
$\hat{\gamma}_{a}$ in the main specification, and $\hat{\gamma}_{a}=\frac{1}{20}\sum_{s}\hat{\gamma}_{as}$
in the augmented specification). A variance decomposition of the point
estimates for the $\gamma_{as}$s in the augmented model (Table \ref{tab:anova-site-specific-gamma})
shows that 92.45\% of the variance of the advertiser-site priors is
explained by advertisers alone. In sum, this analysis largely supports
the assumption that prior beliefs differ predominantly by advertiser,
and that adding private information about publisher heterogeneity
in the advertiser match term has a largely inconsequential effect
on model performance and inference.

\subsection{Estimating Learning Rates (Prior Variances)\label{online-subsec:Estimating-Learning-Rates}}

As an extension to the proposed learning model, one could incorporate
a second parameter, $\kappa$, to determine the prior variance of
$\tilde{c}$, and thus, the rate of advertiser learning. Prior beliefs
in this model are specified as $\tilde{c}|\kappa,\gamma\,\sim\mathrm{Beta}\left(\kappa,\kappa\left(1-\gamma\right)\big/\gamma\right)$,
leading to an updated expected CTR of $\gamma\frac{\kappa+n^{C}}{\kappa+\gamma n^{I}}$.
The value of $\kappa$, however, drops out of the expectation prior
to advertising (i.e., the prior expectation is still $\gamma$ in
this model), and $\kappa$'s effect on the expectation is likely to
be dominated by $n^{C}$ and $n^{I}$ thereafter. Hence, the parameter
$\kappa$ is weakly identified in this setting, so it is normalized
it to one in our main specification. As a robustness check, this section
relaxes that assumption.

We estimate a model in which the prior variance parameter ($\kappa$)
is estimated rather than set to 1. Several findings emerge. First,
the posterior mean is $9.9$, and the inner 90\% interval is $\left[1.4,28.5\right]$
(note: we constrain $\kappa\geq1$). The prior mean for $\kappa$
is $10$; although theoretically identified, the posterior mean differs
little from the prior mean suggesting weak identification in a finite
sample. The $ELPD_{LOO}$ for this model is not better than the model
in which $\kappa$ is set to 1 (worsening by $0.7$, with a standard
error of $0.8$). Second, the posterior means for the $\gamma$s across
the two models are nearly identical, with correlation of $.998$.
Adding $\kappa$ to the model neither improves fit, nor affects inferences
on $\gamma$.

\subsection{Stability of CTRs}

\begin{figure}[!t]
\begin{centering}
\includegraphics[width=1\textwidth]{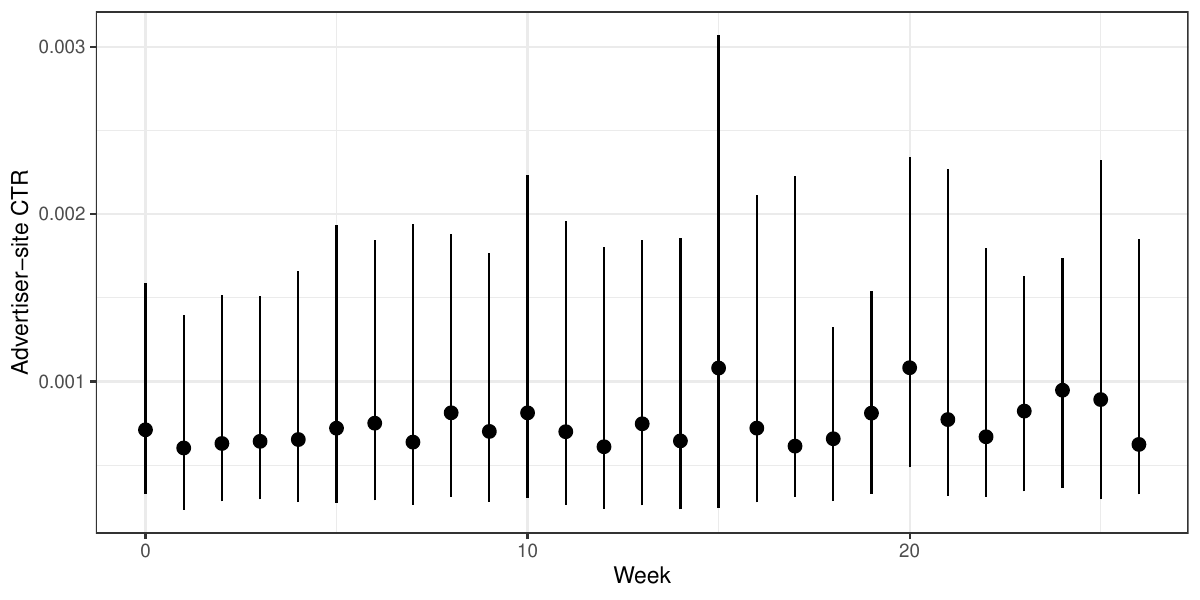}
\par\end{centering}
\caption{Advertiser-site CTRs by Week. Points represent medians and vertical
bars cover the inner 50th percentile. The first 26 weeks of ads served
by the focal cluster of 165 sites are shown for advertisers who placed
at least one ad with a given publisher 26 weeks or more after their
first ad. The timing of ads is normalized relative to when the advertiser
first placed an ad with a given site. \label{fig:ctr_by_week}}
\end{figure}

The model is predicated on an assumption that $c_{as}$ is an invariant
term affecting advertiser-site match. One possibility is that $c_{as}$
should instead be indexed by time to reflect, for example, a long-run
decrease in the marginal value of advertiser $a$'s impressions at
site $s$. We find no evidence in the data supporting the idea that
CTRs vary over time within site-advertiser pair. Using observed impressions
and clicks in the focal cluster of 165 sites, we calculate a weekly
CTR for each advertiser-site pair. To mitigate selective dropout,
we only retain data for advertiser-site pairs with ads running past
26 weeks, and consider how CTRs change over the first 26 weeks. Figure
\ref{fig:ctr_by_week} shows the median and inner 50th percentile
by week for advertiser-site CTRs. The distribution of CTRs in this
group shows no evidence of decline over time.

\end{document}